\documentclass[english,11pt,oneside]{article}
\pdfoutput=1
\usepackage{amsmath}
\usepackage{amsfonts}
\usepackage{amssymb,bbold}
\usepackage{graphicx, rotating}
\usepackage{epstopdf}
\usepackage{epsfig}
\usepackage{latexsym}
\usepackage{multirow}
\usepackage{color}
\usepackage{slashed}
\usepackage[top=3 cm, bottom=3 cm, left=3.1416 cm, right=3.1416 cm]{geometry}
\usepackage{hyperref}
\def\beq{\begin{equation}}
\def\eeq{\end{equation}}
\def\bea{\begin{eqnarray}}
\def\eea{\end{eqnarray}}
\def\ds{\displaystyle}

\newcommand{\no}{\nonumber}

\begin{document}

\title{Resonant Diphoton Phenomenology Simplified}
\author{Giuliano Panico$^a$\footnote{gpanico@ifae.es} , Luca Vecchi$^{b,c}$\footnote{vecchi@pd.infn.it} \ and Andrea Wulzer$^c$\footnote{andrea.wulzer@pd.infn.it}\\
{\small\emph{$^a$ IFAE, Universitat Aut\` onoma de Barcelona, E-08193 Bellaterra, Barcelona, Spain}}\\
{\small\emph{$^b$ SISSA, via Bonomea 265, 34136, Trieste, Italy}}\\
{\small\emph{$^c$ Dipartimento di Fisica e Astronomia, Universit\`a di Padova and}}\\
{\small\emph{INFN, Sezione di Padova, via Marzolo 8, I-35131 Padova, Italy
}}
}
\date{}
\maketitle

\begin{abstract}

A framework is proposed to describe resonant diphoton phenomenology at hadron colliders in full generality. It can be employed for a comprehensive model-independent interpretation of the experimental data. Within the general framework, few benchmark scenarios are defined as representative of the various phenomenological options and/or of motivated new physics scenarios. Their usage is illustrated by performing a  characterization of the $750$~GeV excess, based on a recast of available experimental results.

We also perform an assessment of which properties of the resonance could be inferred, after discovery, by a careful experimental study of the diphoton distributions. These include the spin $J$ of the new particle and its dominant production mode. Partial information on its CP-parity can also be obtained, but only for $J\geq2$. The complete determination of the resonance CP properties requires studying the pattern of the initial state radiation that accompanies the resonant diphoton production.
\end{abstract}

\newpage

\section{Introduction}

The resonant production of a photon pair at hadron colliders is quite a simple process, which we can hope to characterize with
a high degree of generality. To do so, first of all we need to understand the possible initial states that can lead
to the production of the intermediate resonance ${\cal R}$ decaying to $\gamma\gamma$.
If no additional hard objects are present in the final state, which is our working hypothesis, only a few
partonic scattering processes are likely to be relevant, namely the ones involving
gluons ($gg$), quarks ($q{\overline{q}}$, with $q=u,d,c,s,b$) or photons ($\gamma\gamma$).
``Mixed'' situations such as $qg$-initiated production are forbidden by color conservation and by Lorentz symmetry, which requires the heavy resonance $\mathcal{R}$ to have integer spin $J$ (with $J\neq1$ by the Landau--Yang theorem).
Channels of the type $q^\prime{\overline{q}}$ with $q\neq q^\prime$ are strongly disfavored by flavor constraints,
which make very difficult to imagine how a resonance within the energy reach of the LHC might have sizable flavor
non-diagonal couplings to the light quarks. We will thus ignore this possibility in what follows.

Among the other channels that might be considered we can definitely exclude $t{\overline{t}}$, because
$t{\overline{t}}$-initiated production unavoidably comes together with a $t{\overline{t}}$ pair in the final state
from the splitting of the initial gluons, while we choose to limit our analysis to final states with no extra hard objects.
Although similar considerations hold for $b{\overline{b}}$ production, the associated $b$-quarks are typically soft.
Thus they are hard to detect and to identify as $b$-jets and can be easily confused with the radiation
pattern that characterizes the other partonic modes. Still, after the identification of a signal and with large
enough luminosity, checking for the presence of bottom quarks will allow to distinguish the $b{\overline{b}}$
mode from the others. 

Production through Massive Vector Bosons (MVB), namely $W^+W^-$-, $ZZ$- or $\gamma Z$-ini\-tia\-ted processes,
will also be neglected.\footnote{Within the on shell formalism adopted in this paper, MVB production can be  included, but only relying on the Effective $W$ (or $Z$) Approximation (EWA)~\cite{Dawson:1984gx}, which allows to treat the MVB's as partons.}
The MVB processes are marginal to the present study for two reasons. First, they are accompanied by the production
of forward energetic jets from the quark splitting,
which are typically hard enough and not too forward to be detected. MVB production is thus distinguishable from the partonic
processes provided suitable forward jet selection cuts are put in place.
Notice that the situation is different for the $\gamma\gamma$ production mode because the photon is massless
and thus the $p_\bot$ of the emission is only cut-off by the proton mass.
The QCD jets from $\gamma\gamma$ fusion are thus softer than the MVB ones and difficult to detect. Actually, the $\gamma\gamma$ radiation pattern is even softer (and possibly even consist, in the elastic scattering regime,
of just two extremely forward protons) than the one associated with the other partonic processes $gg$ and $q{\overline{q}}$,
giving a possible handle to pin it down~\cite{Harland-Lang:2016qjy}.
The second reason to neglect the MVB processes is the fact that the photon parton distribution function (PDF),
again because of the lack of a hard low-$p_\bot$ cut-off in the photon splitting,
is larger than the MVB one.\footnote{Notice that these considerations are qualitative because the photon PDF, differently from the ones
of MVB's, receives non-perturbative contributions at the QCD scale. A quantitative confirmation comes from
a recent photon PDF calculation \cite{Martin:2014nqa} and (large error) measurements~\cite{Ball:2013hta}.}
This makes MVB processes also quantitatively marginal.
An exception is the situation in which the couplings of $\mathcal{R}$ to MVB's are much larger than
the $\gamma\gamma$ one, in which case, however, resonance searches in MVB final states are much more effective.

On top of the analysis of the possible production channels, a full study of a resonant diphoton process
also requires a characterization of the cross section and kinematical distributions of the signal.
Providing this characterization is the main aim of the present paper. As we will discuss in details, our analysis allows to
derive a simple phenomenological parametrization that can be used to describe resonances with arbitrary
(integer) spin and CP parity, produced in any of the $gg$, $q{\overline{q}}$ and $\gamma\gamma$ partonic channels
described above. For definiteness, although we will discuss our formalism in full generality, for the explicit examples
we will focus on the commonly considered cases of resonances with spin $J=0$ and $J=2$
and on a more exotic possibility, $J=3$, which provides a peculiarly simple collider phenomenology.

Our characterization of the diphoton signal is based on symmetries (see e.g.~\cite{Trueman:1978kh,Dell'Aquila:1985ve} for earlier references and~\cite{Choi:2002jk,Gao:2010qx} for more recent ones) and is not new from the technical point of view, since it closely follows the strategy employed for the experimental studies of the Higgs boson $J^{\textrm{CP}}$ properties (see for instance ref.~\cite{Khachatryan:2014kca,Choi:2012yg}). It however provides a new, simple and comprehensive way to parametrize a possible signal excess in diphoton production, allowing to encompass in a unified way the variety
of theoretical origins of the intermediate resonance ${\cal R}$.~\footnote{To be more specific, the relevance of the basis functions ${\cal D}_{|m|,S}^{(J)}(\theta)$ introduced in eq.~(\ref{D}) was previously appreciated for instance in~\cite{Gao:2010qx,Choi:2012yg} whereas the results of Appendix~\ref{effective_ops} can be recovered as particular limits of the analysis of~\cite{Gao:2010qx}. 
On the other hand, the general expression~(\ref{XS1}) for $in\to{\cal R}\to\gamma\gamma$ as a function solely of the independent probabilities ${\cal{P}}^{in}_{|m|\,S}$, the characterization of the various $in$ channels described in section~\ref{sec:LHCcs}, and all the results of section~\ref{sec:bench} (including the identification of appropriate benchmark models for the diphoton resonance and their analysis in terms of ${\cal{P}}^{in}_{|m|\,S}$) are new. } Our approach is particularly convenient in scenarios that are difficult
to fully describe through explicit models, as could be for a generic spin-$2$ resonance or for higher-spin states
(as for instance $J=3$) which can not be described within an effective Lagrangian formalism. The framework, nevertheless,
remains useful also in the simpler $J=0$ case thanks to its unified treatment of the various production channels.

The first step for the characterization of the signal properties is the classification of the partonic cross sections. Due to the simplicity of the $2 \rightarrow 2$ scattering process, the only relevant kinematic variable at the partonic level is the center-of-mass (COM) scattering angle $\theta$. The form of the partonic cross-section, namely its dependence on $\theta$ is strongly constrained by angular momentum conservation. This  observation allows to parametrize the decay distributions of the resonance in terms of only $5$ basis functions of $\theta$, whose explicit form is dictated by the resonance spin~$J$. The number of independent basis functions decreases to $3$ in the case of $gg{\textrm{/}}\gamma\gamma$ production and to $4$ in the $q{\overline{q}}$ channels. Further simplifications emerge if $J$ is odd and, of course, if $J=0$, in which case the $5$ functions collapse to a constant leading to the well-known result that scalars (or pseudo-scalars) decay in a spherically symmetric way. The second step for the signal characterization is to convolute the partonic cross section with the PDF's which are appropriate for each partonic initial state. The PDF's affect the overall signal normalization through the parton luminosity factor, which is of course very different for the various production modes. Moreover they considerably affect the dependence of the cross-section on the collider energy, which is a crucial information to combine $8$ and $13$~TeV LHC searches. Finally, the PDF's determine the distribution of the  COM rapidity in the laboratory frame, which in turns affects the angular distributions of the final state photons. This opens up the possibility of distinguishing different production modes by diphoton distributions measurements.

The paper is organized as follows. In section~\ref{sec:general_framework} we introduce our framework along the lines mentioned above, in a way that allows semi-analytical (because of the required PDF input) calculations of the signal rate and distributions in terms of the parameters that control the on-shell resonance production and decay Feynman amplitudes. The translation of the latter parameters into effective operator coefficients, which straightforwardly allows to implement our signal in an event generator in order to deal with QCD radiation and detector effects, is reported in Appendix~\ref{effective_ops} for $J=0$ and $J=2$ resonances. In the fully general case, in which all the $7$ production modes are active and no further assumption is made on the resonance couplings, the proliferation of free parameters makes the problem untreatable. Therefore in section~\ref{three} we define a set of representative benchmark scenarios, whose number and variety should be sufficient to provide a wide enough coverage of the various phenomenological options. These scenarios are analyzed by recasting, with a strategy described in Appendix~\ref{app:statistics}, available $8$ and $13$~TeV experimental searches. Rather than aiming to fully quantitative results, which might be only obtained by the experimental collaborations, the goal of this study is to illustrate the usage of our benchmarks to characterize possible signals such as the popular $750$~GeV excesses. Still, we will be able to reach semiquantitative conclusions on the viability of our scenarios. In section~\ref{four} we report our conclusions and a preliminary assessment of the additional information which can be extracted from the study of initial state radiation emission. A complete discussion of the latter point is left for future work.

\section{General framework}\label{sec:general_framework}

\begin{figure}[t]
\begin{center}
\includegraphics[width=8cm]{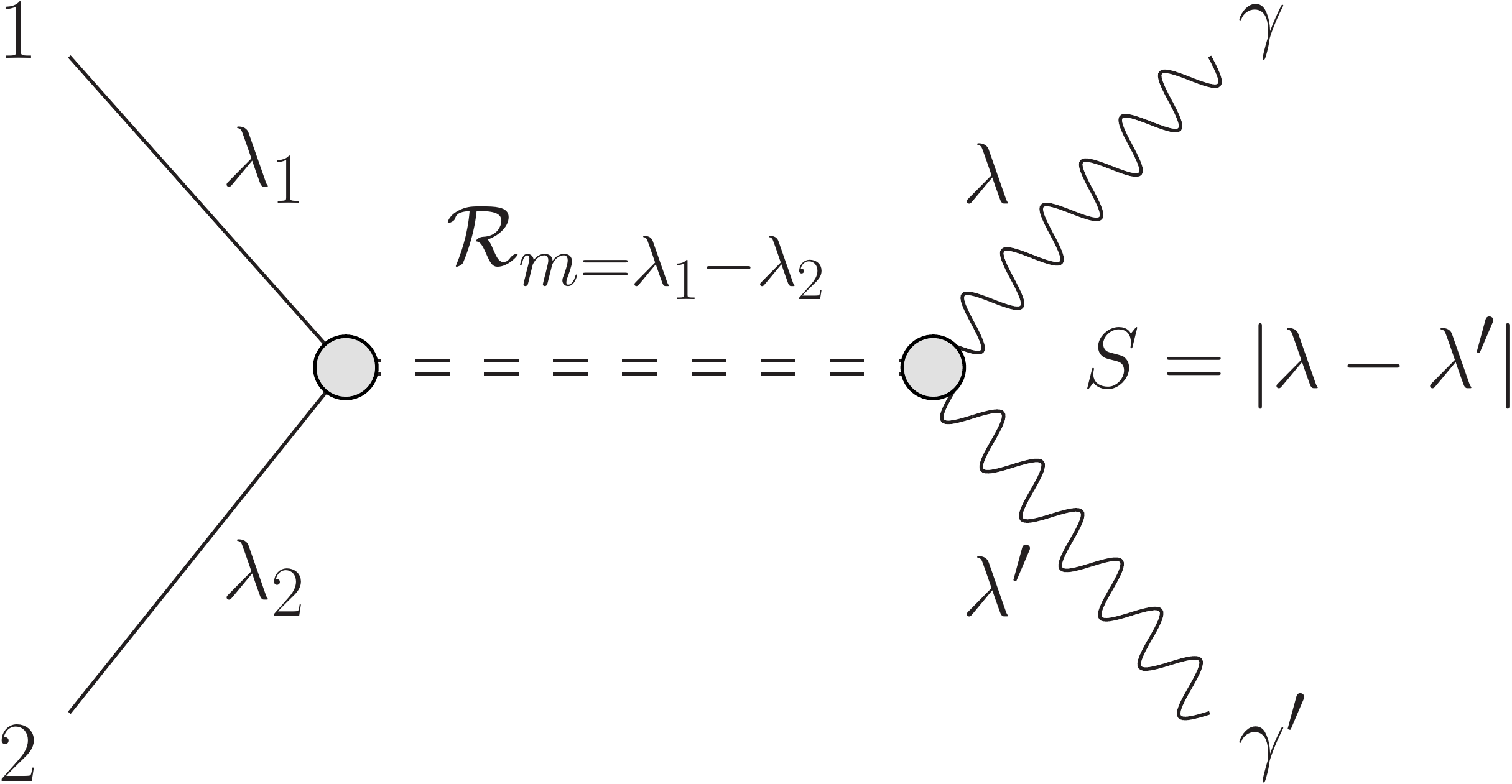}
\caption{Schematic plot of resonant diphoton production. The incoming partons (of helicity $\lambda_{1,2}$) annihilate into a resonance of spin $J$ (and spin-projection $m=\lambda_1-\lambda_2$ along the beam axis) that subsequently decays into two photons with helicities $\lambda$ and $\lambda'$. We denote as $S=|\lambda-\lambda'|$ the absolute value of the spin of the diphoton system along the decay axis.
}\label{Feyn}
\end{center}
\end{figure}

We consider a resonance ${\mathcal{R}}$ of integer spin $J$, produced at the LHC out of a given $2$-partons initial state ${{in}}=\{gg,q{\overline{q}},{\overline{q}}q,\gamma\gamma\}$ and decaying to $\gamma\gamma$ as depicted in figure~\ref{Feyn}.\footnote{Initial partons are ordered by the direction they come from, this is why $q{\overline{q}}$ and ${\overline{q}}q$ are distinct $in$ states.} We start our discussion from the fully polarized scattering process and denote by $\lambda_1$ and $\lambda_2$ the helicities of the incoming partons, $\lambda$ and $\lambda'$ those of the final state photons. These helicities cannot be measured at the LHC and we will eventually have to sum/average over them to obtain the cross-section.\footnote{We assume that it will never be possible to measure photon polarizations at the LHC and we restrict our attention to inclusive $\gamma\gamma$ production. The exclusive case, in which we imagine having access to the radiation from the initial state, is briefly discussed in section~\ref{four}.} Conservation of angular momentum along the beam direction implies that only a single spin component of the resonance can contribute to the partonic process, namely the one with spin projection $m=\lambda_1-\lambda_2$ along the beam axis oriented in the direction of parton ``$1$''. Thus, the resonance production process can be fully described by a set of dimensionless coefficients $A^{in}_{\lambda_1\lambda_2}$ which parametrize the corresponding Feynman amplitudes as
\beq
\ds
{\mathcal{A}}\left([{in}]_{\lambda_1,\lambda_2}\hspace{-3pt}\to{\mathcal{R}}_{m}\right)=M A^{in}_{\lambda_1,\lambda_2}\delta_{m,\lambda_1-\lambda_2}\,,
\eeq
where $M$ is the resonance mass. The helicities $\lambda_{1,2}$ can assume the values $\lambda_{1,2}=\pm 1$ for $in=gg$ or $in=\gamma\gamma$ and $\lambda_{1,2}=\pm 1/2$ for $in=q{\overline{q}}{\textrm{/}}{\overline{q}}q$. Correspondingly, only resonances with $m=0,\pm2$ and $m=0,\pm1$ can be produced, respectively, in the bosonic and fermionic channels. Not all the four complex coefficients $A^{in}_{\lambda_1\lambda_2}$ we have for each production mode are independent. Invariance of the amplitudes under a $\pi$ rotation in a direction orthogonal to the beam implies
\beq\label{rotrel}
\ds
A^{gg{\textrm{/}}\gamma\gamma}_{\lambda_1,\lambda_2}=(-)^JA^{gg{\textrm{/}}\gamma\gamma}_{\lambda_2,\lambda_1}\,,\;\;\;\;\;
A^{{\overline{q}}q}_{\lambda_1,\lambda_2}=(-)^{J} A^{q{\overline{q}}}_{\lambda_2,\lambda_1}\,,
\eeq
where in the first equality we implicitly made use of the fact that the $gg$ and $\gamma\gamma$ states are made of indistinguishable particles.

In the case of the resonance production, which we discussed until now, the incoming partons momenta are completely fixed in the COM frame, thus it is trivial that the amplitudes can be parametrized in terms of few constant coefficients. The situation is different for the resonance decay process, which depends on the kinematical variables of the $\gamma\gamma$ final state and in particular on the COM scattering angle $\theta$. Still, each polarized decay amplitude can be parametrized by a single constant because the angular dependence is completely determined, and encapsulated in the so-called ``Wigner $d$-matrices'' $d^J_{m,m'}(\theta)$~\cite{wiki}. The point is that by a rotation one can connect a $\gamma\gamma$ state with arbitrary polar and azimuthal angles $\theta$ and $\phi$ to a photon pair moving along the beam axis and obtain the angular dependence of the amplitude from the matrix elements of the rotation matrix among the resonance spin eigenstates. The result reads (see for instance \cite{Choi:2002jk})
\beq\label{Wigner}
{\mathcal{A}}({\mathcal{R}}_m\to [\gamma\gamma]_{\lambda \lambda'}\hspace{-2pt})=e^{i(m-\lambda+\lambda')\phi}
{d}^{J}_{m,\lambda-\lambda'}
M\cdot(-)^JA^{\gamma\gamma}_{-\lambda,-\lambda'}\,,
\eeq
where we made use of the CPT symmetry to relate (up to phases, which eventually produce the $(-)^J$ factor) the amplitude coefficients of the ${\mathcal{R}}\to\gamma\gamma$ decay to those associated with the production process $\gamma\gamma\to{\mathcal{R}}$. Therefore describing the resonance decay does not require introducing new parameters.

The set of processes we are considering is thus fully characterized, taking into account the relations in eq.~(\ref{rotrel}), by a rather small number of parameters shown in table~\ref{pars}. Namely, we have in general $4$ complex parameters for the $q{\overline{q}}$ (and ${\overline{q}}q$) production, $3$ complex parameters describing $gg{\textrm{/}}\gamma\gamma$ if $J$ is even and only $1$ complex parameter if $J$ is odd. For $J=0$, the ``$+-$'' and ``$-+$'' amplitudes vanish and we are left with $2$ complex parameters for $gg{\textrm{/}}\gamma\gamma$ and again $2$ for the $q{\overline{q}}$ channels. The case $J=1$ is not worth discussing because the decay to $\gamma\gamma$ (and the production from $gg$) is forbidden by the Landau-Yang theorem, or equivalently by noting that also $a_2^{g{\textrm{/}}\gamma}$ vanishes in this case (see table~\ref{pars}) because of angular momentum conservation. 

It is important to remark that the derivations above are completely model-independent as they only rely on the invariance under rotations and CPT, which are symmetries of any relativistic quantum theory of particles. In particular they do not rely on the CP symmetry, therefore our results hold irregardless of the resonance CP-parity and even of whether CP is at all a symmetry or not. If CP is a symmetry, we get the additional constraint
\beq\label{cprel}
\ds
A^{in}_{\lambda_1,\lambda_2}=\rho_{\textrm{CP}}A^{in}_{-\lambda_2,-\lambda_1}\,,
\eeq
where $\rho_{\textrm{CP}}=\pm1$ is the intrinsic CP-parity of the resonance. Therefore only some of the parameters, denoted as untilded $a$'s in table~\ref{pars}, survive for a CP-even resonance and only tilded ones in the CP-odd case. Sizable tilded and untilded parameters would be simultaneously present only if the CP symmetry was badly broken by the resonance couplings.

\begin{table}[t]
\begin{center}
\begin{tabular}{c|c} 
 $\ds{\bf {J=2k}}$ & $\ds{\bf {J=2k+1}}$\\[0pt]
\hline\\[-10pt] 
${
\begin{array}{l}
A_{++}^{gg{\textrm{/}}\gamma\gamma}\hspace{-4pt}=a_0^{g{\textrm{/}}\gamma}+i\, {\widetilde{a}}_0^{g{\textrm{/}}\gamma}\\[5pt]
A_{--}^{gg{\textrm{/}}\gamma\gamma}\hspace{-4pt}=a_0^{g{\textrm{/}}\gamma}-i\, {\widetilde{a}}_0^{g{\textrm{/}}\gamma}\\[5pt]
A_{+-}^{gg{\textrm{/}}\gamma\gamma}\hspace{-4pt}=\hspace{-2pt}A_{-+}^{gg{\textrm{/}}\gamma\gamma}\hspace{-4pt}=a_2^{g{\textrm{/}}\gamma}\\[5pt]
A_{++}^{q{\overline{q}}}\hspace{-4pt}= a_0^q+i\, {\widetilde{a}}_0^q \\[5pt]
A_{--}^{q{\overline{q}}}\hspace{-4pt}= a_0^q-i\, {\widetilde{a}}_0^q \\[5pt]
A_{+-}^{q{\overline{q}}}\hspace{-4pt}=a_1^{q}\\[5pt] 
A_{-+}^{q{\overline{q}}}\hspace{-4pt}=a_{-1}^{q}
\end{array} 
}$
& 
${
\begin{array}{l}
A_{++}^{gg{\textrm{/}}\gamma\gamma}\hspace{-4pt}=0\\[5pt]
A_{--}^{gg{\textrm{/}}\gamma\gamma}\hspace{-4pt}=0\\[5pt]
A_{+-}^{gg{\textrm{/}}\gamma\gamma}\hspace{-4pt}=\hspace{-2pt}-A_{-+}^{gg{\textrm{/}}\gamma\gamma}\hspace{-4pt}=a_2^{g{\textrm{/}}\gamma}\\[5pt]
A_{++}^{q{\overline{q}}}\hspace{-4pt}= a_0^q+i\, {\widetilde{a}}_0^q \\[5pt]
A_{--}^{q{\overline{q}}}\hspace{-4pt}= a_0^q-i\, {\widetilde{a}}_0^q \\[5pt]
A_{+-}^{q{\overline{q}}}\hspace{-4pt}=a_1^{q}\\[5pt] 
A_{-+}^{q{\overline{q}}}\hspace{-4pt}=a_{-1}^{q}
\end{array} 
}$
\end{tabular}
\caption{Amplitude coefficients expressed in terms of a set of complex parameters ``$a$''. Untilded and tilded parameters are, respectively, CP-even and CP-odd. For shortness, $+1$ and $+1/2$ helicities (which are appropriate for $gg/\gamma\gamma$ and $q{\overline{q}}$ initial states, respectively) are both denoted as ``$+$'' and the same for ``$-$''.
\label{pars}}
\end{center}
\end{table}

We stress that the ``$a$'' (and ``${\widetilde{a}}$'') coefficients in table~\ref{pars} are, in general, complex numbers.\footnote{In spite of the fact that they were erroneously taken real in the first version of the manuscript. We thank R.~Rattazzi for pointing this out to us.} However they become real when the resonance production/decay processes are induced by heavy mediators. Establishing experimentally whether they are real or not would therefore allow us to verify or falsify this hypothesis. 
In order to appreciate this claim, we notice that if the resonance couplings are mediated by the exchange of heavy particles it is possible to integrate them out, giving rise to a set of local operators (contact interactions) that induce resonance production and decay. The heavy-mediator condition can thus be equivalently formulated as the hypothesis that the production/decay amplitudes are well described by a contact interaction at Born level, i.e. by the matrix element of a local Hermitian operator, in which case the CPT symmetry, combined with eq.~(\ref{rotrel}), gives a relation
\beq\label{cptrel}
\ds
A^{in}_{\lambda_1,\lambda_2}=\left[A^{in}_{-\lambda_2,-\lambda_1}\right]^*\,.
\eeq
It is easy to check that this condition implies that the $a$'s in table~\ref{pars} are real. If instead the resonance couplings are due to light particles loops, imaginary parts will arise in the amplitudes, by the optical theorem, due to the propagation of on-shell intermediate states. Establishing whether the $a$'s are real or not would thus give us relevant information on the resonance dynamics. However, this will turn out to be impossible through the measurement of  unpolarized inclusive diphoton production distributions, which are our main target. Indeed, restricting to real $a$'s is enough to span the whole variety of kinematical distributions one would obtain even for general complex $a$'s. We will briefly come back to this point in section~\ref{four}. 

A priori, the parametrization of the resonance production and decay amplitudes provided in table~\ref{pars} might still be redundant because they solely followed from rotation and CPT invariance. In principle, further constraints might arise by requiring invariance of the amplitudes under the complete Lorentz group. This is however not the case, as explicitly shown in Appendix~\ref{effective_ops} for $J=0$ and $J=2$ resonances.\footnote{The case $J=3$ has also been checked, but it is not discussed in the Appendix.} In the Appendix, we classify all the Lorentz-invariant terms, expressed as functions of the $4$-momenta of the resonance and $in$ particles and of their polarization vectors or spinor wave functions, which can appear in the polarized amplitudes. The coefficients of these Lorentz-invariant terms are found to be in one-to-one correspondence with the parameters in table~\ref{pars}, showing that no further restrictions emerge from imposing the full Lorentz symmetry. Moreover, Lorentz-invariant amplitudes are easily mapped to Lorentz- and gauge-invariant operators and therefore another result of the Appendix (see eq.s~(\ref{eq:L0}) and (\ref{eq:L2})) is to relate the phenomenological parameters $a_i, \widetilde a_i$, and in turn the $A_{\lambda_1,\lambda_2}^{in}$'s, to the couplings of a phenomenological Lagrangian.\footnote{Notice that the correspondence among the Lorentz-invariant terms in the (on-shell) amplitude decomposition and the operators is not at all one-to-one. Namely, infinitely many operators reduce, on-shell, to a single term in the amplitude. The simplest set of operators, just sufficient to produce arbitrary on-shell amplitudes, is selected in the Appendix.} This is required for the implementation of our parametrization in a multi-purpose event generator. {Consistently with the discussion following (\ref{cptrel}), if the phenomenological Lagrangian is taken to be Hermitian (i.e. the only phenomenologically relevant states are ${\cal R}, in, \gamma\gamma$) the amplitude coefficients obey eq.~(\ref{cptrel}) and the $a$'s are real, as expected.} In the Appendix we focused on $J=0$ and $J=2$ resonances because higher spin particles are anyhow not implemented in multi-purpose event generators. Complete simulations for $J\geq3$, taking properly into account soft QCD radiation, hadronization and detector effect would thus require a different approach, based on matrix-element reweighting techniques as discussed in the next section.

\subsection{Partonic cross sections}\label{sec:partonic}

We are now in the position of constructing, with the amplitude coefficients as building blocks, the partonic unpolarized cross-section of the complete $2\to2$ reaction $in\to\gamma\gamma$. This will allow us to identify the combinations of amplitude coefficients that appear in the unpolarized cross-section and will suggest a convenient phenomenological parametrization of the signal, to be employed for the experimental characterization of the resonance properties. 

The $in\to\gamma\gamma$ Feynman amplitude is the product of the production and decay amplitudes, times the Breit-Wigner propagator of the resonance
\begin{equation}
{\cal A}(in \rightarrow {\cal R} \rightarrow \gamma\gamma) = \sum_m {\cal A}(in \rightarrow {\cal R}_m)
\frac{1}{\hat{s} - M^2 + i M \Gamma}
{\cal A}({\cal R}_m \rightarrow \gamma\gamma) \,,
\end{equation}
where $\Gamma$ is the total resonance width and ${\hat{s}}$ is the partonic COM energy squared. The Breit-Wigner propagator produces, in the amplitude squared, a factor of $\pi/(\Gamma M)$ times the normalized Breit-Wigner distribution ${\textrm{BW}}(\hat{s})$. The partonic cross-section thus reads
\beq\label{eq:PartonSigma}
\ds
\frac{d\hat\sigma_{in}}{d\cos\theta}=M^2{\textrm{BW}}(\hat{s})\frac{d{\bar{\sigma}}_{in}}{d\cos\theta}\simeq M^2\delta(\hat s-M^2)\frac{d{\bar{\sigma}}_{in}}{d\cos\theta}\,,
\eeq
having reabsorbed in ${\bar{\sigma}}_{in}$ some factors, and in particular the dependence on $\Gamma$. The second equality in the equation holds for a narrow resonance, namely in the limit $\Gamma\hspace{-1pt}/\hspace{-1pt}M\hspace{-2pt}\to\hspace{0pt}0$. In that limit ${\bar{\sigma}}_{in}$ assumes, as we will readily see, the physical meaning of the signal cross-section for unit parton luminosity at the resonance mass, namely for $[\tau\,{d}{\mathcal{L}}_{in}/d\tau]|_{\tau=M^2/s}=1$. A compact expression for $d{\bar{\sigma}}_{in}/d\cos\theta$ (see eq. (\ref{XS}) below) may be obtained as follows.

As previously explained, each polarized $in\to\gamma\gamma$ process is mediated by a single resonance spin $m=\lambda_1-\lambda_2$. Therefore its angular dependence is fixed by the Wigner formula (\ref{Wigner}) to be the square of the associated Wigner matrix, $[{d}^{J}_{m,\lambda-\lambda'}]^2$.
By summing the polarized cross sections over $m$, $\lambda$ and $\lambda'$ we obtain the unpolarized one, expressed as a sum of known functions of the COM scattering angle $\theta$ weighted by the square of the corresponding polarized production and decay amplitudes. 
The polarized production amplitudes can be traded for the resonance production cross section, whereas the decay amplitudes
can be traded for the branching ratios.

In the sum, several terms can be grouped together by proceeding as follows. We first sum over the photons helicities $\lambda$ and $\lambda'$ and notice that the $++$ and $--$ terms in the sum produce the same angular function, $[{d}^{J}_{m,0}]^2$, while the $+-$ and $-+$ ones have identical coefficients $|A^{\gamma\gamma}_{+-}|^2=|A^{\gamma\gamma}_{-+}|^2$ by eq.~(\ref{rotrel}) and can thus be collected in a single term with angular dependence $[{d}^{J}_{m,+2}]^2+[{d}^{J}_{m,-2}]^2$. This allows us to cast the double $\lambda,\,\lambda'$ sum into a single sum over $S=|\lambda-\lambda'|=0,2$, with angular dependence $[{d}^{J}_{m,S}]^2+[{d}^{J}_{m,-S}]^2$. In order to deal with the sum over the initial state polarizations $\lambda_1$ and $\lambda_2$ we exploit the property of Wigner matrices ${d}^{J}_{m,m'}=(-)^{m-m'}{d}^{J}_{-m,-m'}$ to prove that
\beq\label{D}
\ds
[d^J_{m,S}]^2+[d^J_{m,-S}]^2=[d^J_{-m,S}]^2+[d^J_{-m,-S}]^2\equiv\frac2{2J+1}{\cal D}_{|m|,S}^{(J)}(\theta)\,.
\eeq
{The functions ${\cal D}_{|m|,S}^{(J)}(\theta)$ have also appeared in previous work, see e.g.~\cite{Choi:2012yg}. Here we chose to normalize them to unity in the integration domain $\cos\theta\in[0,1]$, which is the appropriate one since the final state photons are indistinguishable particles.} The above equation ensures that terms in the $\lambda_{1,2}$ sum with a given value of $m=\lambda_1-\lambda_2$ have the same angular dependence of those with the opposite value, so that the two can be grouped in a single term. The double sum over the initial state polarization thus becomes a single sum over the absolute value of $m$, $|m|=0,1,2$. 

\begin{table}[t]
\begin{center}
\begin{tabular}{c||l} 
$\ds{\bf J=0}$ & $\ds{\cal D}_{0,0}^{(0)}=1$\\[5pt]
\hline\\[-5pt]
${\bf J=2}$ & $\ds{\cal D}_{|m|,S}^{(2)}=\left[
{\begin{array}{cc}
\ds\frac{5}{4}(3c^2-1)^2 & \ds\frac{15}{8}s^4\\[10pt]
\ds\frac{15}{2}s^2c^2&\ds\frac{5}{4}s^2(1+c^2)\\[10pt]
\ds\frac{15}{8}s^4 & \ds\frac{5}{16}(1+6c^2+c^4)
\end{array}}
\right]$\\[35pt]
\hline\\[-5pt]
${\bf J=3}$ & $\ds{\cal D}_{|m|,S}^{(3)}=\left[{\begin{array}{cc}
\ds\frac{7}{4}c^2(3-5c^2)^2 & \ds\frac{105}{8}s^4c^2\\[10pt]
\ds\frac{21}{16}s^2(5c^2-1)^2 & \ds\frac{35}{32}s^2(1-2c^2+9c^4)\\[10pt]
\ds\frac{105}{8}s^4c^2& \ds\frac{7}{16}(4-15c^2+10c^4+9c^6)
\end{array}}\right]$ \\[35pt]
\end{tabular}
\caption{The ${\cal D}$ functions for $J=0,2,3$. For brevity we defined $s\equiv\sin\theta$ and $c=\cos\theta$.
\label{dD}}
\end{center}
\end{table}

Since $S=0,2$ ranges over two values and $|m|=0,1,2$, six terms are present in the sum, each characterized by its own angular distribution ${\cal{D}}^{(J)}_{|m|,S}(\theta)$. Notice however that only four of the six terms can be simultaneously turned on in a given partonic process because $|m|=0,1$ for $in=q{\overline{q}}$ and $|m|=0,2$ for $in=gg{\textrm{/}}\gamma\gamma$. Nevertheless we will momentarily retain the six of them for a more concise exposition. 

\begin{figure}[t]
\begin{center}
\includegraphics[width=7.5cm]{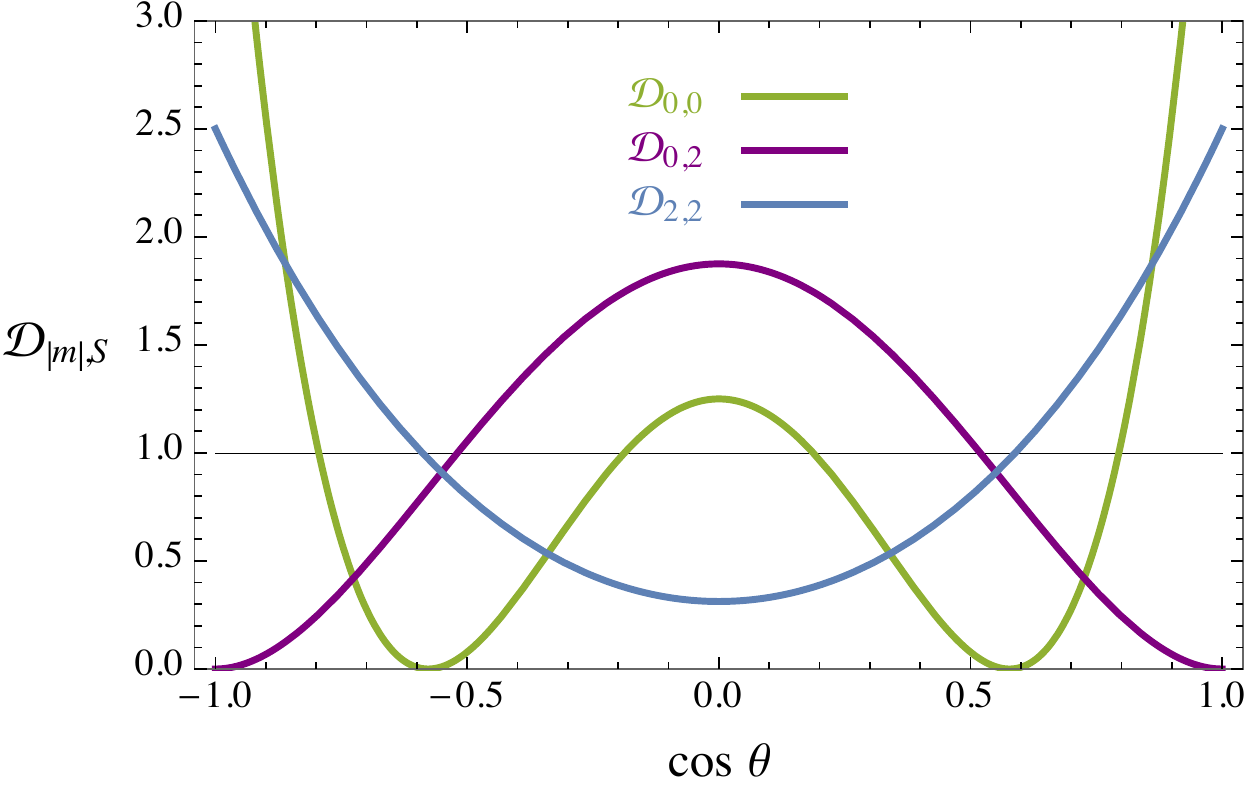}~~~~\includegraphics[width=7.5cm]{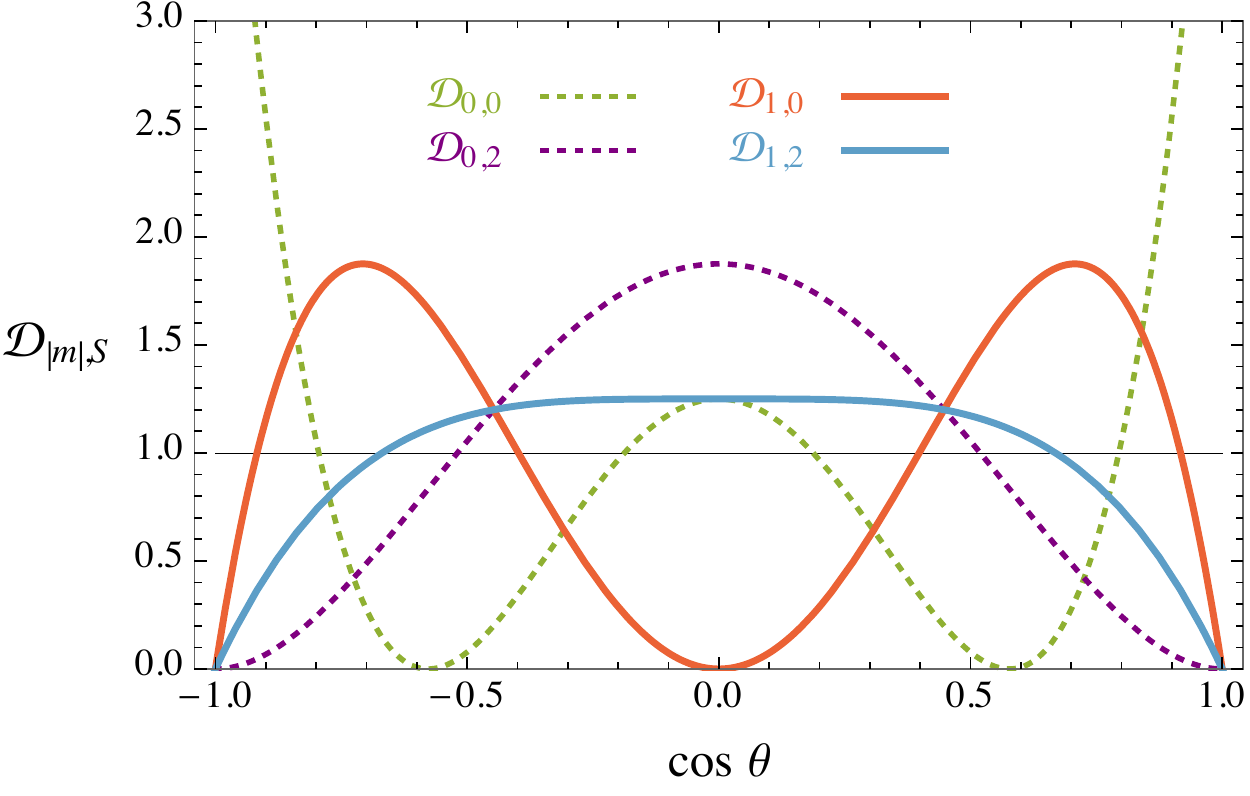}
\caption{The ${\cal{D}}$ distributions relevant for $gg$/$\gamma\gamma$ (left) and $q{\overline{q}}$ (right) production at $J=2$. Notice that ${\cal{D}}^{(2)}_{0,0}$ and ${\cal{D}}^{(2)}_{0,2}$ can appear in both production modes.}\label{DPLOT}
\end{center}
\end{figure}

The unpolarized cross section can finally be written as
\beq
\label{XS}
\ds
\frac{d\bar\sigma_{in}}{d\cos\theta}
=\sum_{|m|,S}\bar\sigma({ in}\to{\cal R}_{|m|}){\cal D}_{|m|,S}^{(J)}~{\rm BR}({\cal R}\to[\gamma\gamma]_S)\,.
\eeq
The explicit form of the ${\cal{D}}$'s is reported in table~\ref{dD} for $J=0$, $J=2$ and $J=3$. The result is trivial for $J=0$, where $m=S=0$ and the angular distribution is flat, while already for $J=2,3$ all the ${\cal{D}}$'s are non-vanishing and non-trivial. Notice however that ${\cal D}_{2,0}^{(J)}={\cal D}_{0,2}^{(J)}$, leading to only five independent distributions. Moreover, since the only viable values of $|m|$ are $0,2$ for $gg{\textrm{/}}\gamma\gamma$ production and $0,1$ for $q{\overline{q}}$, only three distributions are present in the former case and four in the latter.\footnote{Further simplifications emerge for $J=3$ as discussed in section~\ref{three}.}
For $J=2$, the distributions relevant for $gg{\textrm{/}}\gamma\gamma$ and for $q{\overline{q}}$ are displayed in the plots in figure~\ref{DPLOT}. We see they have considerably different shapes so that it should be possible to distinguish them even with moderate experimental accuracy. 

The cross sections and branching ratios appearing in eq.~(\ref{XS}) are defined as follows. The $\bar\sigma$'s are the total  cross sections (at unit parton luminosity) for the production of the resonance with spin $m=|m|$ plus, if $m\neq0$, the one with $m=-|m|$. Namely
\beq
\ds\label{sbar}
\bar\sigma({ in}\to{\cal R}_{|m|})=\frac{\pi}{4M^2c_{in}}|{ A}^{in}_{|m|}|^2\,,
\eeq
where $c_{in}=1,3,8$ are the color factors for, respectively, $in=\gamma\gamma, q\overline{q}, gg$ and
\bea\ds
|{ A}^{{in}}_{0}|^2&=&|{ A}^{{in}}_{++}|^2+|{ A}^{{in}}_{--}|^2 \no\\
\ds
|{A}^{q\overline{q}}_{1}|^2&=&|{ A}^{q\overline{q}}_{+-}|^2+|{ A}^{q\overline{q}}_{-+}|^2 \,,\label{A2}\\
\ds
|{ A}^{gg{\textrm{/}}\gamma\gamma}_{2}|^2&=&|{ A}^{gg{\textrm{/}}\gamma\gamma}_{+-}|^2+|{ A}^{gg{\textrm{/}}\gamma\gamma}_{-+}|^2\,.\no
\eea
The cross-section for $in=\overline{q}q$ need not to be discussed explicitly because it is just identical to the $q\overline{q}$ one by the second relation in eq.~(\ref{rotrel}).
The BR's in eq.~(\ref{XS}) are those for the resonance decaying to a polarized diphoton pair with equal helicities $\lambda=\lambda'=\pm1$ for $S=0$ and with opposite helicities for $S=2$, i.e.
\beq\ds\label{br}
{\rm BR}({\cal R}\to[\gamma\gamma]_S)=\frac{1}{32\pi(2J+1)}\frac{M}{\Gamma}|{ A}^{\gamma\gamma}_{S}|^2\,,
\eeq
with ${A}^{\gamma\gamma}_{S}$ again as defined in eq.~(\ref{A2}).
Notice that the fact of having two distinct decay channels ($++$ and $--$) for $S=0$ and only one ($+-$, which is indistinguishable from $-+$ after angular integration) for $S=2$ compensates for the fact that the $\pm\pm$ states are made of indistinguishable particles and thus they have to be integrated over half of the solid angle. Furthermore, the branching ratios, as apparent from the notation, do not depend on the resonance spin $m$ because of rotational invariance.

Eq.s~(\ref{sbar}), (\ref{A2}) and (\ref{br}) provide the required map among the amplitude coefficients and the potentially observable quantities ($\overline{\sigma}$'s and ${\textrm{BR}}$'s) that parametrize the partonic cross section in eq.~(\ref{XS}). We see that the observables depend on few combinations of the $a$ and $\widetilde{a}$ parameters that control the amplitude coefficients through table~\ref{pars}. 
In particular, no information can be extracted on whether the $a$'s are real or complex, namely on whether eq.~(\ref{cptrel}) is satisfied or not, as previously mentioned.

The cross section parametrization in eq.~(\ref{XS}) can be directly employed for the comparison with experiments or, as we find convenient to do for the analysis in section~\ref{three}, rewritten in a ``probabilistic'' format by factoring out the total resonance production cross section times the total branching ratio to an unpolarized photon pair (${\rm BR}$), namely
\beq
\label{XS1}
\ds
\frac{d\bar\sigma_{in}}{d\cos\theta}
=\bar\sigma_{in}\hspace{-3pt}\times\hspace{-2pt}{\textrm{BR}}\sum_{|m|,S} {\cal{P}}^{in}_{|m|\,S}{\cal D}_{|m|,S}^{(J)}\,.
\eeq
Here 
\beq\ds\label{eq:P}
{\cal{P}}^{in}_{|m|\,S}=\frac{\bar\sigma({ in}\to{\cal R}_{|m|}){\rm BR}({\cal R}\to[\gamma\gamma]_S)}{\bar\sigma_{in}\hspace{-3pt}\times\hspace{-2pt}{\textrm{BR}}}
=\frac{|{A}^{in}_{|m|}|^2|{ A}^{\gamma\gamma}_{S}|^2}{\sum\limits_{|m|,S}|{A}^{in }_{|m|}|^2|{ A}^{\gamma\gamma}_{S}|^2}
\in[0,1]\,,
\eeq
is the probability for the produced resonance to have spin equal to $|m|$ in absolute value and to decay to a state of spin $S$.
The last identity in eq.~(\ref{eq:P}) has been obtained using eq.s~(\ref{sbar}) and (\ref{br}).
The probabilistic format is useful as it allows to disentangle the total signal rate from the normalized angular distribution, encapsulated in the ${\cal{P}}$'s. Notice that the ${\cal{P}}$'s, precisely because they are probabilities, sum up to one.

\subsection{LHC cross sections and distributions}
\label{sec:LHCcs}

It is conceptually straightforward to go from the partonic cross section, characterized by the $\sigma\hspace{-3pt}\times\hspace{-2pt}{\textrm{BR}}$ and ${\cal{P}}$ parameters as in eq.~(\ref{XS1}) (or by eq.~(\ref{XS})), to LHC differential cross sections or to event samples to be compared with the experimental data. The result will consist in a linear combination of distributions or in an admixture of event samples, each generated with its own ``${\cal{D}}$'' partonic distribution and weighted by the corresponding ``${\cal{P}}$'' probability. Such event samples could be obtained in two ways. Either by direct simulations, from the Lagrangian in Appendix~\ref{effective_ops} implemented in {\sc{MadGraph}} \cite{Alwall:2014hca}, turning on at each time the couplings associated with a given ``${\cal{D}}$'', or by matrix element reweighting, starting from the simulation of a scalar and reweighting each event, with partonic scattering angle $\theta$, by ${\cal{D}}(\theta)$. This latter approach is the only viable one for $J>2$, where no multi-purpose event generator implementation is available as previously mentioned.

For an accurate comparison with the data, properly taking into account soft QCD radiation, hadronization and detector effects one of the two strategies described above should be adopted. For the illustrative purpose of the present paper, however, it is sufficient to stick to purely leading order predictions, on top of which experimental effects will be attached as overall efficiency factors as described in the next section. This simple approach has the advantage of producing semi-analytical formulas for the distributions from which we can get an idea of which aspects of the signal properties are easier to extract from data.

The cross section, differential in the cosine of the scattering angle in the COM and in the boost of the COM frame, reads
\beq\label{eq:pp}\ds
\frac{d\sigma}{dy\, d{\hspace{-1pt}}\cos\theta}=\sum_{in}\tau\frac{d{\cal{L}}_{in}}{d\tau}\frac{dP_{in}}{dy}\frac{d\bar\sigma_{in}}{d\cos\theta}\,,
\eeq
having made use of the right hand side of eq.~(\ref{eq:PartonSigma}), that holds in the narrow resonance limit $\Gamma/M\to0$.
In the above equation, $\tau=M^2/s$, with ``$s$'' the collider energy squared, $\tau{d}{\cal{L}}_{in}/d\tau$ is the differential parton luminosity and ${dP_{in}}/{dy}$ is the distribution of the COM boost $y$.
These functions are related to the initial state PDF's $f$ by
\bea\label{PD}\ds
\frac{d{\cal{L}}_{q\overline{q}}(\tau)}{d\tau}\frac{dP_{q\overline{q}}(\tau,y)}{dy}&=&f_q(\sqrt{\tau}e^{-y})f_{\overline{q}}(\sqrt{\tau}e^{y})+f_{\overline{q}}(\sqrt{\tau}e^{-y})f_q(\sqrt{\tau}e^{y})\,,
\\\no\ds
\frac{d{\cal{L}}_{gg{\textrm{/}}\gamma\gamma}(\tau)}{d\tau}\frac{dP_{gg{\textrm{/}}\gamma\gamma}(\tau,y)}{dy}&=&f_{g{\textrm{/}}\gamma}(\sqrt{\tau}e^{-y})f_{g{\textrm{/}}\gamma}(\sqrt{\tau}e^{y})\,,
\eea
where
\beq\ds
\int\limits_{\frac12\log\tau}^{-\frac12\log\tau}\hspace{-11pt}dy\;\frac{dP_{q\overline{q}}(\tau,y)}{dy}=1\,.
\eeq
The variables $y$ and $\cos\theta$ are related to the rapidity of the two photons and to their $p_\bot$ as
\bea\ds
y&=&\frac{\eta+\eta'}2\,,\no\\
\cos\theta&=&\tanh\frac{|\eta-\eta'|}2=\sqrt{1-\frac{4p_\bot^2}{M^2}}\,.
\eea
Notice that $\cos\theta$ ranges from $0$ to $1$ as the photons are indistinguishable.

\begin{figure}[t]
\begin{center}
\includegraphics[width=7.5cm]{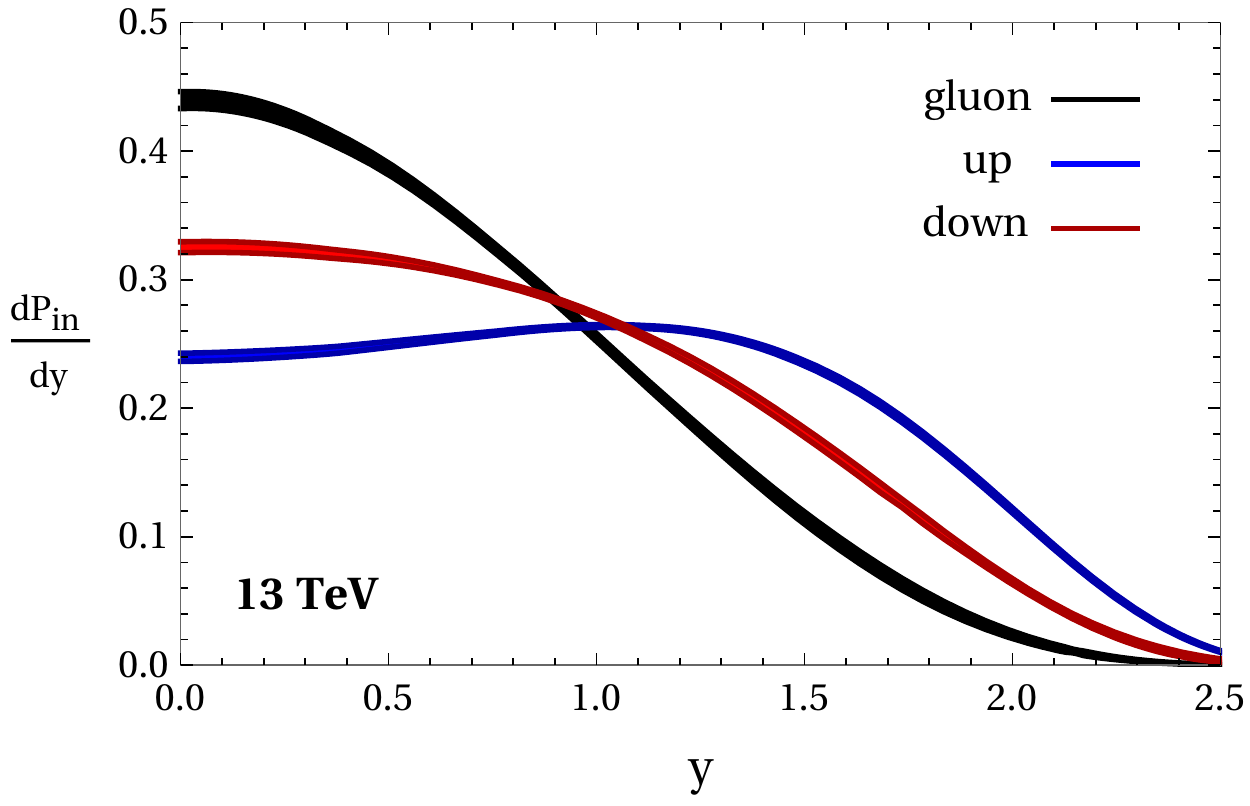}~~~~\includegraphics[width=7.5cm]{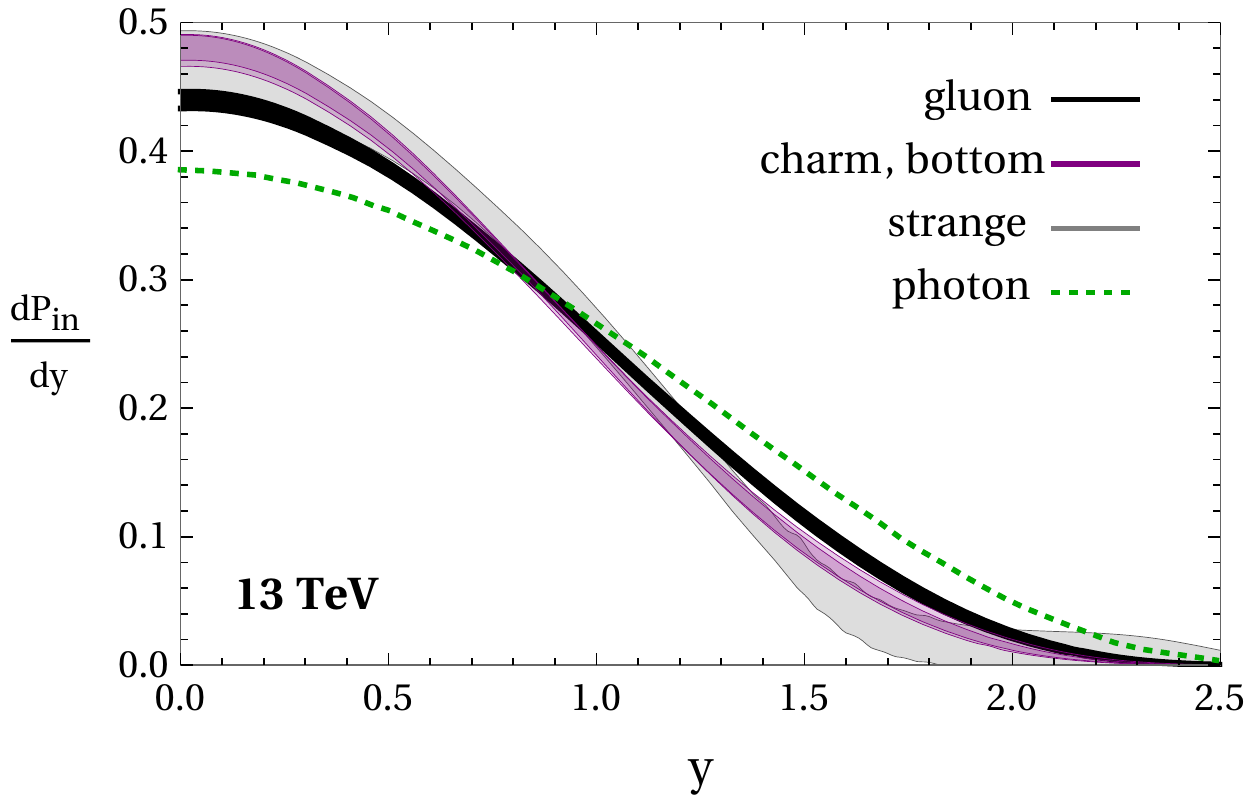}
\caption{Differential parton luminosities ${dP_{in}}/{dy}$, as defined in eq.~(\ref{PD}). The left plot shows $gg$, $u{\overline{u}}$ and $d{\overline{d}}$ initial states while $c{\overline{c}}$, $b{\overline{b}}$, $s{\overline{s}}$, $\gamma\gamma$ (and again $gg$ for comparison) are displayed on the right. The $1\,\sigma$ bands are obtained as described in the text. 
}\label{PDensities}
\end{center}
\end{figure}

Both $\cos\theta$ and $y$ are measurable and both the $\cos\theta$ and $y$ differential distributions contain interesting and, to large extent, complementary information about the signal. Namely, the $\cos\theta$ distribution gives us direct access, at least  if only one ``$in$'' channel is active in eq.~(\ref{eq:pp}), to the partonic differential cross section, which in turn is related to the resonance spin as previously discussed. It also provides partial information about the production mode, given that the $\cos\theta$ distributions, i.e. the ${\cal{D}}$ functions, can be different if the resonance is produced by the $gg{\textrm{/}}\gamma\gamma$ or by the $q{\overline{q}}$ initial state.\footnote{However they can also be equal, since we saw in the previous section that ${\cal{D}}^{(2)}_{0,0}$ and ${\cal{D}}^{(2)}_{0,2}$ can appear in both $gg{\textrm{/}}\gamma\gamma$ and in $q{\overline{q}}$ production. If this is the case, distinguishing the two channels requires looking at the $y$ distribution as we will readily discuss.} It is instead unable to distinguish, for instance, $q{\overline{q}}=u{\overline{u}}$ from $q{\overline{q}}=d{\overline{d}}$ as the ${\cal{D}}$'s are the same in the two cases. The situation is basically reversed for the differential distribution in $y$, which is insensitive to the details of the partonic cross section and is entirely dictated by the production mode, which determines the shape of $dP/dy$. Whether or not and how easily this may be exploited to distinguish different production mechanisms depends on how much different the $dP/dy$'s are in the different cases. This is quantified in figure~\ref{lum}, for a resonance mass of $M=750$~GeV (chosen in preparation for the discussion of the next section) and $\sqrt{s}=13$~TeV. We see that the two valence quarks have slightly different distributions, allowing in principle to distinguish $u{\overline{u}}$ from  $d{\overline{d}}$. All the sea quark distributions are instead very similar, or even identical within the uncertainties, and not far from the ones for $gg$ and $\gamma\gamma$. The plots in figure~\ref{lum} are obtained by the NNPDF23\_nnlo\_as\_0119\_qed set of NNPDF2.3~\cite{Ball:2013hta} with a factorization scale of $750$~GeV. The uncertainties are obtained from the variance over the PDF replicas provided in the PDF set. Scale uncertainties, quantified by varying the factorization scale, are found to be negligible. This is valid for the ``ordinary'' partons $g$ and $q$, but not for the photon, whose PDF measurement is too bad to extract any quantitative information. The $\gamma\gamma$ luminosity is thus taken from ref.~\cite{Harland-Lang:2016qjy}, where it has been estimated from the theoretical calculation of the photon PDF presented in refs.~\cite{Martin:2014nqa,Harland-Lang:2016apc}. Uncertainties in $dP_{\gamma\gamma}/dy$ are not reported in ref.~\cite{Harland-Lang:2016qjy} and consequently they do not appear in our plot.

\begin{table}[t]
\begin{center}
\begin{tabular}{c||c|c|c|c|c|c|cc} 
\rule[-.5em]{0pt}{1.2em} & ${gg}$ & ${u\overline{u}}$ & ${d\overline{d}}$ & ${s\overline{s}}$ & ${c\overline{c}}$ & ${b\overline{b}}$ & ${\gamma\gamma}$~\cite{Harland-Lang:2016qjy} & ${\gamma\gamma}$~\cite{Ball:2013hta}\\
\hline\hline
\rule[-.5em]{0pt}{1.6em}$[\tau d{\cal{L}}/d\tau]_{13}$ &$5.5$ & $0.78$ & $0.48$ & $0.051$ & $0.028$ & $0.012$ & $1.2\times10^{-3}$&$(2.4\pm1)\times10^{-3}$  \\
\hline
\rule[-.5em]{0pt}{1.6em}$[\tau d{\cal{L}}/d\tau]_{8}$ &$1.1$ & $0.30$ & $0.18$ & $0.011$ & $0.0054$ & $0.0021$ &  $0.43\times10^{-3}$ & $(1.2\pm1)\times10^{-3}$ \\
\hline
\rule{0pt}{1.1em}$r$ & $4.8$ & $2.6$ & $2.7$ & $4.8$ & $5.2$ & $5.7$ & $2.9$ & $(2\pm0.5)$ 
\end{tabular}
\caption{\small Parton luminosities $\tau d{\cal{L}}/d\tau$ at $\sqrt{s}=8, 13$~TeV and gain $r=[\tau d{\cal{L}}/d\tau]_{13}/[\tau d{\cal{L}}/d\tau]_{8}$ for $M=750$~GeV and factorization scale equal to the resonance mass. The uncertainty from scale variation is of order $10\%$. 
\label{lum}}
\end{center}
\end{table}

We saw that $\cos\theta$ and $y$ differential distributions provide complementary information about the signal, however because of the photon acceptance cuts it is not clear that the two distributions can actually be disentangled experimentally and measured separately. While performing separate measurements (possibly unfolding the experimental effects) would facilitate the interpretation, allowing for instance to compare directly the $\cos\theta$ distribution with the shape of the ${\cal{D}}$ functions in figure~\ref{DPLOT}, notice that the exact same amount of information could be extracted from the study of the doubly differential distribution.

Before concluding this section and in preparation for the next one, where we will use our framework for a first characterization of the $750$~GeV excess, we report in table~\ref{lum} the total parton luminosity at $M=750$~GeV at the $13$ and $8$~TeV LHC and the gain, defined as the ratio of the $13$ and $8$~TeV cross sections, for each production mode. Contrary to $dP/dy$, the uncertainties are now dominated by scale variation and is of order $10\%$ (up to $15\%$ for the gluon, and below $6\%$ for quarks). Two set of results are reported in the table concerning the $\gamma\gamma$ channel. The first one is based on the theoretical prediction from Ref.~\cite{Harland-Lang:2016qjy}, which we will employ in what follows. The second one, subject to a large error, is obtained with the NNPDF2.3~\cite{Ball:2013hta} PDF set.

\section{Benchmark scenarios}\label{three}
\label{sec:bench}

In the previous section we saw how the production of a resonance of arbitrary spin decaying to $\gamma\gamma$ is conveniently parametrized, for each given $in=gg{\textrm{/}}\gamma\gamma{\textrm{/}}q{\overline{q}}$ production channel, in terms of a rather small number of phenomenological parameters with a sharply defined and intuitive physical meaning. However, being completely agnostic about the resonance couplings would require taking all the production channels into account simultaneously, with independent free parameters for each of the $7$ (i.e., $gg{\textrm{/}}\gamma\gamma{\textrm{/}}q{\overline{q}}=\{u{\overline{u}},d{\overline{d}},c{\overline{c}},s{\overline{s}},b{\overline{b}}\}$) $in$ states. This proliferation of parameters makes the problem untreatable in full generality and obliges us to make additional assumptions in order to reduce the dimensionality of the parameter space. A set of plausible restrictive assumptions is defined in the present section, producing a set of alternative benchmark scenarios. Each of these benchmarks contains a small enough number of free parameters to be experimentally tested in full generality. The variety of benchmarks should provide a sufficient (but still unavoidably partial) coverage of the phenomenology. Additional benchmarks can be defined, if needed, within our general framework.

The benchmark scenarios can be used for exclusions, producing limits on $\sigma\hspace{-2pt}\times\hspace{-2pt}{\textrm{BR}}$ which are more general and easier to reinterpret in specific models than those obtainable with the habitual benchmarks of a scalar or of a $J=2$ ``RS graviton'' resonance. More interestingly, they can be used to characterize the properties of a new resonance that we might happen to discover in the diphoton final state. In the latter case, the SM $p$-value and other statistical quantities aimed at assessing the actual existence and viability of the signal, could be reported on the benchmark model parameter space. This will select the signal hypothesis that best fits the data and will give us information about the resonance spin and (see section~\ref{four}) CP properties. At a later stage, with enough data, it will be possible to measure the parameters of the benchmark models, namely those that control the signal kinematical distribution and the total $\sigma\hspace{-2pt}\times\hspace{-2pt}{\textrm{BR}}$. The model-independent nature of our parametrization will straightforwardly allow to translate these measurements into whatever the ``true'' resonance model turns out to be.

A good fraction of the program outlined above is slightly premature, as a discovery still has to come. However the $M=750$~GeV excess reported by ATLAS~\cite{ATLAS13} and CMS~\cite{CMS:2015dxe} with $13$~TeV LHC data gives us the opportunity to practice, at least on some aspects of the signal characterization strategy.\footnote{Provided that the signal originates from a single
resonance decaying in a photon pair rather than a pair of axions decaying into highly collimated photons as suggested in
ref.~\cite{Aparicio:2016iwr} (see also refs.~\cite{others}).}
We will do so by recasting ATLAS $13$~TeV~\cite{ATLAS13}, CMS $13$~TeV~\cite{CMS:2015dxe}, ATLAS $8$~TeV~\cite{Aad:2015mna} and CMS $8$~TeV~\cite{Khachatryan:2015qba} experimental searches, with a procedure described in Appendix~\ref{app:statistics} in detail. It suffices here to say that the recast is performed by reconstructing, in the Gaussian approximation, the likelihoods associated to each experimental search from the background-only $p$-value and the observed limit. The four searches are eventually treated as statistically independent in the combination. The intrinsic inaccuracy of our statistical method and our approximate treatment of the experimental efficiencies make our results not fully quantitative. Moreover, the experimental searches we use are not optimized to provide information about the angular distributions of the putative signal and thus they are poorly sensitive to the resonance spin and production mode. Consequently our results will often show a rather limited discriminating power within the parameter space of each benchmark and among different benchmarks. Most of what we will be able to tell will come from the combination of $8$ and $13$~TeV searches because of the slightly different gain factors $r$ (see table~\ref{lum}) in the total signal rate. Notice however that the situation would substantially improve with dedicated experimental analysis and/or more data.

In view of the considerations above, we warn the reader that the results that follow should be mostly regarded as a pragmatic illustration of the usage of our benchmarks. Still, it will be interesting to see that in some cases the various analyses do display slightly different acceptances for the same signal shape, merely due to the slightly different selection cuts. This produces, in the combination, some discriminating power among the different hypotheses and indicates that progress in the signal characterization should be relatively easy to achieve with a dedicated analysis.

\subsection{Scalar resonance}

As a first case we consider the simplest scenario, that is the model with a scalar resonance. This case is rather peculiar since
the angular distribution of the two photons in the COM frame is completely flat. Indeed, as we saw in the previous
section, the only contribution to the production comes from the $m = S = 0$ mode, which is described by the angular
function ${\cal D}^{(0)}_{0,0} = 1$ (see table~\ref{dD}).
The only model-dependence is encoded in the relative strengths of the various production channels,
which can be parametrized through the partonic production cross sections $\bar \sigma_{in}$.
Such a parametrization characterizes the possible scenarios
in a way that is completely independent of the details of the experimental searches, in particular of the COM energy of the collider.
From a practical point of view, however, this does not seem a convenient choice. Due to the extremely different parton
luminosities (see table~\ref{lum}), partonic cross sections of similar size give rise
to signal cross sections for the various production channels that can differ by more than one order of magnitude.
For instance, the production modes through quarks or photons can be comparable to the $gg$ one only if their partonic
cross sections $\bar \sigma_{q\bar q/\gamma\gamma}$ are much larger than $\bar \sigma_{gg}$.
Therefore, we find more convenient to adopt a parametrization that allows to efficiently treat cases
in which various production modes lead to comparable signal yields. Of course a parametrization of this kind is necessarily
collider dependent, since it must take into account the parton luminosities.
A possible choice, which we will adopt in the following, is to use the ratios of signal cross sections for the various channels
for a collider energy of $13$~TeV. In particular we define the quantities
\begin{equation}\label{eq:Ri_def}
{R}_{in} \equiv \frac{\sigma^{13\,\textrm{TeV}}_{in}}{\sigma^{13\,\textrm{TeV}}_{tot}}\,,
\end{equation}
where $\sigma^{13\,\textrm{TeV}}_{in}$ is the $13$~TeV production cross section in the $in$ channel, whereas $\sigma^{13\,\textrm{TeV}}_{tot}$ is the total production cross section.\footnote{The branching ratio into diphotons is
clearly the same for all channels and drops out in the ratio of the signal cross sections.}
The $R_{in}$ parameters directly encode the relative importance of the contributions to the signal cross section from
the various production channels. Since they are normalized to the total production cross section, the ${R}_{in}$ parameters
sum up to unity, \mbox{$\sum_{in} {R}_{in} = 1$}.
The relative strengths of the production channels at $8$~TeV can be easily related to the $13$~TeV ones by taking into
account the change in the partonic cross sections listed in table~\ref{lum}.

From the experimental point of view, the various production channels are characterized by the different gain factors
between the $8$ and $13$~TeV cross sections and
by different signal acceptances for the experimental searches, possibly corresponding to different event selection categories.
As can be seen from the numerical values in
table~\ref{tab:eff_scalar}, the geometric acceptances for the various production channels are quite similar to each other. The most
important differences, of the order of $\sim20\%$, are present for the CMS analysis, which explicitly presents the results
in two categories: barrel-barrel (EBEB), which includes events with both photons in the central detector region,
and barrel-endcap (EBEE), in which one photon is central while the second falls in the detector endcap.
As we discussed before, the various production channels lead to slightly different rapidity distributions for the final photons,
thus giving rise to different acceptances for the two CMS categories. Obviously this property cold be used to differentiate the
production channels, although at present the experimental sensitivity is limited. We will discuss better this aspect in Appendix~\ref{sec:categories}.

\begin{table}
\centering
\begin{tabular}{c|c|c|c|c}
Production & ATLAS $13$ & CMS $13$ (EBEB, EBEE) & ATLAS $8$ & CMS $8$\\
\hline
\hline
\rule{0pt}{1.05em}$u\overline u$ & $0.57$ & $0.40$\hspace{2.em} $0.29$ & $0.80$ & $0.68$ \\
\rule{0pt}{1.05em}$d\overline d$ & $0.58$ & $0.49$\hspace{2.em} $0.27$ & $0.83$ & $0.70$ \\
\rule{0pt}{1.05em}$gg$ ($s \overline s$, $c \overline c$, $b \overline b$, $t \overline t$) & $0.59$ & $0.59$\hspace{2.em} $0.24$ & $0.86$ & $0.71$ \\
\rule{0pt}{1.05em}$\gamma \gamma$ & $0.56$ & $0.48$\hspace{2.em} $0.25$ & $0.80$ & $0.68$
\end{tabular}
\caption{Acceptances for the scalar resonance case. The numerical results are derived for the following analyses:
ATLAS $13$~TeV~\cite{ATLAS13}, CMS $13$~TeV (split into the two categories barrel-barrel (EBEB)
and barrel-endcap (EBEE))~\cite{CMS:2015dxe},
ATLAS $8$~TeV~\cite{Aad:2015mna} and CMS $8$~TeV~\cite{Khachatryan:2015qba}. The efficiencies for the $gg$ case
also apply to the $s \overline s$, $c \overline c$, $b \overline b$, $t \overline t$, since the differences among all these cases
are $\lesssim 2\%$.}\label{tab:eff_scalar}
\end{table}

Let us start the description of the numerical results by considering the quark and gluon production modes.
The difference between the $gg$ mode and the production through sea quarks ($s\overline s$, $c\overline c$,
$b\overline b$) is very small. All these channels have comparable gain factors (see table~\ref{lum}) and
similar signal acceptances (see table~\ref{tab:eff_scalar}). This is not unexpected, since
the parton luminosities for these production channels are quite similar (see fig.~\ref{PDensities}).
For this reason in our recast we only consider the $gg$ channel,
which provides also a good approximation of the sea quark ones. Significant differences, instead, are present with respect to
the valence quark modes ($u\overline u$ and $d\overline d$), mostly due to the gain factors that are much smaller than
in the $gg$ case.
In addition to the acceptances we also included some reconstruction efficiency factors for the signal, which we take from the
experimental papers. The numerical values are $70\%$ for ATLAS $13$~TeV, $81\%$ and $77\%$ for the EBEB and EBEE
categories of the CMS $13$~TeV analysis, $56\%$ for ATLAS $8$~TeV and $81\%$ for CMS $8$~TeV.
Finally, in our numerical analyses we assume the resonance to have a small width, below the experimental resolution
$\sim 7~\textrm{GeV}$.

\begin{figure}[t]
\centering
\includegraphics[width=.485\textwidth]{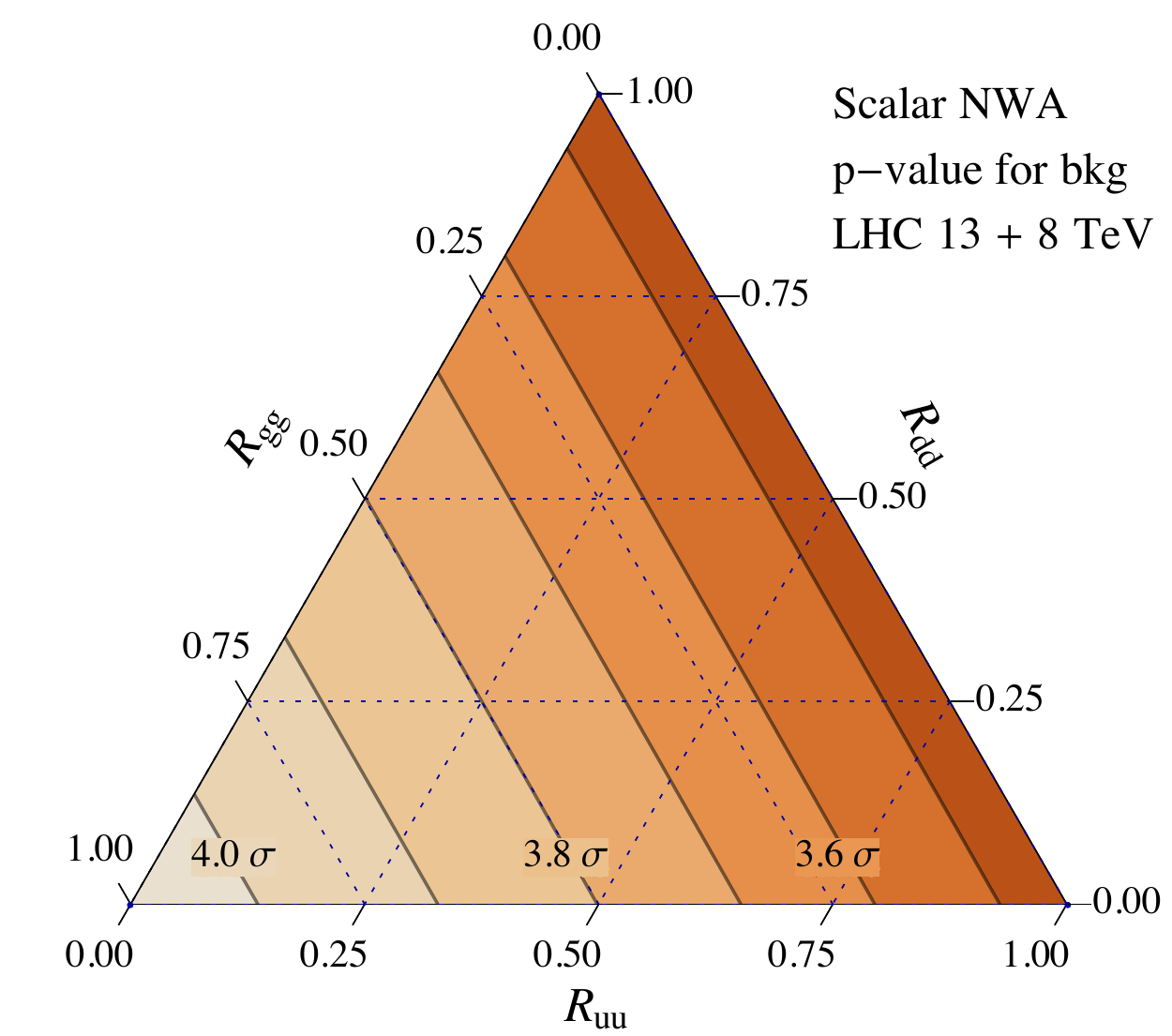}
\hfill
\includegraphics[width=.485\textwidth]{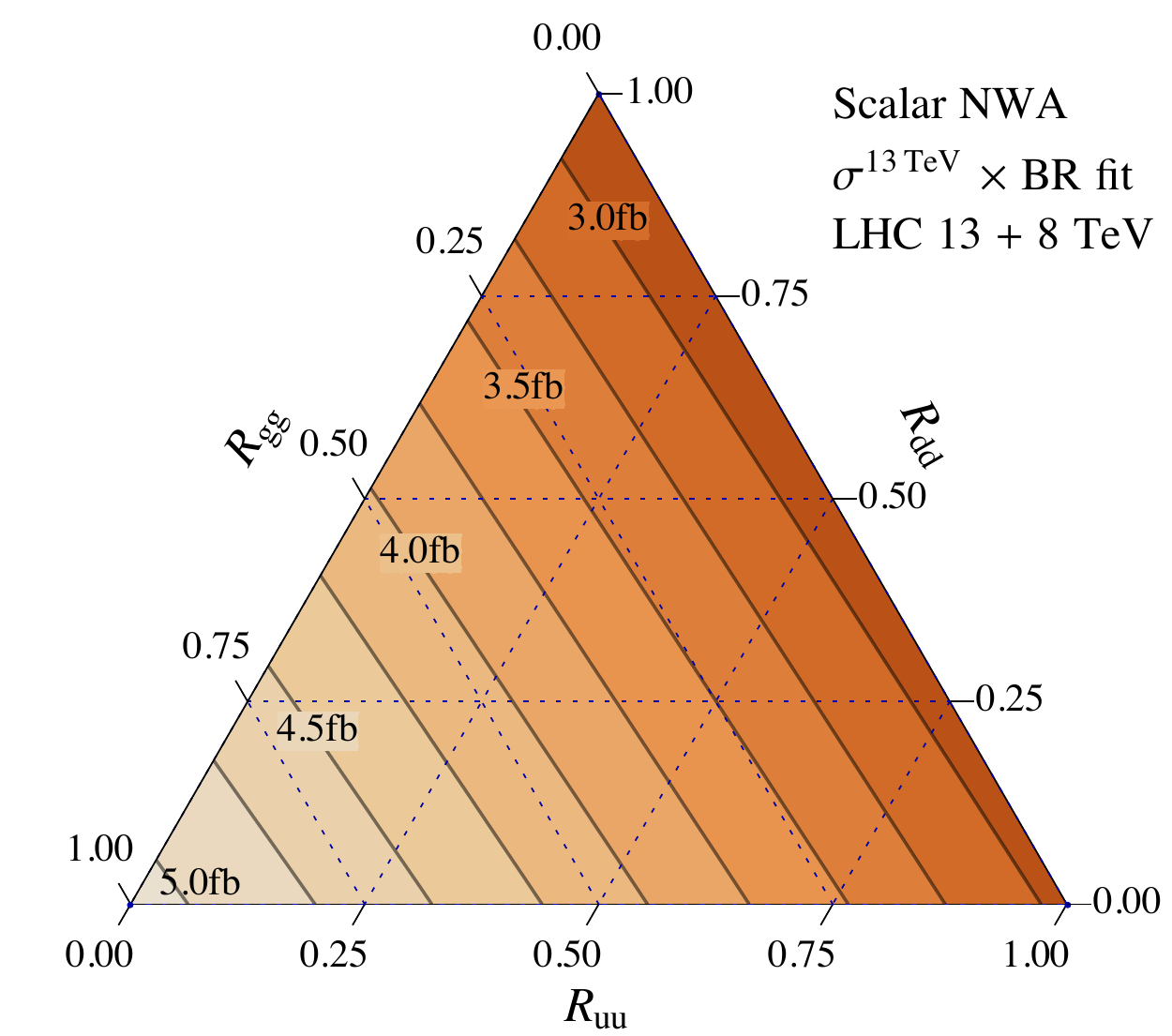}
\caption{On the left panel we show the reconstructed p-value for the background-only hypothesis in the case of a narrow scalar
resonance produced in the $u\overline u$, $d\overline d$ and $gg$ channels. The results are obtained by combining the
ATLAS and CMS $13$ and $8$~TeV searches. On the right panel we show the best fit of the $13$~TeV signal cross
section.}\label{fig:scalar_1}
\end{figure}

The local significance of the diphoton excess is shown in the left panel of fig.~\ref{fig:scalar_1} as a function of
the ${R}_{gg}$, ${R}_{uu}$ and ${R}_{dd}$ parameters. Since these tree parameters
sum up to one, it is convenient to present the results in a ``triangle'' plot. One can see that the local p-value is
sensitive almost exclusively to ${R}_{gg}$, ranging from $4\,\sigma$ in the case with purely $gg$-initiated production
(${R}_{gg} = 1$) to $3.5\,\sigma$ in the cases with ${R}_{gg} = 0$.
The dependence on the other two parameters is quite limited,
since the gain and efficiencies for the $u\overline u$ and $d\overline d$ modes are similar. The best fit of the signal
cross section is shown in the right panel of fig.~\ref{fig:scalar_1} and ranges from $5~\textrm{fb}$ for the
${R}_{gg} = 1$ case to $3~\textrm{fb}$ for ${R}_{gg} = 0$.

In fig.~\ref{fig:scalar_2} we show the combined goodness of fit (see Appendix~\ref{app:statistics} for more details).
One can see that the compatibility of the various searches is never high. In the best case ${R}_{gg} = 1$, the
compatibility is only $\sim 9\%$, while it drops below $1\%$ in the $u\overline u$ and $d\overline d$-initiated modes.
Analyzing the breakdown of the likelihood in each experimental search, one finds that the major source of tension
is the ATLAS $13$~TeV search, which favors a quite large signal cross sections $\sim10~\textrm{fb}$, to be compared with the
much smaller ones $\sim 2~\textrm{fb}$ preferred by the other three searches. On the other hand, the two CMS searches and the
ATLAS $8$~TeV one show a very good compatibility ($\gtrsim 30\%$). Obviously the better global agreement found
in the case of $gg$-initiated production is due to the larger gain factor between the $8$~TeV and $13$~TeV cross sections.

\begin{figure}[t]
\centering
\includegraphics[width=.485\textwidth]{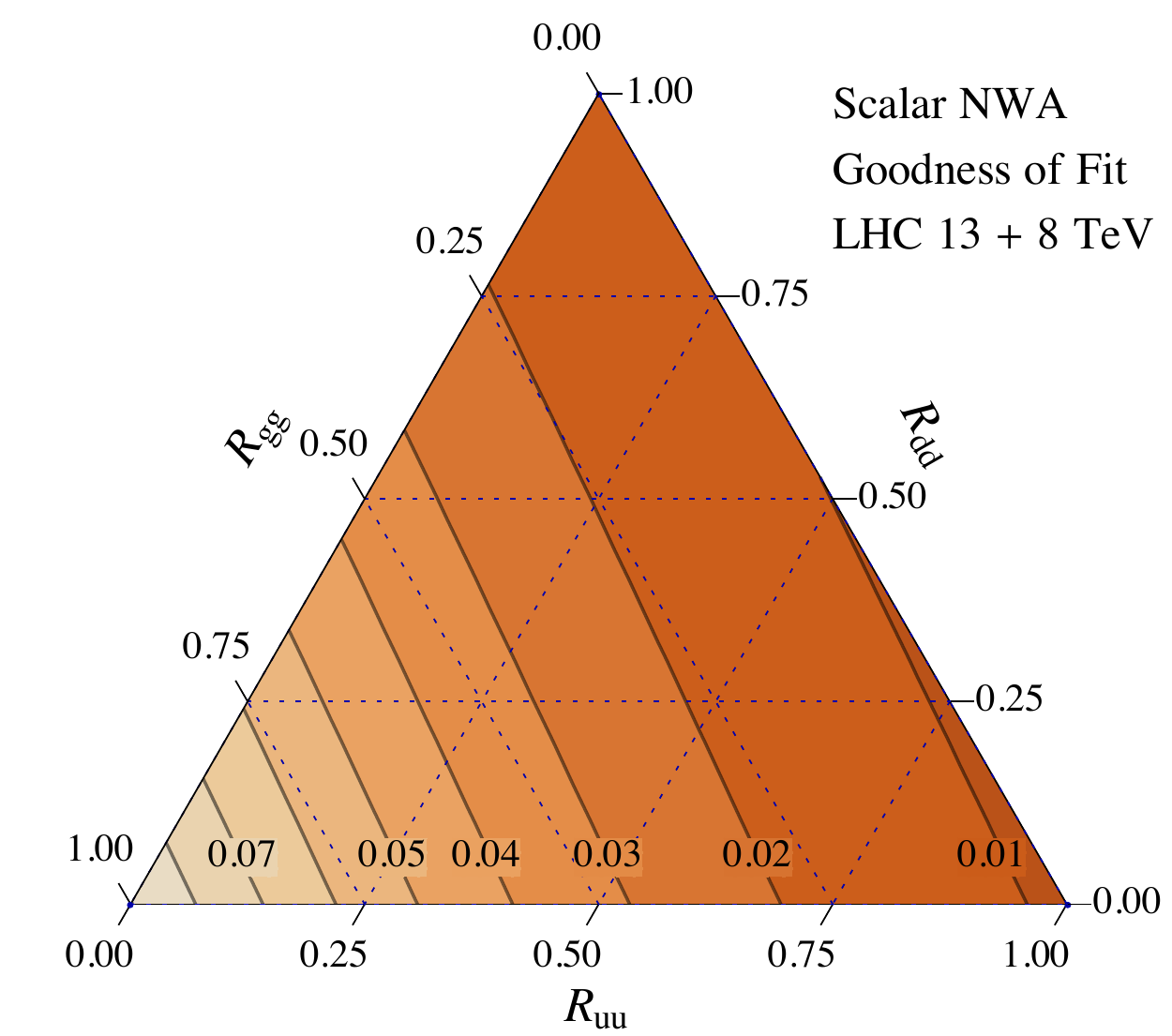}
\caption{Goodness of fit in the case of a narrow scalar resonance produced in the $u\overline u$, $d\overline d$ and $gg$ channels.
The results are obtained by combining the ATLAS and CMS $13$ and $8$~TeV searches.}\label{fig:scalar_2}
\end{figure}

As a final case we consider the scenario in which the scalar resonance is produced exclusively through the $\gamma\gamma$
mode.\footnote{For phenomenological analyses of this scenario see ref.~\cite{photon-fusion}.}
In this case there is no free parameter and the scenario is fully characterized by the gain factor $r_\gamma = 2.9$ and by the
efficiencies given in table~\ref{tab:eff_scalar}. The efficiencies are quite similar to the ones for the $d\overline d$ initiated
mode, thus we expect the overall features of this scenario to be comparable to the case ${R}_{dd} = 1$.
As we already discussed, the relatively small gain factor increases the tension between the ALTAS $13$~TeV results
and the other searches, thus the photon-initiated mode is not favored by the present data.
The goodness of fit is indeed $1\%$ and also the statistical significance of the excess is relatively small, $3.5\,\sigma$.
The best fit of the $13$~TeV signal cross section is $3$~fb.

\subsection{\boldmath Spin-$2$ resonances}

Let us now move to the case of spin-$2$ states. As we did for the scalar resonances, we will adopt here a broad perspective
and we will consider a generic new state without imposing any restriction on its production modes and on its
decay distributions. When looking for a physics interpretation of these scenarios, it must be however kept in mind that
resonances of spin $J\geq2$ have a typical interpretation as composite states. Therefore, the hypothesis that a new state
of this kind is within the reach of the LHC requires an exotic strong dynamics not far above the TeV scale. Such a framework
is considerably constrained by a variety of experimental tests, which limit the number of realistic benchmark scenarios.

As for the scalars, the production modes can be encoded in the ${R}_{in}$ parameters defined as in eq.~(\ref{eq:Ri_def}).
In the present case, however, additional free parameters are needed to take into account the angular distribution of the
decay products. As we explained in section~\ref{sec:general_framework}, the decay distributions in the COM
frame are a combination of a limited number of functional forms, which depend on the production channel (three forms
for the $gg$ and $\gamma\gamma$ mode and four for the quark-initiated channels).
The total number of free parameters is thus significantly greater than in the scalar-resonance case.
It is thus unpractical to keep all of them free in an analysis, but instead it is reasonable to consider a few benchmark scenarios.
In the following we will describe some of them. In particular we will focus on single production modes, namely the $gg$ initiated channel
and the quark production modes. In addition we will also discuss a benchmark that parametrizes a very specific, but
well motivated scenario, the Randall-Sundrum graviton.\footnote{Extensions of this minimal framework have been recently discussed in~\cite{spin2}.}

\subsubsection{\boldmath $gg$-initiated production}

The most straightforward way to couple an exotic strong dynamics to the SM is via gauge interactions.
This is typically realized whenever the constituents of the resonance are charged under the SM gauge symmetry.
If the strong sector is charged under QCD, the leading production modes at hadron collider is expected to be the one
involving gluons. Another interesting possibility is the case in which the resonance is produced from  photons.

\begin{table}[t]
\begin{center}
\begin{tabular}{c|c|c|c|c} 
\rule{0pt}{1.2em}
 & ATLAS 13 & CMS 13 (EBEB, EBEE) & ATLAS 8 & CMS 8 \\
\hline
\rule{0pt}{1.1em}${\cal P}^{gg}_{00}$ & $0.38$ & $0.39$\hspace{2.em} $0.23$ & $0.71$ & $0.41$\\
\rule{0pt}{1.1em}${\cal P}^{gg}_{02}+{\cal P}^{gg}_{20}$ & $0.87$ & $0.78$\hspace{2em} $0.14$ & $0.94$ & $0.94$\\
\rule{0pt}{1.1em}${\cal P}^{gg}_{22}$ & $0.32$ & $0.40$\hspace{2.em} $0.32$ & $0.77$ & $0.47$\\
\end{tabular}
\caption{\small Acceptances for $gg$-initiated spin-$2$ diphoton resonances. }\label{ACCgg}
\end{center}
\end{table}

\begin{table}[t]
\begin{center}
\begin{tabular}{c|c|c|c|c} 
\rule{0pt}{1.2em}
 & ATLAS 13 & CMS 13 (EBEB, EBEE) & ATLAS 8 & CMS 8 \\
\hline
\rule{0pt}{1.1em}${\cal P}^{\gamma\gamma}_{00}$ & $0.38$ & $0.31$\hspace{2.em} $0.23$ & $0.64$ & $0.40$\\
\rule{0pt}{1.1em}${\cal P}^{\gamma\gamma}_{02}+{\cal P}^{\gamma\gamma}_{20}$ & $0.84$ & $0.64$\hspace{2.em} $0.17$ & $0.91$ & $0.90$\\
\rule{0pt}{1.1em}${\cal P}^{\gamma\gamma}_{22}$ & $0.30$ & $0.32$\hspace{2.em} $0.31$ & $0.68$ & $0.45$\\
\end{tabular}
\caption{\small Acceptances for $\gamma\gamma$-initiated spin-$2$ diphoton resonances. }\label{ACCgamma}
\end{center}
\end{table}

\begin{figure}[t!]
\centering
\includegraphics[width=.485\textwidth]{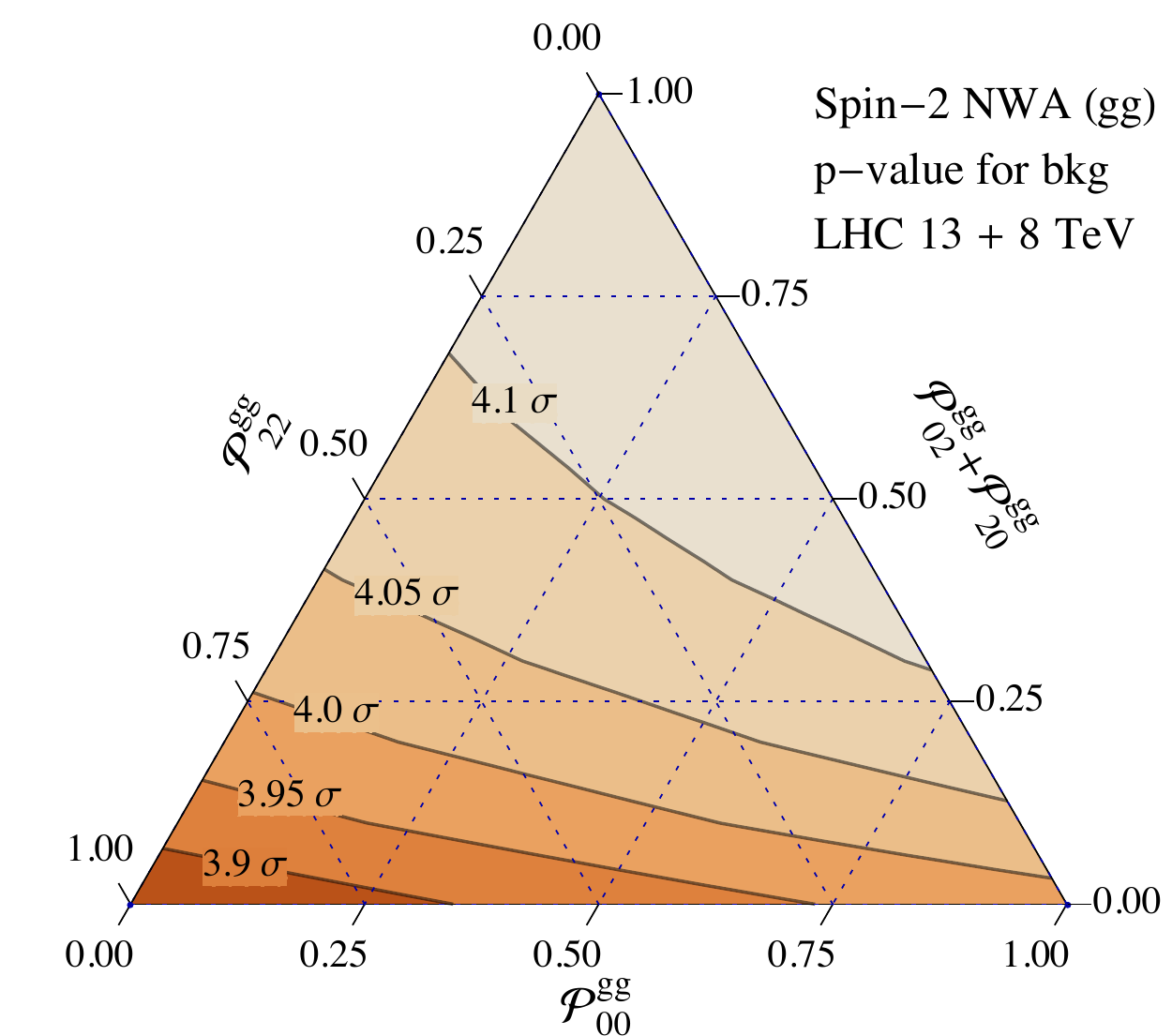}
\hfill
\includegraphics[width=.485\textwidth]{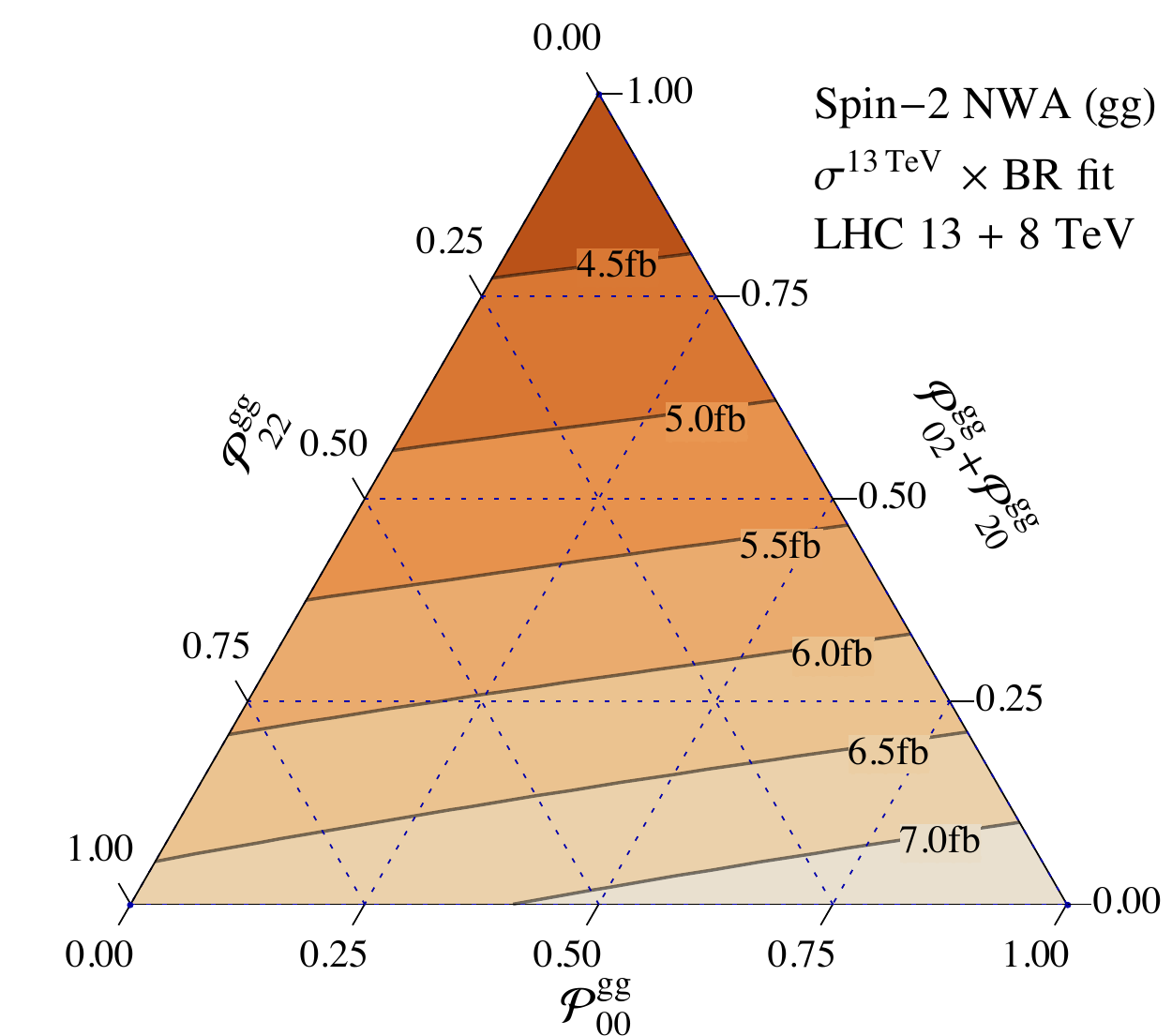}
\caption{On the left panel we show the reconstructed p-value for the background-only hypothesis in the scenario with a
narrow spin-$2$ resonance produced in the $gg$ channel. The results are presented as a function of the parameters ${\cal P}_{00}$,
${\cal P}_{02} + {\cal P}_{20}$ and ${\cal P}_{00}$, which encode the angular distribution of the final-state photons
(see eq.~(\ref{eq:spin2_gg_param}). The numerical values are obtained by combining the
ATLAS and CMS $13$ and $8$~TeV searches. On the right panel we show the best fit of the $13$~TeV signal cross
section.}\label{fig:spin2_1}
\end{figure}

Applying the results of section~\ref{sec:general_framework}, it is straightforward to check that the
COM angular distribution of the
decay products is a combination of the functions ${\cal D}^{(2)}_{0, 0}$, ${\cal D}^{(2)}_{0, 2}$, ${\cal D}^{(2)}_{2, 0}$
and ${\cal D}^{(2)}_{2, 2}$.\footnote{This result trivially follows from the fact that the gluons and the photons can only have
helicities $\pm 1$, thus they give rise to a combined state with $m = +2, 0 -2$.}
However, since ${\cal D}^{(2)}_{0, 2} = {\cal D}^{(2)}_{2, 0}$ (see table~\ref{dD}),
we are left with just three possible functional forms. We can thus fully parametrize the differential cross section as
\begin{equation}\label{eq:spin2_gg_param}
\frac{d\bar\sigma_{in}}{d\cos\theta} = \bar\sigma_{in} \left[{\cal D}^{(2)}_{0,0}\, {\cal P}_{00} + {\cal D}^{(2)}_{0,2} \left({\cal P}_{02} + {\cal P}_{20}\right)
+ {\cal D}^{(2)}_{2,2}\, {\cal P}_{22}\right]\,,
\end{equation}
as a function of three free quantities, $ {\cal P}_{00}$,  $ {\cal P}_{02} + {\cal P}_{20}$ and $ {\cal P}_{22}$, which
are normalized such that they sum up to unity.

As a representative example, we recast the experimental searches for a diphoton resonance in the scenario with a
narrow spin-$2$ resonance produced exclusively from $gg$. The case of $\gamma\gamma$ production is similar, however,
analogously to the scalar case, it is disfavored by the current data because of the small cross section gain between
$8$ and $13$~TeV.

\begin{figure}[t]
\centering
\includegraphics[width=.485\textwidth]{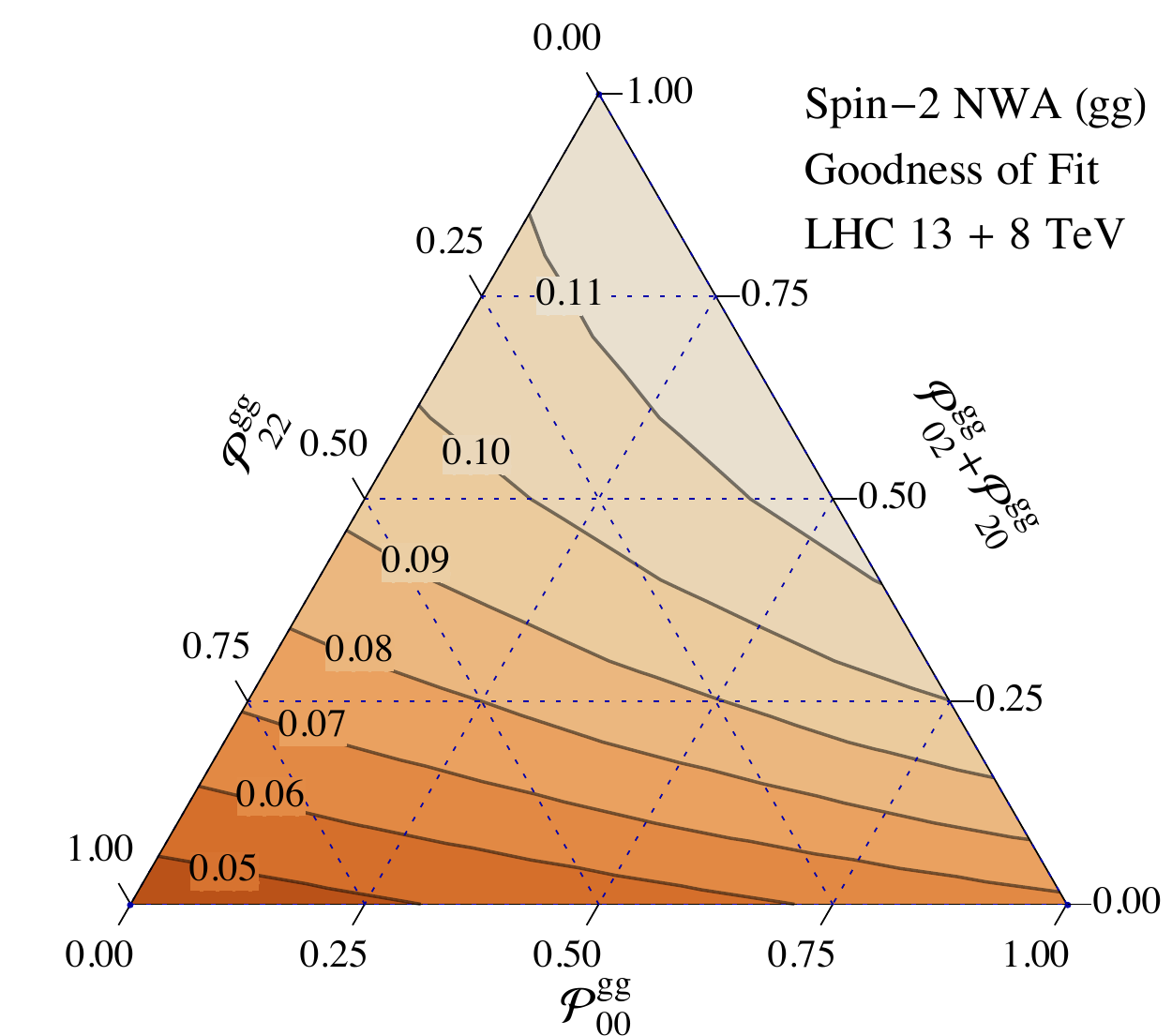}
\caption{Goodness of fit in the case of a narrow spin-$2$ resonance produced in the $gg$ channel.
The results are obtained by combining the ATLAS and CMS $13$ and $8$~TeV searches.}\label{fig:spin2_2}
\end{figure}

The geometric acceptances for the various experimental searches are listed in table~\ref{ACCgg} (see table~\ref{ACCgamma}
for the acceptances in the $\gamma\gamma$ channel).
In fig.~\ref{fig:spin2_1} we show the signal significance and the best fit of the cross section for the $gg$ mode as a function of the three
free parameters, ${\cal P}_{00}$, ${\cal P}_{02} + {\cal P}_{20}$ and ${\cal P}_{00}$. The goodness of the fit is instead shown in
fig.~\ref{fig:spin2_2}. We find that the signal significance is around $4\,\sigma$ and is slightly higher for a resonance
decaying in the ${\cal D}^{(2)}_{0,2}$ and ${\cal D}^{(2)}_{2,0}$ modes. The goodness of fit in the ${\cal P}_{02} + {\cal P}_{20} = 1$
corner is $\sim 12\%$ and is significantly higher that in the other configurations, in particular for
${\cal P}_{22} = 1$ we find a compatibility around $4\%$. The best fit of the signal cross section varies from
$\sim 4~\textrm{fb}$ in the configurations with ${\cal P}_{02} + {\cal P}_{20} = 1$ to $\sim 7~\textrm{fb}$ in the cases
${\cal P}_{00} = 1$ and ${\cal P}_{22} = 1$.

\subsubsection{\boldmath $q\overline{q}$-initiated production}

A spin-$2$ resonance can have sizable couplings to quarks if some of the latter mix significantly with fermionic composites
of the exotic strong dynamics. We can thus envisage a scenario in which a spin-$2$ resonance is produced mainly through
the $q\overline q$ channel. In this set-up the initial partons can have $m = \pm1$ or $m = 0$. The latter spin, however,
is only generated by interactions suppressed by a chirality flip, which thus are expected to give rise to smaller
contributions than the $|m| = \pm 1$ channel. For this reason we will neglect the $m = 0$ case in what follows.
In the $m = \pm1$ channel, the decay distribution can be parametrized in terms of two quantities,
${\cal P}_{10}$ and ${\cal P}_{12}$, so that
\begin{equation}\label{eq:spin2_qq_param}
\frac{d\bar\sigma_{in}}{d\cos\theta} = \bar\sigma_{in} \left[{\cal D}^{(2)}_{1,0}\, {\cal P}_{10} + {\cal D}^{(2)}_{1,2} {\cal P}_{12}\right]\,.
\end{equation}

\begin{table}[t]
\begin{center}
\begin{tabular}{c|c|c|c|c} 
\rule{0pt}{1.2em}
 & ATLAS 13 & CMS 13 (EBEB, EBEE) & ATLAS 8 & CMS 8 \\
\hline
\rule{0pt}{1.1em}${\cal P}^{u\overline{u}}_{10}$ & $0.40$ & $0.32$\hspace{2em} $0.41$ & $0.80$ & $0.62$\\
\rule[-.5em]{0pt}{1.6em}${\cal P}^{u\overline{u}}_{12}$ & $0.70$ & $0.47$\hspace{2em} $0.28$ & $0.86$ & $0.80$\\
\hline
\rule{0pt}{1.1em}${\cal P}^{d\overline{d}}_{10}$ & $0.42$ & $0.41$\hspace{2em} $0.38$ & $0.83$ & $0.64$\\
\rule[-.5em]{0pt}{1.6em}${\cal P}^{d\overline{d}}_{12}$ & $0.71$ & $0.57$\hspace{2em} $0.25$ & $0.89$ & $0.82$\\
\hline
\rule{0pt}{1.1em}${\cal P}^{\rm sea~\overline{\rm sea}}_{10}$ & $0.41$ & $0.50$\hspace{2em} $0.35$ & $0.85$ & $0.64$\\
\rule[-.5em]{0pt}{1.6em}${\cal P}^{\rm sea~\overline{\rm sea}}_{12}$ & $0.72$ & $0.69$\hspace{2em} $0.21$ & $0.91$ & $0.84$\\
\hline
\rule{0pt}{1.1em}${\cal P}^{\rm univ}_{10}$ & $0.40$ & $0.36$\hspace{2em} $0.40$ & $0.81$ & $0.63$\\
\rule{0pt}{1.1em}${\cal P}^{\rm univ}_{12}$ & $0.70$ & $0.52$\hspace{2em} $0.26$ & $0.87$ & $0.81$
\end{tabular}
\caption{\small Acceptances for $q\overline{q}$-initiated spin-2 diphoton resonances. The acceptances for the sea quarks $s,c,b,t$ differ by less than $5\%$ and have been combined in a single class. }\label{ACCqq}
\end{center}
\end{table}

\begin{figure}[t]
\centering
\includegraphics[width=.5\textwidth]{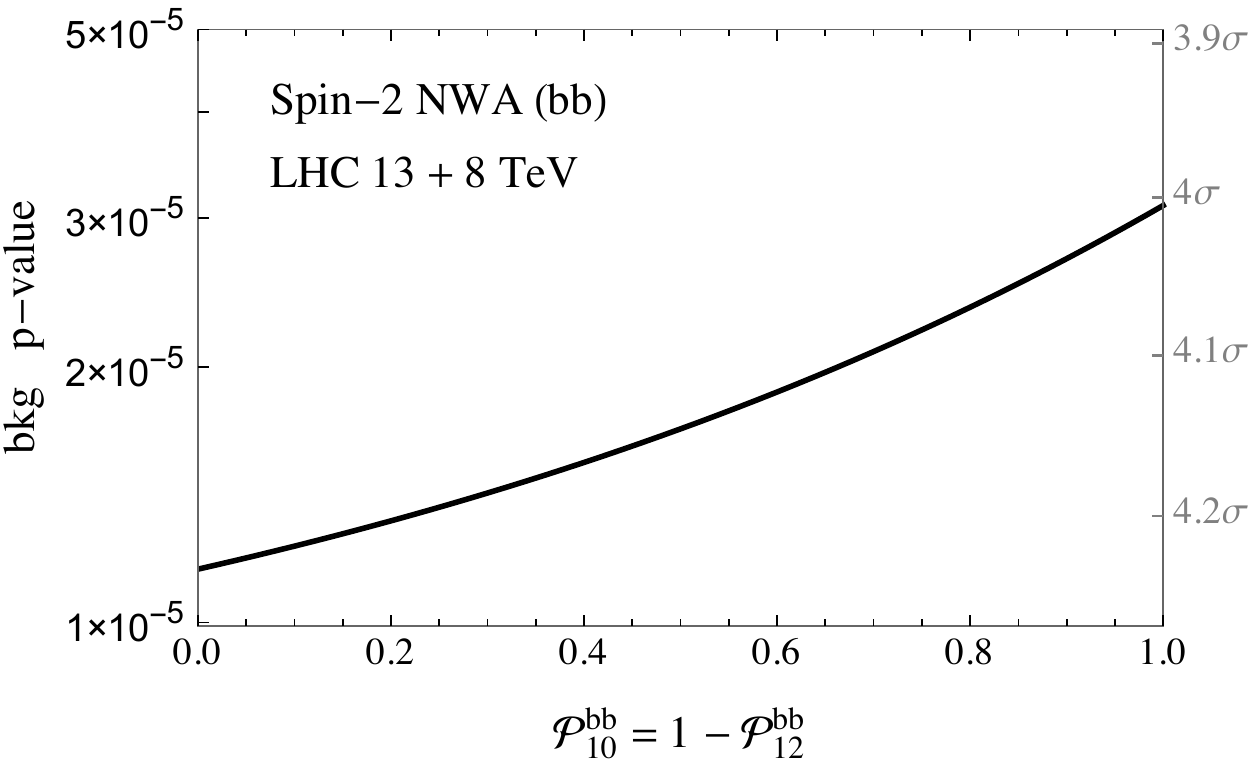}
\hfill
\includegraphics[width=.455\textwidth]{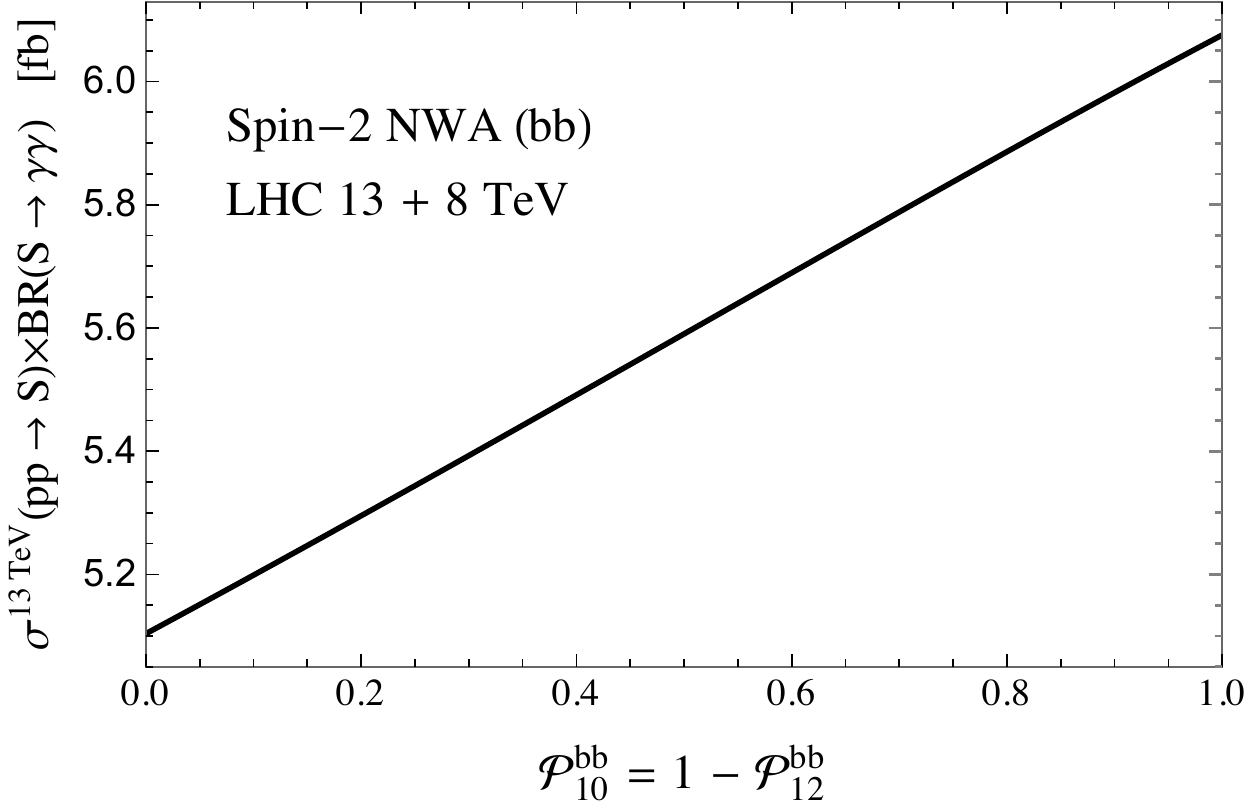}\\
\vspace{.5em}
\includegraphics[width=.455\textwidth]{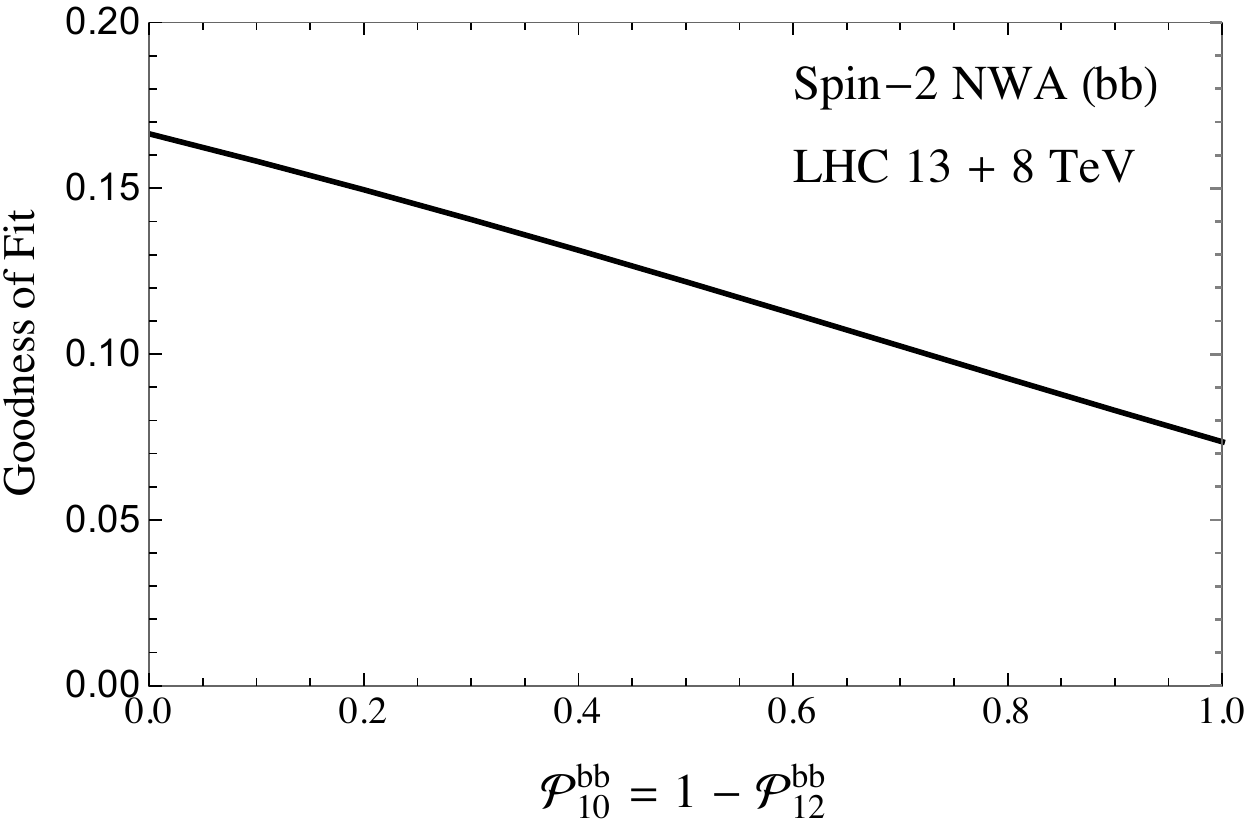}
\caption{On the upper left panel we show the reconstructed p-value for the background-only hypothesis in the scenario with a
narrow spin-$2$ resonance produced in the $bb$ channel. The results are presented as a function of the parameter
${\cal P}_{10} = 1- {\cal P}_{12}$, which encodes the angular distribution of the final-state photons
(see eq.~(\ref{eq:spin2_qq_param}). The numerical values are obtained by combining the
ATLAS and CMS $13$ and $8$~TeV searches. On the upper right panel we show the best fit of the $13$~TeV signal cross
section. In the lower panel we plot the goodness of fit.}\label{fig:spin2_bb}
\end{figure}

\begin{figure}[t]
\centering
\includegraphics[width=.5\textwidth]{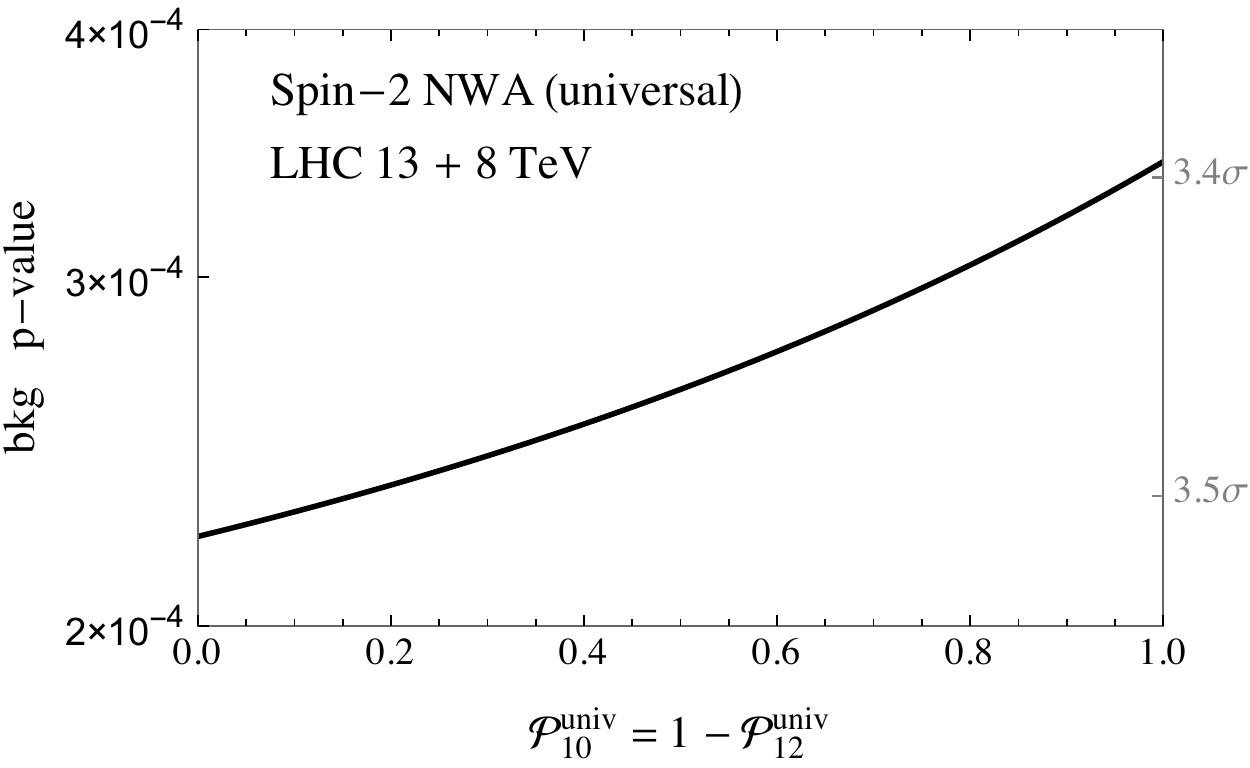}
\hfill
\includegraphics[width=.452\textwidth]{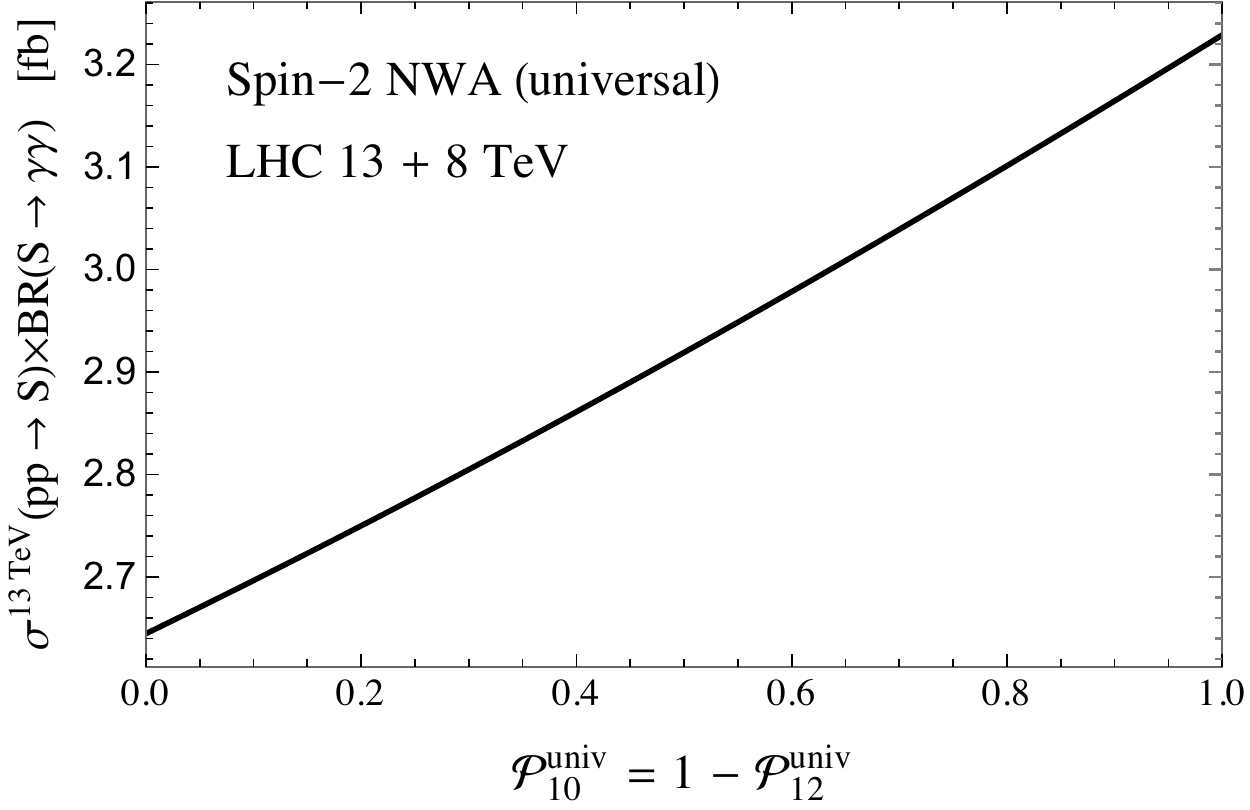}\\
\vspace{.5em}
\includegraphics[width=.475\textwidth]{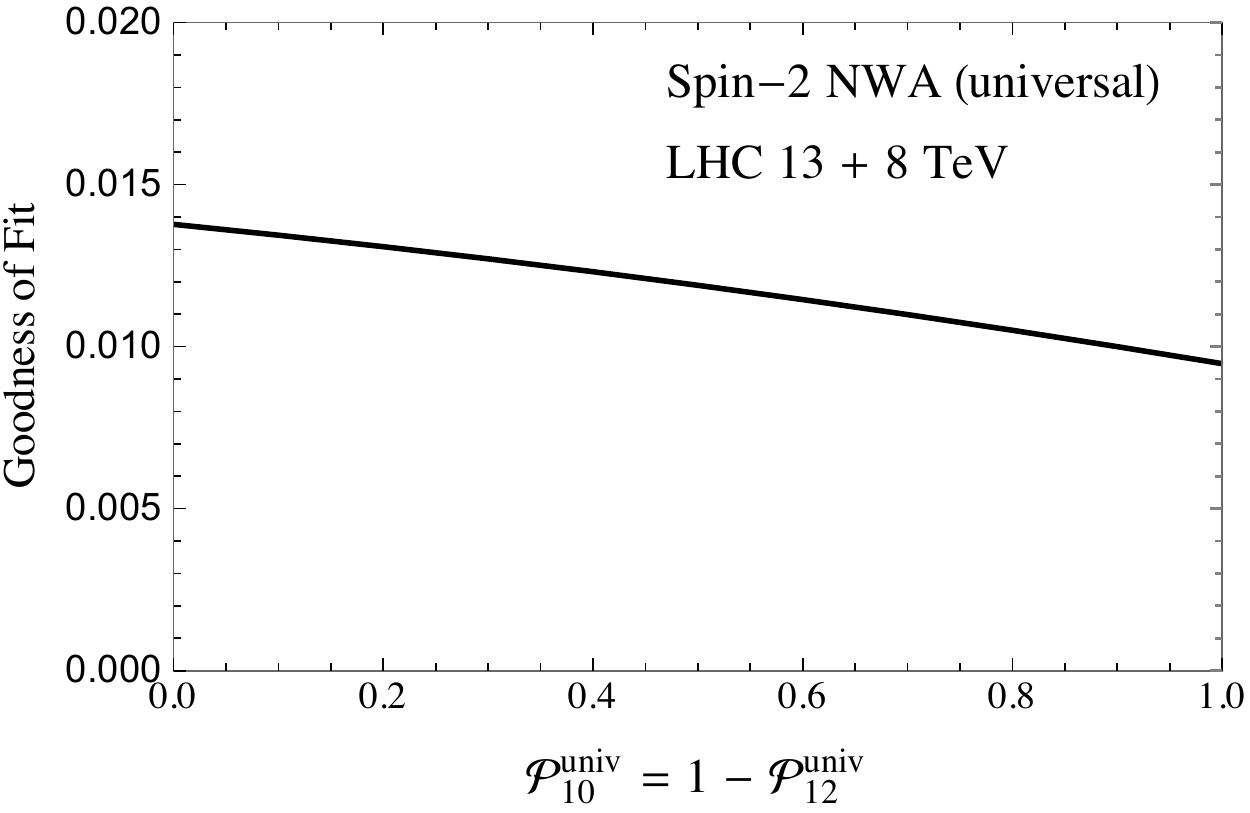}
\caption{On the upper left panel we show the reconstructed p-value for the background-only hypothesis in the scenario with a
narrow spin-$2$ resonance produced in the scenario with universal couplings to the quarks.
On the upper right panel we show the best fit of the $13$~TeV signal cross section. In the lower panel we plot the goodness of fit.
}\label{fig:spin2_univ}
\end{figure}

In principle all quarks could couple to the new resonance. However in order to avoid tensions with existing bounds 
we assume the resonance has negligible flavor-violating couplings to the light quarks. Plausible scenarios may be constructed if the coupling is either family-universal or dominantly with the heavy quarks (in particular with the third generation). In the following we will thus consider two benchmark scenarios. In the first the spin-$2$ resonance couples
dominantly to the bottom quark. In the second scenario it has a family-universal coupling with a single quark representation
(as, for instance, the right-handed up-type quarks). We will assume that in both scenarios the unavoidable
coupling to gluons can be neglected.

The geometric acceptances for the various quark production channels are listed in table~\ref{ACCqq}.
The signal significance and the best fit of the cross section for the $b\overline b$ production channel is shown in
fig.~\ref{fig:spin2_bb} as a function of the ${\cal P}_{10} = 1- {\cal P}_{12}$ parameter. One can see that the significance
is around $4\,\sigma$ and the cross section best fit is $\simeq 5.5~\textrm{fb}$. This scenario provides a
good compatibility among the experiments, at the level of $10 - 15\%$. The dependence on ${\cal P}_{10}$
is relatively mild due to the limited statistical precision currently available. As can be seen from table~\ref{ACCqq}, the
acceptances for the two different angular distributions differ significantly, thus they could allow to better differentiate
the various scenarios when more data will be available.

The results for the scenario with universal couplings to the fermions are shown in fig.~\ref{fig:spin2_univ}. In this case
the cross section gain factor is mostly determined by the one of the valence quarks and is given by $r_{\rm univ} = 2.9$.
Due to the relatively small gain factor the significance in this scenario is lower than in the $b\overline b$ channel,
namely it is around $\sim 3.5\,\sigma$. The best fit for the signal cross section is $\sim 3~\textrm{fb}$.
In this scenario the compatibility among the experiments is rather poor, at the $1\%$ level.

\subsubsection{The RS graviton}

We conclude the discussion of the spin-$2$ resonances by considering a well-known scenario that includes
a new state of this kind, the Randall-Sundrum (RS1) model.
In this case the spin-$2$ state is identified with the massive graviton of RS1, which is coupled to the stress tensor of the SM.
The particular form of the coupling implies a peculiar relation between $q\overline{q}$ and $gg$ production modes, namely
$\sigma_{q\overline{q}}=\sigma_{\overline{q}q}=\frac{2}{3}\sigma_{gg}$. It also completely fixes the angular distributions of the diphoton final state.
In the notation introduced in the previous section one gets ${\cal P}^{gg}_{22}={\cal P}_{12}^{q\overline{q}}=1$.
On top of fixing the properties of the diphoton final state, the RS1 scenario also determines the relative importance of
the other decay channels of the massive graviton. In particular it possesses large branching ratios into leptons,
which suggests that diphoton searches are probably not competitive with di-leptons for this specific scenario.
Taking into account the implications of the other final states, however, goes beyond the scope of this paper,
thus we just concentrate on the diphoton channel.

\begin{table}[t]
\begin{center}
\begin{tabular}{c|c|c|c|c} 
\rule{0pt}{1.2em}
& ATLAS 13 & CMS 13 (EBEB, EBEE) & ATLAS 8 & CMS 8 \\
\hline
\rule{0pt}{1.1em}RS1-graviton & $0.41$ & $0.43$\hspace{2.em} $0.31$ & $0.81$ & $0.60$
\end{tabular}
\caption{\small Acceptances for an RS1 graviton into photon pairs. }\label{ACCrs1}
\end{center}
\end{table}

The geometric acceptance for the RS1 graviton are listed in table~\ref{ACCrs1}, while the gain factor between the
$8$ and $13$~TeV cross section is mostly determined by the $gg$ production mode and is equal to $r_{RS} = 4.1$.
We find that the available searches imply a statistical significance of $3.8\,\sigma$ for a graviton with a mass $750~\textrm{GeV}$,
with a compatibility of the different searches at the $3\%$ level. The best fit of the signal cross section is $5~\textrm{fb}$.

\subsection{\boldmath Spin-$3$ resonances}

As a last benchmark scenario we discuss the case of spin-$3$ resonances. From table~\ref{pars}, one sees that for a resonance of odd spin produced through
$gg$ or $\gamma\gamma$ only the channels $|m| = 2$ are allowed. Therefore the most general $gg/\gamma\gamma$ cross sections can be written in terms of a single parameter
\begin{equation}\label{2n+1VV}
\frac{d\bar\sigma_{in}}{d\cos\theta}=\bar\sigma_{in} {\cal D}^{(2n+1)}_{2,2}\,.
\end{equation}
Quark production, analogously to the even-spin case, can instead occur via both $m=0$ and $|m| = 1$:
\begin{equation}\label{2n+1qq}
\frac{d\sigma}{d\cos\theta} = \sigma \left[{\cal D}^{(2n+1)}_{0,2} {\cal P}_{02} + {\cal D}^{(2n+1)}_{1,2}{\cal P}_{12}\right]\,.
\end{equation}


\begin{table}[t]
\begin{center}
\begin{tabular}{c|c|c|c|c} 
\rule{0pt}{1.2em}
Production & ATLAS 13 & CMS 13 (EBEB, EBEE) & ATLAS 8 & CMS 8 \\
\hline
\rule{0pt}{1.1em}gg & $0.66$ & $0.57$\hspace{2.em} $0.16$ & $0.80$ & $0.65$
\end{tabular}
\caption{\small Acceptances for a spin-$3$ resonance produced in the $gg$ channel. }\label{ACCspin3}
\end{center}
\end{table}

Here, for simplicity, we will focus on the case of a spin-$3$ resonance produced in the $gg$ channel. The acceptances for
this scenario are listed in Table~\ref{ACCspin3}. From our recast of the experimental searches we find that the hypothesis
of a resonance with a mass of $750~\textrm{GeV}$ has a statistical significance of $4.2\,\sigma$ with a best fit of the
signal cross section $5.6~\textrm{fb}$. The compatibility of the various experimental results is $14\%$.
The higher significance and better compatibility between the various searches comes from the fact that the decay
distribution of the two photons (controlled by ${\cal D}^{(3)}_{2,2}$) is quite central (similarly to ${\cal D}^{(2)}_{0,2}$ for the analogous spin-2 benchmark model). This implies a larger geometric acceptance for the ALTAS $13$~TeV search and a slightly lower acceptance for the other searches. This difference mitigates the preference for higher signal
strengths implied by the ATLAS $13$~TeV data.

\section{Conclusions and outlook}\label{four}

In this paper we provided a general characterization of the resonant diphoton production at hadron colliders. Our main result is the derivation of a new, simple phenomenological parametrization that can be used to describe resonances with arbitrary (integer) spin and CP parity, produced in any of the $gg$, $q{\overline{q}}$ and $\gamma\gamma$ partonic channels. By exploiting angular momentum conservation, the decay distributions of the resonance can be expressed as a combination of a small number of basis function that encode the angular distributions of the diphoton pair in the COM frame. The form of the basis functions is fully determined by the spin of the resonance. Their relative importance in the signal distributions, as well as the relative importance of the various production channels, are controlled by polarized resonance cross sections and decay branching ratios.

An important advantage of our parametrization is the fact that it does not depend on any assumption about the underlying theory describing the resonance. In particular it can be used even if the resonance dynamics can not be encoded into a local effective Lagrangian, which could be the case if it emerges from a strongly-coupled QCD-like dynamics. Our approach is thus completely model-independent and particularly suitable to describe in an unbiased way a possible signal observed in the diphoton channel.
Although mainly aimed at characterizing a possible signal, our parametrization can also be used to express exclusions in the case of a measurement compatible with the background-only hypothesis.

As an example of the use of our results, we performed a simple recast of the ATLAS and CMS resonant diphoton searches, which recently reported an excess around an invariant mass $M = 750$~GeV. These recasts should not be interpreted as fully quantitative results, but rather as an illustration of the usage of our parametrization. For definiteness we focused on a few benchmark scenarios with resonances of spin $J = 0$, $J = 2$ and $J = 3$.

The $J=0$ case is particularly simple, since the diphoton angular distribution is fixed to be completely flat in the COM frame. The properties of the resonance thus only depend on the relative importance of the various partonic production channels. Each channel is characterized by the gain ratio between the $8$ and $13$~TeV production cross section and by the acceptances in the various searches, which depend on the $y$ distribution. We found that the $gg$, $s\overline s$, $c\overline c$ and $b\overline b$ channels are quite similar and difficult to distinguish experimentally. The situation is instead different for the $u\overline u$, $d\overline d$ and $\gamma\gamma$ channels, which have a significantly smaller gain ratio with respect to the $gg$ mode. The present data show some degree of tension between the $8$~TeV results and the $13$~TeV ones, in particular the ATLAS analysis, which prefers large gain ratios. As a consequence the $gg$ or heavy-quarks production modes are favored. In this cases a good signal significance, $\sim4\sigma$, is found with a $~9\%$ compatibility among the various searches. The compatibility is instead poor, around $1\%$, in the case of the light-quarks or $\gamma\gamma$ production modes.

Since they can lead to different non-trivial angular distributions for the diphoton pair, spin-$2$ resonances are characterized by a more varied phenomenology. Also in this scenario production channels, as the $gg$ one, with large gain ratios are preferred. Moreover more central angular distributions are slightly favored since they lead to a higher acceptance, especially in the $13$~TeV searches (see table~\ref{ACCgg}). In the most favorable case, namely $gg$ production with the ${\cal D}_{0,2}^{(2)}$ angular distribution, a signal significance of $\sim 4.2\sigma$ is found with a compatibility of $12\%$ between the various searches. Another scenario that has a good compatibility with the data is the case of $b\overline b$-initiated production, which can lead to an overall compatibility of $15\%$. Another spin-$2$ benchmark we considered is the case of a Randall-Sundrum massive graviton. In this scenario the production mode is dominantly $gg$ and the angular distribution is described by the function ${\cal D}_{2,2}^{(2)}$, which leads to a more forward diphoton distribution. This property implies a not so good compatibility with the data, at the $3\%$ level.

As a final scenario we considered a spin-$3$ resonance produced in the $gg$ channel. This set-up is particularly simple since it is characterized by a single angular function, ${\cal D}_{2,2}^{(3)}$. In this case we find a good significance $\sim 4.2\sigma$ and a good compatibility among the various searches, $\sim 14\%$.

Besides providing the general framework within which benchmark scenarios can be defined, the phenomenological analysis presented in section~\ref{sec:general_framework} allows us to draw interesting conclusions concerning which properties of the resonance (once it is discovered) could be extracted from a careful experimental study of the resonant diphoton signal. Namely, we saw that the resonance spin and production mode could be established, barring peculiar degeneracies which we have identified, from the combined measurement of the $\cos\theta$ and $y$ distributions. Within a given hypothesis for the resonance spin and production mode, the $\cos\theta$ distribution also gives us information about the resonance CP-parity. Indeed non-vanishing $A_{\pm\mp}$ amplitudes (recall that the $a$'s in table~\ref{pars} are CP-even while the ${\widetilde{a}}$'s are CP-odd), which we could detect through the presence of a ${\cal{D}}_{1,S}$ or ${\cal{D}}_{2,S}$ component in the angular distribution, would imply either that the resonance is CP-even or that CP is badly broken by the resonance couplings. If instead $A_{\pm\mp}$ were to vanish, we would not be able to distinguish a CP-odd ${\cal R}$ from a CP-even resonance with accidentally vanishing $a_{1,-1,2}$. The only way to achieve this would be to measure ${{a}}_0$ and ${\widetilde{a}}_0$ separately, but this is impossible since only a combination of the two enters, through eq.~(\ref{A2}), in the differential cross section. This problem is particularly severe for $J=0$, where $A_{\pm\mp}=0$ by spin conservation and thus the resonance CP-parity cannot be measured.

A possible way out is to study, as pointed out in refs.~\cite{Harland-Lang:2016qjy,Kaidalov:2003fw} for the $J=0$ case, the structure of the forward initial state radiation (ISR) that unavoidably accompanies the hard resonance production process. Consider the emission of two forward ISR jets \footnote{For $\gamma\gamma$-initiated processes, the objects produced by ISR might not be jets, but the single protons that elastically emitted the initial state photons~\cite{Harland-Lang:2016qjy}.} emitted in the forward and backward direction, respectively,  and denote by $\varphi_{1}$ and $\varphi_2$ their azimuthal angles. For $p_\bot(j_{1,2})\ll M$, the Feynman amplitude for the complete $2\to4$ process takes the form~\cite{Borel:2012by}
\bea
{\cal{A}}(in\to j_1j_2\gamma\gamma)\propto\sum_{\lambda_1\lambda_2}{g}_{\lambda_1}{\hspace{-2pt}}(x_1){g}_{\lambda_2}{\hspace{-2pt}}(x_2)e^{-i\lambda_1\varphi_1+i\lambda_2\varphi_2}A_{\lambda_1\lambda_2}^{in}e^{i\phi(\lambda_1-\lambda_2-\lambda+\lambda')}d^J_{\lambda_1-\lambda_2,\lambda-\lambda'}
A^{\gamma\gamma}_{-\lambda,-\lambda'}\,,
\eea
where $\phi$ is the azimuthal angle of the hard scattering plane, i.e. the one of the diphoton pair appearing in eq.~(\ref{Wigner}). The $e^{i\lambda\varphi}$ factors from the parton splittings are dictated by momentum conservation, as discussed in ref.~\cite{Borel:2012by} for the case of effective massive vector bosons splittings. The ${g}_{\lambda_{1,2}}$'s are given functions, specific of the ISR splitting process at hand, of the incoming partons momentum fractions $x_{1,2}$. The above formula illustrates that by studying the kinematical distributions of the ISR jets one can get more information about the polarized resonant production amplitudes than that obtainable from the $2\to2$ process. Taking for simplicity the soft limit, in which the most singular $g$-functions (i.e., those for $gg$, $qg$, and $qq$ splittings) become independent of $\lambda$, one easily obtains an approximate formula for the complete $2\to4$ process cross section. Such a cross section, differential in the azimuthal angular difference between the two jets, $\varphi_{12}=\varphi_1-\varphi_2$, and integrated over all other variables, reads
\bea
\frac{d\bar\sigma_{gg{\textrm{/}}\gamma\gamma}}{d\varphi_{12}}&\propto&2|A_{++}||A_{--}|\cos(2\varphi_{12}+\delta)+|A_{++}|^2+|A_{--}|^2+2|A_{+-}|^2\,,\no\\
\frac{d\bar\sigma_{q\overline{q}}}{d\varphi_{12}}&\propto&2|A_{++}||A_{--}|\cos(\varphi_{12}+\delta)+|A_{++}|^2+|A_{--}|^2+|A_{+-}|^2+|A_{-+}|^2\,,\no
\eea
for, respectively, $gg{\textrm{/}}\gamma\gamma$ and $q{\overline{q}}$ hard production. We defined $\delta={\rm arg}(A_{++}/A_{--})$. Because according to table~\ref{pars} a CP-even (CP-odd) resonance has $\delta=0$ ($\delta=\pi$), we see that measuring the $\varphi_{12}$ distribution one would be able to infer the resonance CP-parity, even for $J=0$. The distribution can also tell us if the $a$'s in table~\ref{pars} are complex, which would mean that the resonance interactions are mediated by loops of light particles as we discussed around eq.~(\ref{cptrel}). Indeed, it might allow to extract the ration $|A_{++}|/|A_{--}|$, which is necessarily equal to one if the $a$'s and ${\widetilde{a}}$'s are real. However $|A_{++}|/|A_{--}|=1$ is also ensured by the CP symmetry, therefore observing $|A_{++}|/|A_{--}|\neq1$ would also mean that CP is broken.

Another process which is worth considering, because of its larger rate, is the emission of a single detectable forward jet, with azimuthal angle $\varphi_j$. In this case one must study the doubly differential distribution in $\cos\theta$ and in $\varphi=\varphi_j-\phi$, i.e. the angle between the jet and the diphoton plane. The angular dependence, focusing once again on the soft/collinear limit and assuming for simplicity a heavy mediator (real $a$'s), is controlled by
\bea
&&\frac{d^2\bar\sigma_{gg/\gamma\gamma}}{d\varphi\cos\theta}\propto\sum_{S=0,2}
{\rm BR}_S
\left\{2[(a_0^{g{\textrm{/}}\gamma})^2+({\widetilde{a}}_0^{g{\textrm{/}}\gamma})^2](d_{0, S}^2+d_{0,-S}^2)+2(a_2^{g{\textrm{/}}\gamma})^2(d_{2, S}^2+d_{2,-S}^2)\right.\no\\
&&+\left.
a_2^{g{\textrm{/}}\gamma}
\left[a_0^{g{\textrm{/}}\gamma}\cos2\varphi+{\widetilde{a}}_0^{g{\textrm{/}}\gamma}\sin2\varphi\right][d_{0,S}(d_{2,S}+d_{-2,S})+d_{0,-S}(d_{2,-S}+d_{-2,-S})]\right\}\no
\eea
for $gg{\textrm{/}}\gamma\gamma$ hard production, and
\bea
&&\frac{d^2\bar\sigma_{q{\overline{q}}}}{d\varphi\cos\theta}\propto\sum_{S=0,2}
{\rm BR}_S
\left\{2[(a_0^{q})^2+({\widetilde{a}}_0^{q})^2](d_{0, S}^2+d_{0,-S}^2)+[(a_1^{q})^2+(a_{-1}^{q})^2](d_{2, S}^2+d_{2,-S}^2)\right.\no\\
&&+\left.
\left[a_0^{q}\cos\varphi+{\widetilde{a}}_0^{q}\sin\varphi\right][d_{0,S}(a_1^q d_{1,S}+a_{-1}^q d_{-1,S})+d_{0,-S}(a_{1}^q d_{1,-S}+a_{-1}^q d_{-1,-S})]\right\}\no
\eea
for $q\overline{q}$ (a similar result holds for $\overline{q}q$). Here the $d$'s are the Wigner matrices for a generic spin $J$ and ${\rm BR}_S$ is the polarized branching ratio of eq.~(\ref{br}). Differently from the $2$-jets emission previously discussed, studying the single ISR jet distribution does not  furnish conclusive information about the resonance CP-parity at $J=0$ because the dependence on $\varphi$ disappears in the scalar case. Still, the measurement of this process gives access to different parameter combinations which do not appear in the fully inclusive $2\to2$ reaction and thus it is nevertheless worth studying.

A detailed analysis of the ISR radiation pattern, and its potential implications for the experimental characterization of the resonance properties, is left for future work.

\section*{Acknowledgments}

We are indebted with R.~Rattazzi for point out to us an hidden assumption which was present in the first version of the manuscript. We thank R. Franceschini for discussions on the uncertainties in the photon PDF. The work of G.~P. has been partly supported by the Spanish Ministry MEC grant FPA2014-55613-P,
by the Generalitat de Catalunya grant 2014-SGR-1450 and by the Severo Ochoa excellence program
of MINECO (grant SO-2012-0234). A.~W.~acknowledges the MIUR-FIRB grant \sloppy\mbox{RBFR12H1MW} and the ERC Advanced Grant no.267985 ({\emph{DaMeSyFla}}). The work of L.~V. is supported by {\emph{DaMeSyFla}}. We thank J. Rojo and R.~Torre for discussions.

\appendix

\section{On-shell amplitudes}\label{effective_ops}

In this appendix we will derive the effective couplings that parametrize the on-shell dynamics of a spin-$0$
or spin-$2$ resonance decaying into a photon pair. We will first compute the on-shell amplitudes for the production of ${\cal R}$, from which the $A_{\lambda_1\lambda_2}^{in}$ immediately follow. The analogous amplitudes for ${\cal R}\to\gamma\gamma$ can be straightforwardly obtained from them. In section~\ref{sec:eff_Lagrangian} we will then present an effective Lagrangian that may be employed to implement the relevant processes into a Montecarlo generator. 

The amplitudes for $in\to{\cal R}$ depend only on a few basic quantities.
First of all they are a function of the $4$-momenta of the initial partons, which we denote by
$p_1^\mu$ and $p_2^\mu$. For later convenience, we introduce the notation $p^\mu=p_1^\mu+p_2^\mu$ for the resonance
momentum and $q^\mu=p_1^\mu-p_2^\mu$ for the other independent combination of the initial momenta.
The only non-trivial Lorentz scalar is given by the resonance mass, $p^2=-q^2=M^2$. The amplitudes also depend on the polarization vectors $\epsilon^\mu_{1,2}$ (for $gg/\gamma\gamma$ production) and the spinors $u_1$ and $v_2$ of the SM quarks (for $q\overline q$ production).
In the case of a spin-$2$ resonance, an additional tensor $t_{\mu\nu}$ is present, that describes the polarization
of ${\cal R}$.

Since the external states are on-shell, the equations of motion may be used to simplify the expressions.
Specifically, the polarization tensor $t_{\mu\nu}$ of a spin-$2$ resonance is required to satisfy the same conditions
as an on-shell spin-$2$ field, {\it i.e.}~to be transverse and symmetric-traceless. This implies that its contraction with
$p^\mu$ vanishes and the only non-trivial terms can be obtained by contractions with $q^\mu$ and $\epsilon_{1,2}^\mu$.
Similarly, the equations of motion for the SM quarks can be used
to remove factors of $\gamma_\mu p^\mu_i$ from the amplitudes relevant to $q\overline q$ production. Furthermore, because of the transversality of the gauge bosons $\epsilon_i^\mu p_{i\,\mu}=0$ may be used to simplify the expressions for $\gamma\gamma, gg$ production and $\gamma\gamma$ decay. Finally, Lorentz invariance constrains the form of the amplitudes for $gg, \gamma\gamma\to{\cal R}$ (and analogously ${\cal R}\to\gamma\gamma$) in a non-trivial way, by forcing them to be
invariant under a shift $\epsilon_i^\mu\to\epsilon_i^\mu+p_i^\mu$, where $p_i^\mu$ is on-shell~\cite{WeinbergBook}. Amplitudes consistent with Lorentz invariance therefore automatically satisfy the on-shell Ward identities.

Interestingly, we find that the amplitudes for $gg, \gamma\gamma\to{\cal R}$ derived following the above recipe can be unambiguously uplifted to expressions valid for {\emph{off-shell}} gauge bosons (and not necessarily transversely polarized) and respecting the {\emph{off-shell}} Ward identities. This allowed us to perform a sum over gauge boson helicities in the familiar fashion $\sum\epsilon^\mu(\epsilon^\nu)^*\to-g^{\mu\nu}$ and check that the squared amplitudes thus obtained agree with the ones derived from the helicity amplitudes and the general formalism of section~\ref{sec:general_framework}. For completeness we will present our results in this off-shell form.
 
In the following we will specialize the discussion to the $J=0$ and $J=2$ cases, although resonances with higher spin can be
treated analogously. Our results agree with ref.~\cite{Gao:2010qx} up to phase conventions.

\subsection{Spin-0 resonance}
\label{VV}

As a first case we consider the production amplitude for a spin-$0$ resonance. In the $gg$ or $\gamma\gamma$ channels
we find
\bea\label{eq:VV0}
{\cal A}(gg/\gamma\gamma\to{\cal R})&=&2\frac{a^{g/\gamma}_0}{M}\left[(\epsilon_1p_2)(\epsilon_2p_1)-(\epsilon_1\epsilon_2)(p_1p_2)\right]\\\no
&+&2\frac{\widetilde a^{g/\gamma}_0}{M}~\epsilon_{\mu\nu\alpha\beta}\epsilon_1^\mu p_1^\nu\epsilon_2^\alpha p_2^\beta\,.
\eea
For $in=q\overline{q}$ instead we obtain
\bea\label{eq:qq0}
{\cal A}({q\overline{q}}\to{\cal R})=-{a^q_0}\,\overline{v}_2u_1+{\widetilde a^q_0}\,i\overline{v}_2\gamma^5u_1\,.
\eea
In the above equations, the CP-even (CP-odd) coefficients $a_0, b_0$ are all dimensionless
quantities (in general complex).

The helicity amplitudes $A_{\lambda_1\lambda_2}$ can be straightforwardly derived from eq.s~(\ref{eq:VV0}) and (\ref{eq:qq0}). For the polarization tensors (including both $J=1,2$) we use the conventions of~\cite{Hagiwara:2008jb}, whereas the spinors are taken from~\cite{Peskin}. The result reads
\bea\label{ab0}
{A}^{gg/\gamma\gamma}_{++}=a^{g/\gamma}_0+i\widetilde a^{g/\gamma}_0\,,
~~~~~~&&~~~~~~
{A}^{q\overline{q}}_{++}=a^q_0+i\widetilde a^q_0,\\\no
{A}^{gg/\gamma\gamma}_{--}=a^{g/\gamma}_0-i\widetilde a^{g/\gamma}_0\,,
~~~~~~&&~~~~~~
{A}^{q\overline{q}}_{--}=a^q_0-i\widetilde a^q_0,
\eea
in agreement with table~\ref{pars}.

\subsection{Spin-2 resonance}

We can now discuss the scenario with a spin-$2$ resonance. In this case there are two differences with respect
to the spin-$0$ state. First, the amplitude will depend on the polarization tensor of the resonance $t_{\mu\nu}$.
Second, the amplitudes for $gg/\gamma\gamma\to{\cal R}$, which according to the rules described above may contain terms that are anti-symmetric in the exchange of the incoming state particles, must be symmetrized since the two gauge bosons are indistinguishable. 

Starting with $gg,\gamma\gamma\to{\cal R}$, we obtain
\bea\label{eq:VV2}
{\cal A}(gg/\gamma\gamma\to{\cal R})&=&\sqrt{6}\frac{a^{g/\gamma}_0}{M^3}t_{\mu\nu}q^\mu q^\nu\left[(\epsilon_1p_2)(\epsilon_2p_1)-(\epsilon_1\epsilon_2)(p_1p_2)\right]\\\no
&+&\frac{a^{g/\gamma}_2}{M}t_{\mu\nu}\left[-(\epsilon_1p_2)q^\mu \epsilon_2^\nu+(\epsilon_2p_1)q^\mu \epsilon_1^\nu+2(p_1p_2) \epsilon^\mu_1\epsilon_2^\nu-\frac{1}{2}(\epsilon_1\epsilon_2)q^\mu q^\nu\right]\no\\
&+&\sqrt{6}\frac{\widetilde a^{g/\gamma}_0}{M^3}~\epsilon_{\mu\nu\alpha\beta}\epsilon_1^\mu p_1^\nu\epsilon_2^\alpha p_2^\beta\left(t_{\rho\sigma}q^\rho q^\sigma\right).\no
\eea
To arrive at these expression we used the Schouten identity to eliminate all CP-odd structures in which $t_{\mu\nu}$
is contracted with $\epsilon^\mu_i$ or with the Levi-Civita tensor. These either vanish identically or are equivalent
to a renormalization of the vertices in eq.~(\ref{eq:VV2}). In the case $in=q\overline{q}$ we find
\bea\label{eq:qq2}
{\cal A}({q\overline{q}}\to{\cal R})&=&-\sqrt{\frac{3}{2}}\frac{a^q_0}{M^2}t_{\mu\nu}\overline{v_2}u_1q^\mu q^\nu\\\no
&+&t_{\mu\nu}q^\nu\left[\frac{a^q_1}{M}\overline{v_2}\left(\frac{1+\gamma^5}{2}\right)\gamma^\mu u_1+\frac{a^q_{-1}}{M}\overline{v_2}\left(\frac{1-\gamma^5}{2}\right)\gamma^\mu u_1\right]\\\no
&+&\sqrt{\frac{3}{2}}\frac{\widetilde a^q_0}{M^2}\overline{v_2}i\gamma^5u_1(t_{\mu\nu}q^\mu q^\nu).
\eea

From eq.s~(\ref{eq:VV2}) and (\ref{eq:qq2}) one can derive the corresponding helicity amplitudes
\bea
\left\{
\begin{array}{l}
{A}^{gg/\gamma\gamma}_{++}=a^{g/\gamma}_0+i\widetilde a^{g/\gamma}_0\\
{A}^{gg/\gamma\gamma}_{--}=a^{g/\gamma}_0-i\widetilde a^{g/\gamma}_0\\
\rule{0pt}{1.5em}{A}^{gg/\gamma\gamma}_{+-}={A}_{-+}^{gg/\gamma\gamma}=a^{g/\gamma}_2
\end{array}
\right.,
~~~
\left\{
\begin{array}{l}
{A}^{q\overline{q}}_{++}=a^q_0+i\widetilde{a}_0^q\\
{A}^{q\overline{q}}_{--}=a^q_0-i\widetilde{a}_0^q\\
\rule{0pt}{1.25em}{A}^{q\overline{q}}_{+-}=a^q_1\\
\rule{0pt}{1.25em}{A}^{q\overline{q}}_{-+}=a^q_{-1},
\end{array}
\right..
\eea
again in agreement with table~\ref{pars}.

\subsection{On-shell Lagrangian}\label{sec:eff_Lagrangian}

Finally we present two effective Lagrangians that may be used to simulate spin-$0$ and spin-$2$ diphoton resonances through Montecarlo generators. 

The various terms appearing in the production amplitudes in eq.s~(\ref{eq:VV0}), (\ref{eq:qq0}) and (\ref{eq:VV2}), (\ref{eq:qq2})
may be thought of as effectively arising from the following set of effective operators
\bea\label{eq:L0}
{\cal L}^{(J=0)}&=&{\cal R}\left[-\frac{a^{g/\gamma}_0}{2M}F^{\mu\alpha}F_{\mu\alpha}+\frac{\widetilde a^{g/\gamma}_0}{2M}F_{\mu\alpha}\widetilde F^{\mu\alpha}\right]\\\no
&+&{\cal R}\left[-a^q_0\overline{q}q+i\widetilde a^q_0\overline{q}\gamma^5q\right]\,,
\eea
in the case of a spin-$0$ resonance, and
\bea\label{eq:L2}
{\cal L}^{(J=2)}&=&{\cal R}_{\mu\nu}\left[\frac{a^{g/\gamma}_2}{M}F^{\mu\alpha}F^{\nu}_\alpha-\sqrt{6}\frac{a^{g/\gamma}_0}{M^3}\partial^\mu F_{\alpha\beta}\partial^\nu F^{\alpha\beta}+\sqrt{6}\frac{\widetilde a^{g/\gamma}_0}{M^3}\partial^\mu F_{\alpha\beta}\partial^\nu \widetilde F^{\alpha\beta}
\right]\\\no
&+&{\cal R}_{\mu\nu}\left[\frac{a^q_1}{M}i\overline{q}\left(\frac{1+\gamma^5}{2}\right)\gamma^\mu \partial^\nu q+\frac{a^q_{-1}}{M}i\overline{q}\left(\frac{1-\gamma^5}{2}\right)\gamma^\mu \partial^\nu q+{\rm hc}\right]\\\no
&+&{\cal R}_{\mu\nu}\left[-4\sqrt{\frac{3}{2}}\frac{a^q_0}{M}\partial^\mu\overline{q}\partial^\nu q+4\sqrt{\frac{3}{2}}\frac{\widetilde a^q_0}{M}i\partial^\mu\overline{q}\gamma^5 \partial^\nu q\right]\,,
\eea
in the case of a spin-$2$ state. In our notation $F_{\mu\nu}$ is the field strength of either photons or gluons
and $\widetilde F_{\mu\nu}=\frac{1}{2}\epsilon_{\mu\nu\alpha\beta} F^{\alpha\beta}$.

We warn the reader that eq.s~(\ref{eq:L0}) and (\ref{eq:L2}), where in general $a, \widetilde a$ are {\emph{complex}}, are not meant to describe the off-shell dynamics of these models. They just represent a practical parametrization of the {\emph{on-shell}} couplings relevant for resonant production. Even if ${\cal R}$ ultimately arises from a theory consistent with Lorentz and gauge symmetries that does not admit a Lagrangian formulation, the effective description~(\ref{eq:L0}) (\ref{eq:L2}) can be used to parametrize the diphoton resonant production and decay.

\section{Statistical treatment}\label{app:statistics}

In this appendix we briefly describe the statistical procedure we used to obtain the numerical results for the benchmark
scenarios presented in section~\ref{three}. For our recast we followed a simple strategy to reconstruct the profile
likelihood ratio of the various searches ($q(\mu)$, where $\mu$ is the signal strength parameter)
by exploiting the available data,
namely the p-value of the background-only hypothesis and the exclusion limits on the signal cross section.

The p-value is directly connected to the value of the profile likelihood ratio for a vanishing signal hypothesis $q(\mu = 0)$.
In the asymptotic limit, $q$ follows a half-$\chi^2$ distribution~\cite{Cowan:2010js},
\begin{equation}
f(q) = \frac{1}{2} \delta(q) + \frac{1}{2}\frac{1}{\sqrt{2 \pi}} \frac{1}{q} e^{-q/2}\,,
\end{equation}
and the background-only p-value corresponds to the cumulative distribution starting at $q(\mu = 0)$.\footnote{Notice that
the ATLAS collaboration in the analysis of the $13$~TeV data used a slightly different procedure, the uncapped p-value.
This definition, however, coincide with the usual one if the best fit of the cross section is for $\mu >0$. This is always the case
if there is an excess in the data, as it happens in the available ATLAS and CMS $8$ and $13$~TeV searches
for $m_{\gamma\gamma} \sim 750~\textrm{GeV}$.}
From the exclusion limits, instead, one can reconstruct the value of the cross signal strength $\mu$ for which
the cumulative distribution is equal to the exclusion threshold.  These elements are enough to reconstruct the profile likelihood
if we assume that a Gaussian approximation is valid. In this case the profile likelihood is just a quadratic polynomial in $\mu$,
that is always positive and vanishes in a single point, i.e.~for $\mu$ equal to the signal strength best
fit $\hat \mu$.\footnote{This procedure is correct in the case in which an excess is present in the data,
in which case necessarily $\hat \mu >0$. In the case of a deficit of events the signal strength best fit would be negative
$\hat \mu < 0$, but  the profile likelihood is defined in such a way to vanish for $\mu = 0$. In this case the knowledge of the
background-only p-value and of the exclusion limit is not enough to fully reconstruct the likelihood ratio.}
Notice that with our procedure we are only able to extract the global likelihood ratio for each experiment, but we have no
access to the likelihood for the single event categories used in the experimental analysis. For instance in the
CMS $13$~TeV analysis two categories are considered which could give some information on the angular distribution of the
diphoton events. Due to the limited statistics, however, this information is not extremely significant and our approximate
results are reliable. We will discuss this point quantitatively in subsection~\ref{sec:categories}.

\begin{figure}[t]
\centering
\includegraphics[width=.485\textwidth]{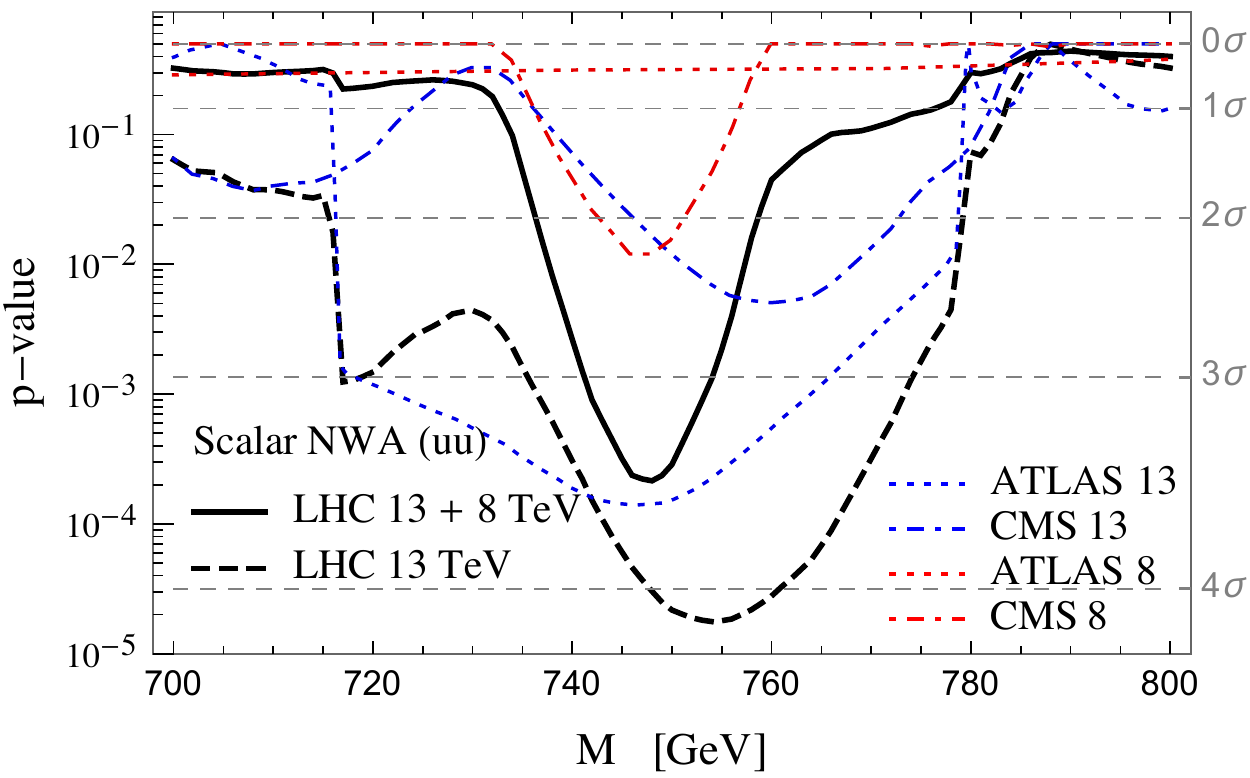}
\hfill
\includegraphics[width=.485\textwidth]{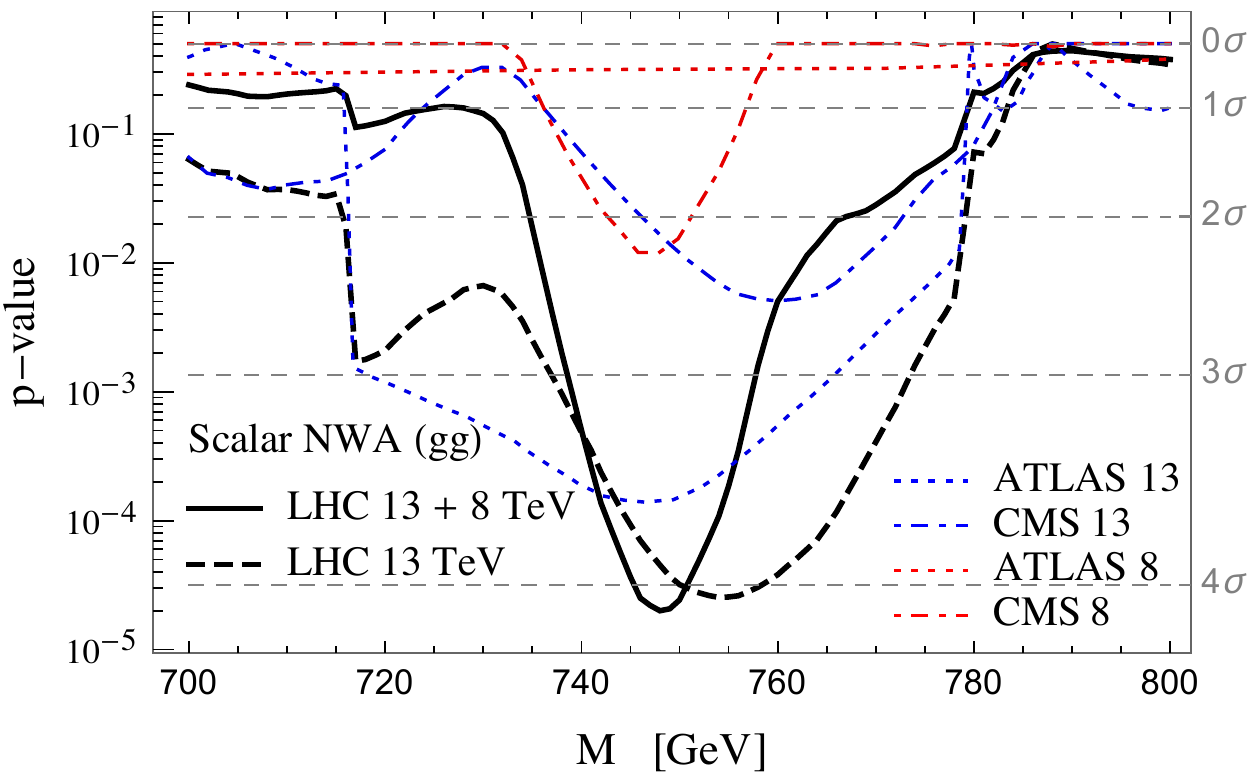}\\
\vspace{.5em}
\includegraphics[width=.485\textwidth]{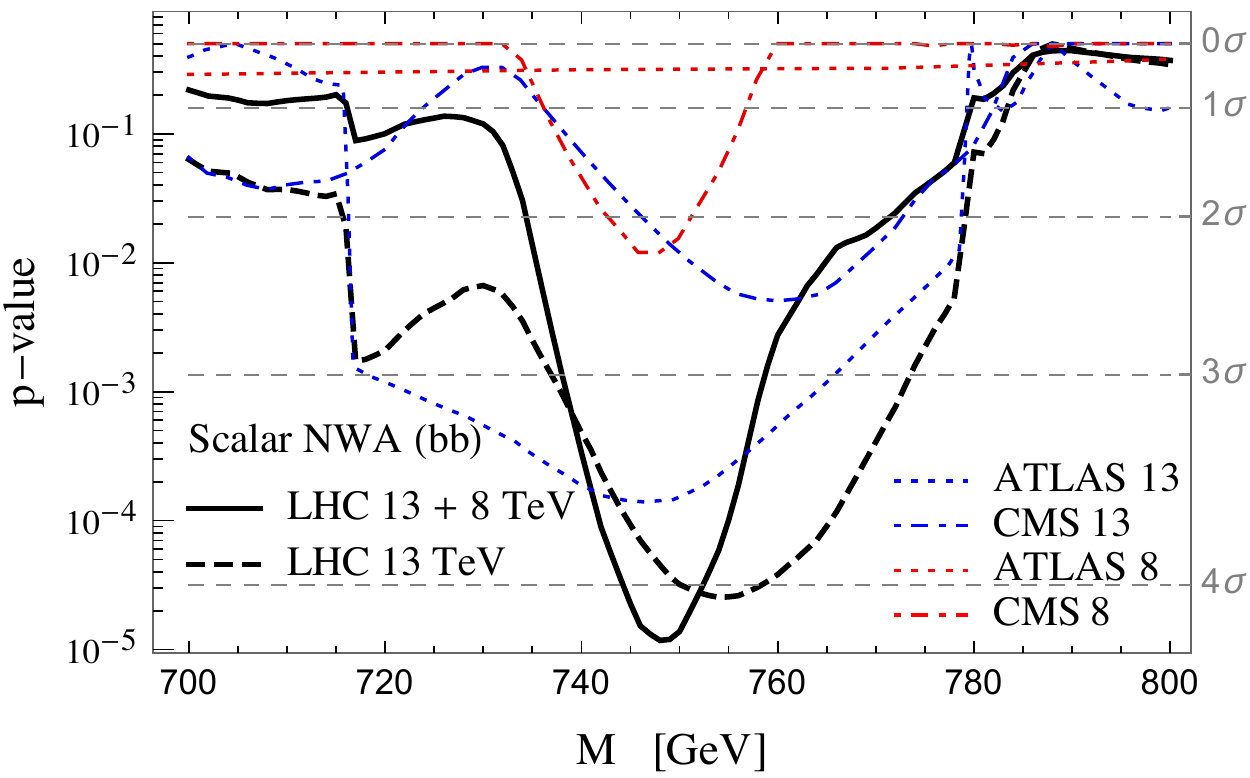}
\caption{Reconstructed p-value for the background-only hypothesis for the scenario with a narrow scalar resonance.
The results are shown as a function of the resonance mass. The upper plots correspond to the cases with
$u\overline u$ and $gg$ production modes, while the lower plot corresponds to the $b\overline b$ channel.
}\label{fig:pvalue_mass}
\end{figure}

The above procedure can be straightforwardly applied to the CMS $8$ and $13$~TeV analyses and to the ATLAS $13$~TeV one,
which provide the p-value and the exclusion limits as a function of the diphoton system invariant mass for the case
of a narrow-width resonance. In the recast it is important to take into account the fact that the CMS results are provided
for a scenario with a RS graviton, while ATLAS consider the case of a scalar resonance. This implies different acceptances
and reconstruction efficiencies as we discussed in section~\ref{three}. For the ATLAS $8$~TeV search, on the other hand,
only the exclusion limits are publicly available. In this case we reconstructed the profile likelihood by estimating the
width of the $1\,\sigma$ and $2\,\sigma$ bands from the expected exclusion limits. Since in this search the data show
only a very mild (below $1\,\sigma$) excess, our estimate is expected to be fairly accurate. We also checked
that this procedure is correct by using the CMS $8$~TeV data, in which case we find that the reconstructed likelihood
is very close to the one obtained with the other method we used.

Once the likelihood ratios for the various searches are reconstructed, it is straightforward to use them to extract the
best fit of the signal strength $\hat \mu$ and the combined signal significance, {\it i.e.}~the p-value of the background-only
hypothesis. Another interesting quantity that can be computed is the compatibility among the various searches, also
known as the ``goodness'' of the fit~\cite{Maltoni:2003cu}. To extract this quantity one compares the likelihood
for the best fit of the cross section with the one obtained by assuming independent signal strengths for each experimental search.
The resulting likelihood follows a $\chi^2$ distribution with a number of degrees of freedom equal to the number of experiments
minus one.

As an application of our recast procedure we show in fig.~\ref{fig:pvalue_mass} the statistical significance of the signal
for the scenario with a
narrow scalar resonance produced in the $u\overline u$, $gg$ and $b\overline b$ channels. In the plots the p-values
for the single searches are shown as a function of the resonance mass, together with the result for the
combination of the $13$~TeV searches only and the full combination of the $8$ and $13$~TeV data. One can see that
the significance of the full combination for $M\simeq 750$~GeV
is quite close to the one of the $13$~TeV only searches if the resonance
is produced in channels with a large cross section gain between $8$ and $13$~TeV, namely the $gg$ and $b\overline b$
modes. This shows that in these scenarios the agreement between the $8$~TeV and the $13$~TeV data is reasonably good.
On the contrary, in the $u\overline u$ case, the p-value for the full combination is significantly smaller than the one
for the $13$~TeV searches only, implying a sizable degree of tension among the experimental searches. These results
confirm what we found in section~\ref{three}.

\begin{figure}[t]
\centering
\includegraphics[width=.46\textwidth]{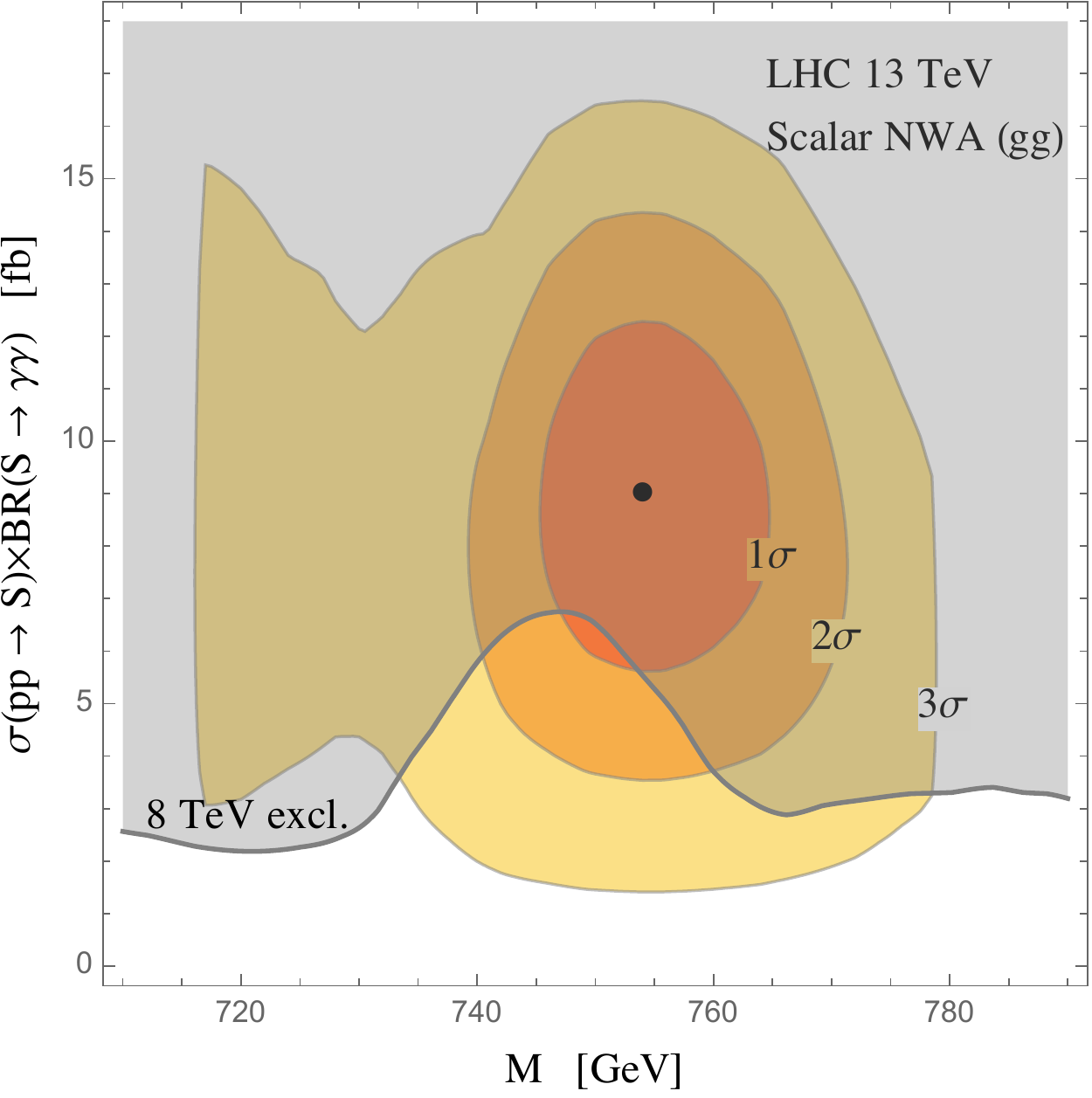}
\hfill
\includegraphics[width=.462\textwidth]{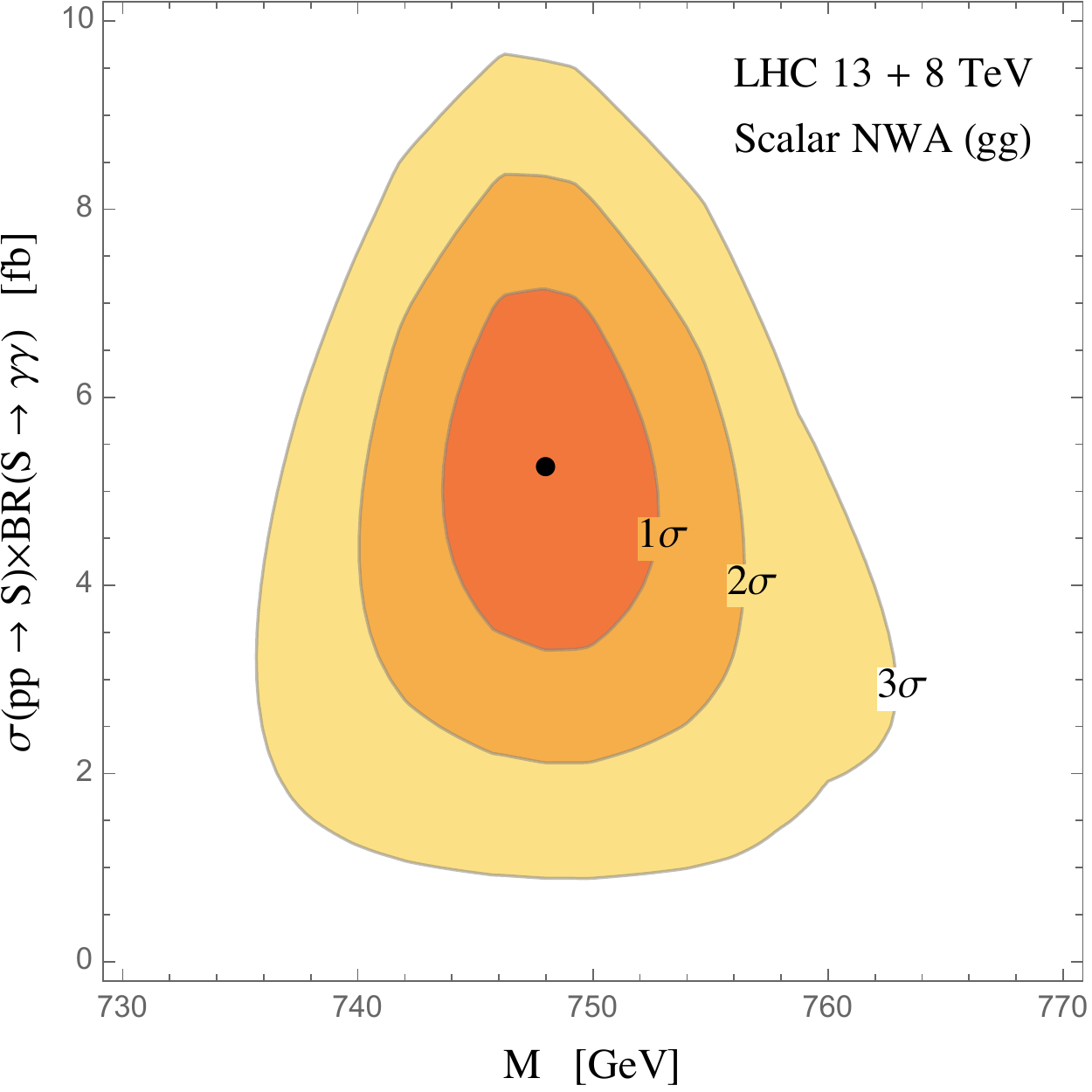}
\caption{Fit of the signal cross section for the scenario with a narrow scalar resonance produced in the $gg$ channel.
The left plot shows the fit obtained by combining only the $13$~TeV searches, with the $8$~TeV bound overlapped
as a shaded area. The right plot shown the fit from the full combination of the $8$ and $13$~TeV searches.
}\label{fig:xsection_mass_gg}
\end{figure}

\begin{figure}[t]
\centering
\includegraphics[width=.46\textwidth]{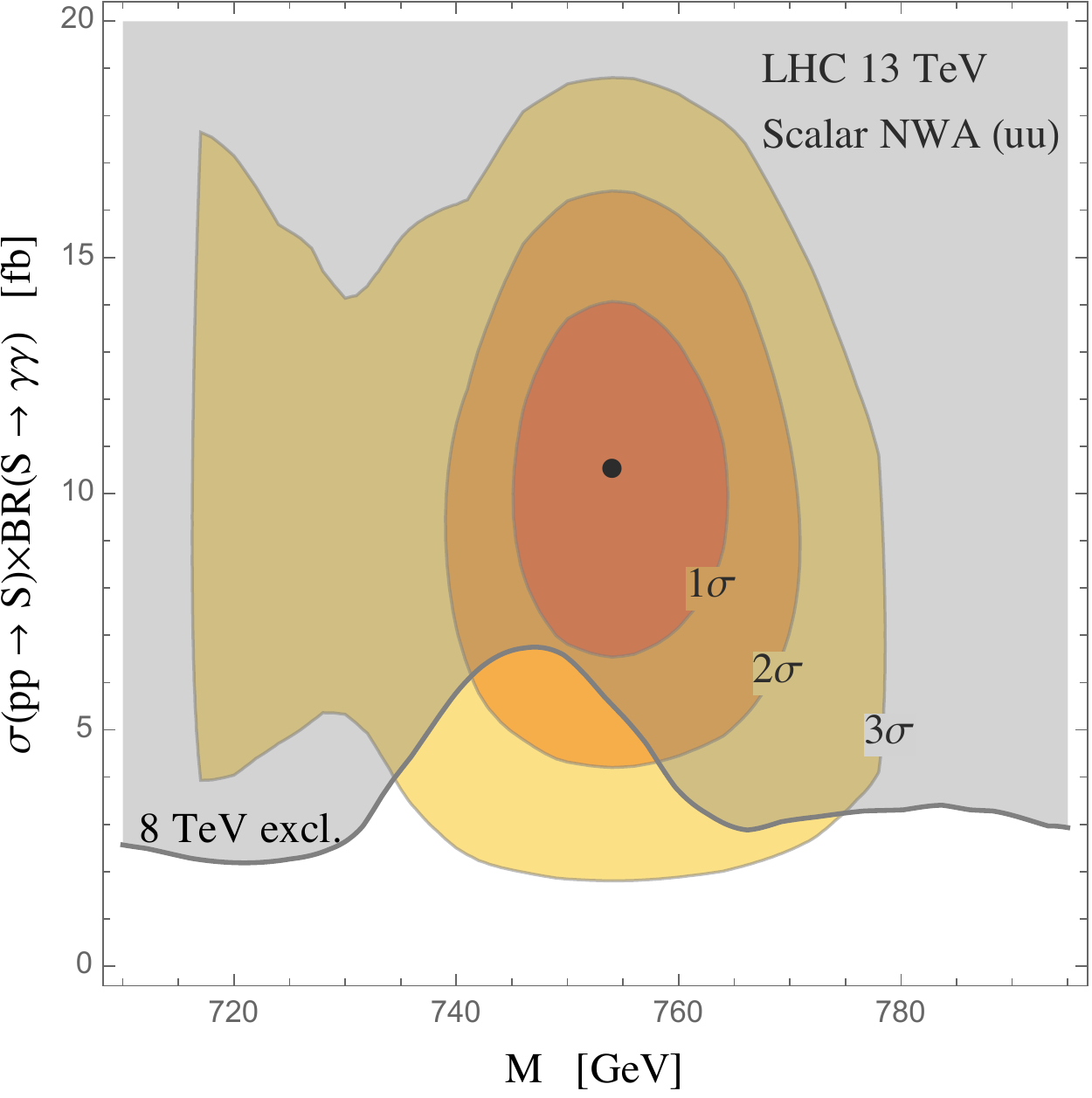}
\hfill
\includegraphics[width=.454\textwidth]{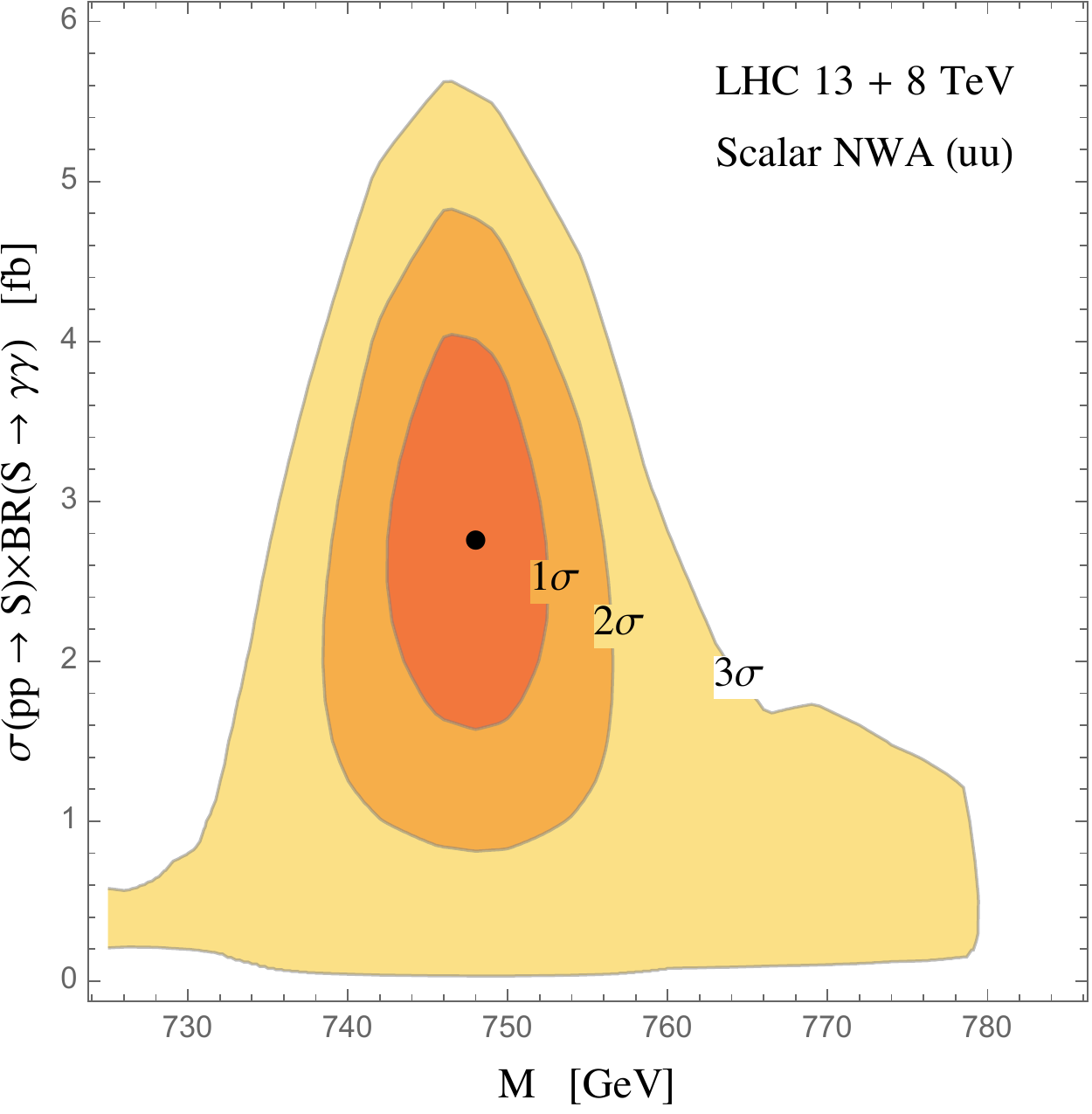}
\caption{Fit of the signal cross section for the scenario with a narrow scalar resonance produced in the $u\overline u$ channel.
The left plot shows the fit obtained by combining only the $13$~TeV searches, with the $8$~TeV bound overlapped
as a shaded area. The right plot shown the fit from the full combination of the $8$ and $13$~TeV searches.
}\label{fig:xsection_mass_uu}
\end{figure}

Finally in figs.~\ref{fig:xsection_mass_gg} and \ref{fig:xsection_mass_uu}, we provide the fit of the signal for the scenario with
narrow scalar resonance produced in the $gg$ and $u\overline u$ channels. The results are presented as a function of the
mass of the resonance.\footnote{See for instance refs.~\cite{bl} for other works resenting a combination of the experimental
results and a fit of the signal cross section and significance in the scalar resonance scenario.}

\subsection{Impact of the CMS $13$~TeV categories}\label{sec:categories}

As we saw in the section~\ref{three}, the impact of the angular distribution on the various searches can be significant,
even if we only consider the total number of events without explicitly looking at the distributions. The reason for this
dependence is the fact that relatively hard cuts are imposed on the signal, with the aim of selecting events in which the
final-state photons are central. As a consequence angular distributions that enhance the signal in the central region of the
detector have larger acceptances than the ones that give rise to a more forward signal.

More details on the angular distribution
can in principle be obtained by looking at the different signal categories used in the experimental analyses. In particular the
CMS $13$~TeV study splits the events in two categories: the EBEB in which both photons are in the barrel of the detector
($|\eta| < 1.44$) and EBEE in which one photon is in the barrel while the second is in the endcap
($|\eta| \in [1.57, 2.5]$).\footnote{The CMS $8$~TeV analysis also considers $4$ categories separating events
in which the photons are in the barrel and in the endcap. However, the distribution of the events in each category is not
provided in the experimental paper, so that we can not fully recast the analysis as we are dong for the $13$~TeV case.}
The two categories allow to get a rough information on the angular distribution, thus improving the discrimination
power for the different benchmark models.
Unfortunately, the procedure we described in the previous section to recast the experimental analyses
did not allow us to take into account
separately the different categories. We now want to estimate how drastic this simplification is and how much a full analysis
could help in discriminating the angular distribution of the diphoton signal.

For this purpose, here we implement a simple recast of the CMS $13$~TeV search.\footnote{For a similar recast applied to the
ALTAS $13$~TeV results see ref.\cite{binned}.}
We reconstruct the likelihoods associated to the two event categories
by using the distributions of events provided in fig.~3 of ref.~\cite{CMS:2015dxe}. We assume that in each bin the events
follow a Poisson distribution and we construct the total likelihood by multiplying the likelihoods for each bin.
We model the background using the functional form given in the experimental paper
\begin{equation}
f_0(m_{\gamma\gamma}) = {\cal N} e^{- p_1 m_{\gamma\gamma}} m_{\gamma\gamma}^{-p_2}\,,
\end{equation}
where $p_{1,2}$ are free parameters and ${\cal N}$ is the overall normalization which we fit together with the other parameters.
For simplicity we only focus on the
narrow resonance scenario, and we model the signal by a Gaussian distribution with a half width equal to the experimental
resolution ($10~\textrm{GeV}$ for the EBEB category and $16~\textrm{GeV}$ for the EBEE category). The signal and the
background are fitted simultaneously for each signal strength hypothesis. The test statistics we use is based
on the profile likelihood ratio and the background-only p-value is computed by assuming that the distribution is asymptotically
equal to a half-$\chi^2$ distribution with one degree of freedom.

\begin{figure}[t]
\centering
\includegraphics[width=.6\textwidth]{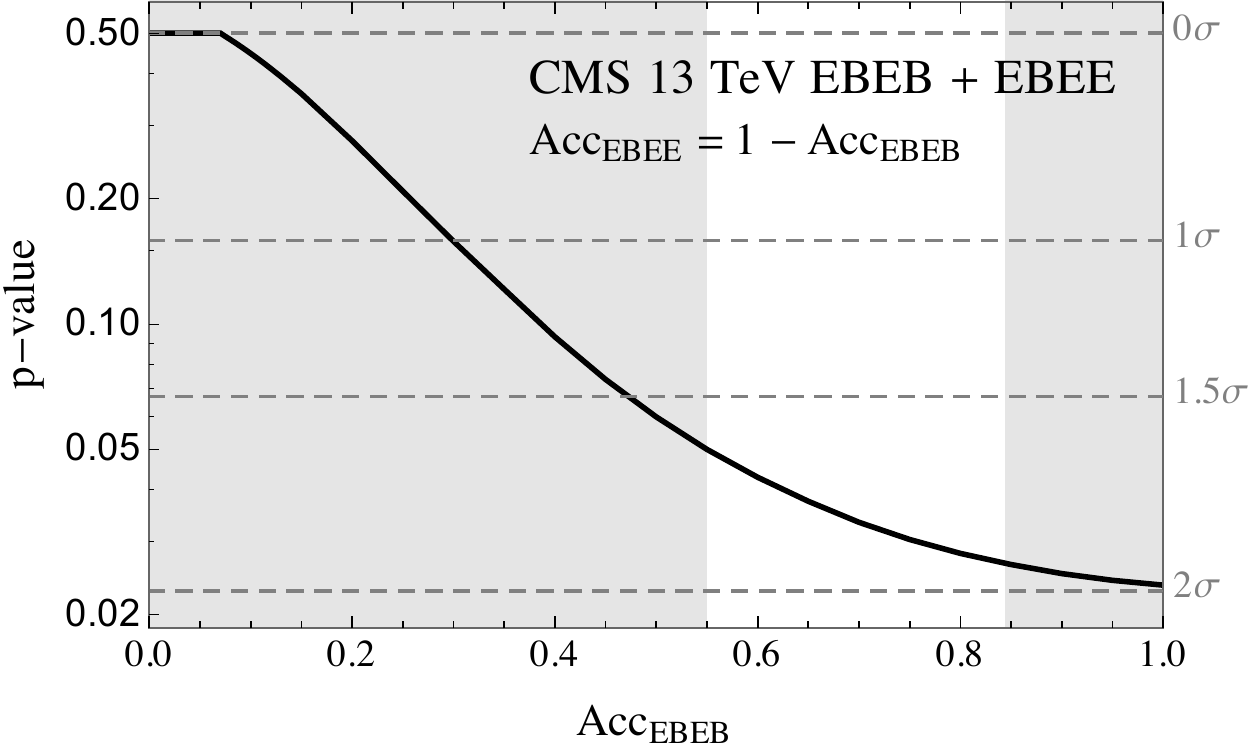}
\caption{Reconstructed p-value for the background-only hypothesis obtained by exploiting the EBEB and EBEE event
categories of the CMS $13$~TeV analysis. The results are shown as a function of the relative signal acceptance in
the EBEB category (${\rm Acc}_{\rm EBEB}$), while the acceptance in the EBEE category is ${\rm Acc}_{\rm EBEE} = 1 - {\rm Acc}_{\rm EBEB}$.
The unshaded region denotes values of ${\rm Acc}_{\rm EBEB}$ that can be obtained in the various benchmark models we discussed
in the main text.}\label{fig:BBvsBE}
\end{figure}

The result of our recast is shown in Fig.~\ref{fig:BBvsBE}, where we plot the significance of the signal as a function of the
acceptance in the EBEB category ${\rm Acc}_{\rm EBEB}$ under the assumption that the total acceptance is equal to $1$.
One can see that the statistical significance of the signal has a non-negligible dependence on the angular distribution.
In particular the signal is mostly present in the EBEB category, so that models with a more central signal distribution
are preferred. The present experimental sensitivity, however, is not very large, so that the impact on the fit in
the benchmark models we considered is mild. This justifies our approximation of combining the
two CMS $13$~TeV categories.\footnote{Notice that the signal significance
we find in our recast is always lower than the one found by the CMS collaboration. This can be explained by the fact that
CMS made a fit by using the unbinned event distributions, whereas we only have access to the binned ones.}



\begin{thebibliography}{99}

\bibitem{Dawson:1984gx}
  See for instance: S.~Dawson,
  Nucl.\ Phys.\ B {\bf 249} (1985) 42.

\bibitem{Harland-Lang:2016qjy}
  L.~A.~Harland-Lang, V.~A.~Khoze and M.~G.~Ryskin,
  arXiv:1601.07187 [hep-ph].
 
  \bibitem{Martin:2014nqa}
  A.~D.~Martin and M.~G.~Ryskin,
  Eur.\ Phys.\ J.\ C {\bf 74} (2014) 3040
  [arXiv:1406.2118 [hep-ph]].

\bibitem{Ball:2013hta} 
  R.~D.~Ball {\it et al.} [NNPDF Collaboration],
  Nucl.\ Phys.\ B {\bf 877}, 290 (2013)
  [arXiv:1308.0598 [hep-ph]].
  
\bibitem{Trueman:1978kh} 
  T.~L.~Trueman,
  Phys.\ Rev.\ D {\bf 18}, 3423 (1978).
  
\bibitem{Dell'Aquila:1985ve} 
  J.~R.~Dell'Aquila and C.~A.~Nelson,
  Phys.\ Rev.\ D {\bf 33}, 80 (1986).
  
   \bibitem{Choi:2002jk}
  S.~Y.~Choi, D.~J.~Miller, M.~M.~Muhlleitner and P.~M.~Zerwas,
  Phys.\ Lett.\ B {\bf 553} (2003) 61
  [hep-ph/0210077].
  
\bibitem{Gao:2010qx}
  Y.~Gao, A.~V.~Gritsan, Z.~Guo, K.~Melnikov, M.~Schulze and N.~V.~Tran,
  Phys.\ Rev.\ D {\bf 81} (2010) 075022
  [arXiv:1001.3396 [hep-ph]].
  \\
  S.~Bolognesi, Y.~Gao, A.~V.~Gritsan, K.~Melnikov, M.~Schulze, N.~V.~Tran and A.~Whitbeck,
  Phys.\ Rev.\ D {\bf 86}, 095031 (2012)
  doi:10.1103/PhysRevD.86.095031
  [arXiv:1208.4018 [hep-ph]].


\bibitem{Khachatryan:2014kca}
  V.~Khachatryan {\it et al.} [CMS Collaboration],
  Phys.\ Rev.\ D {\bf 92} (2015) 1,  012004
  [arXiv:1411.3441 [hep-ex]].
    D.~J.~Miller, S.~Y.~Choi, B.~Eberle, M.~M.~Muhlleitner and P.~M.~Zerwas,
    Phys.\ Lett.\ B {\bf 505} (2001) 149
    [hep-ph/0102023].
    C.~P.~Buszello, I.~Fleck, P.~Marquard and J.~J.~van der Bij,
    Eur.\ Phys.\ J.\ C {\bf 32} (2004) 209
    [hep-ph/0212396].
    
\bibitem{Choi:2012yg} 
  S.~Y.~Choi, M.~M.~Muhlleitner and P.~M.~Zerwas,
  Phys.\ Lett.\ B {\bf 718}, 1031 (2013)
  doi:10.1016/j.physletb.2012.11.050
  [arXiv:1209.5268 [hep-ph]].

\bibitem{wiki}
A few Wigner functions are listed on {\sc{WikipediA}} at this link (references to more exhaustive collections can be found therein):\\
\href{https://en.wikipedia.org/wiki/Wigner_D-matrix}{\color{blue}{https://en.wikipedia.org/wiki/Wigner\_D-matrix}}.







\bibitem{Alwall:2014hca}
  J.~Alwall {\it et al.},
  JHEP {\bf 1407} (2014) 079
  [arXiv:1405.0301 [hep-ph]].
 
\bibitem{Harland-Lang:2016apc}
  L.~A.~Harland-Lang, V.~A.~Khoze and M.~G.~Ryskin,
  arXiv:1601.03772 [hep-ph].
 
\bibitem{ATLAS13}
  The ATLAS collaboration,
  ``Search for resonances decaying to photon pairs in 3.2 fb$^{-1}$ of $pp$ collisions at $\sqrt{s}$ = 13 TeV with the ATLAS detector,''
  ATLAS-CONF-2015-081.

\bibitem{CMS:2015dxe}
  CMS Collaboration [CMS Collaboration],
  ``Search for new physics in high mass diphoton events in proton-proton collisions at 13TeV,''
  CMS-PAS-EXO-15-004.
   
\bibitem{Aparicio:2016iwr}
    L.~Aparicio, A.~Azatov, E.~Hardy and A.~Romanino,
    arXiv:1602.00949 [hep-ph].
    
\bibitem{others}
  P.~Agrawal, J.~Fan, B.~Heidenreich, M.~Reece and M.~Strassler,
  arXiv:1512.05775 [hep-ph].
  J.~Chang, K.~Cheung and C.~T.~Lu,
  arXiv:1512.06671 [hep-ph].
  M.~Chala, M.~Duerr, F.~Kahlhoefer and K.~Schmidt-Hoberg,
  Phys.\ Lett.\ B {\bf 755} (2016) 145
  [arXiv:1512.06833 [hep-ph]].
  X.~J.~Bi {\it et al.},
  arXiv:1512.08497 [hep-ph].    
  
\bibitem{Aad:2015mna}
  G.~Aad {\it et al.} [ATLAS Collaboration],
  Phys.\ Rev.\ D {\bf 92} (2015) 3,  032004
  [arXiv:1504.05511 [hep-ex]].
  
\bibitem{Khachatryan:2015qba}
  V.~Khachatryan {\it et al.} [CMS Collaboration],
  Phys.\ Lett.\ B {\bf 750} (2015) 494
  [arXiv:1506.02301 [hep-ex]].       
  
\bibitem{photon-fusion}  
  S.~Fichet, G.~von Gersdorff and C.~Royon,
  arXiv:1512.05751 [hep-ph].
  C.~Csaki, J.~Hubisz and J.~Terning,
  Phys.\ Rev.\ D {\bf 93} (2016) 3,  035002
  [arXiv:1512.05776 [hep-ph]].
  C.~Csaki, J.~Hubisz, S.~Lombardo and J.~Terning,
  arXiv:1601.00638 [hep-ph].
  S.~Abel and V.~V.~Khoze,
  arXiv:1601.07167 [hep-ph].
  L.~A.~Harland-Lang, V.~A.~Khoze and M.~G.~Ryskin,
  arXiv:1601.07187 [hep-ph]. 
  
  
  \bibitem{spin2}   
  C.~Han, H.~M.~Lee, M.~Park and V.~Sanz,
  arXiv:1512.06376 [hep-ph].
  M.~R.~Buckley,
  arXiv:1601.04751 [hep-ph].
  A.~Martini, K.~Mawatari and D.~Sengupta,
  arXiv:1601.05729 [hep-ph].
  C.~Q.~Geng and D.~Huang,
  arXiv:1601.07385 [hep-ph].
  S.~B.~Giddings and H.~Zhang,
  arXiv:1602.02793 [hep-ph].
  J.~Bernon, A.~Goudelis, S.~Kraml, K.~Mawatari and D.~Sengupta,
  arXiv:1603.03421 [hep-ph].
  
 \bibitem{Kaidalov:2003fw}
   A.~B.~Kaidalov, V.~A.~Khoze, A.~D.~Martin and M.~G.~Ryskin,
   Eur.\ Phys.\ J.\ C {\bf 31} (2003) 387
   [hep-ph/0307064].
   

   
   
    
\bibitem{Borel:2012by}
 P.~Borel, R.~Franceschini, R.~Rattazzi and A.~Wulzer,
  JHEP {\bf 1206} (2012) 122
  [arXiv:1202.1904 [hep-ph]].
  
 \bibitem{WeinbergBook}
   S.~Weinberg,
   ``The Quantum theory of fields. Vol. 1: Foundations''.
 
\bibitem{Hagiwara:2008jb} 
  K.~Hagiwara, J.~Kanzaki, Q.~Li and K.~Mawatari,
  Eur.\ Phys.\ J.\ C {\bf 56}, 435 (2008)
  [arXiv:0805.2554 [hep-ph]].
  
  \bibitem{Peskin} 
  M.~E.~Peskin and D.~V.~Schroeder,
    ``An Introduction to quantum field theory''.
  
\bibitem{Cowan:2010js}
  G.~Cowan, K.~Cranmer, E.~Gross and O.~Vitells,
  Eur.\ Phys.\ J.\ C {\bf 71} (2011) 1554
  [arXiv:1007.1727 [physics.data-an]].

\bibitem{Maltoni:2003cu}
  M.~Maltoni and T.~Schwetz,
  Phys.\ Rev.\ D {\bf 68} (2003) 033020
  [hep-ph/0304176]. 

\bibitem{bl}
  R.~Franceschini {\it et al.},
    arXiv:1512.049
    33 [hep-ph].
  R.~S.~Gupta, S.~Jager, Y.~Kats, G.~Perez and E.~Stamou,
    arXiv:1512.05332 [hep-ph].
J.~Ellis, S.~A.~R.~Ellis, J.~Quevillon, V.~Sanz and T.~You,
  arXiv:1512.05327 [hep-ph].
  A.~Falkowski, O.~Slone and T.~Volansky,
  JHEP {\bf 1602} (2016) 152
  [arXiv:1512.05777 [hep-ph]].
  J.~S.~Kim, K.~Rolbiecki and R.~R.~de Austri,
  arXiv:1512.06797 [hep-ph].
  M.~R.~Buckley,
  arXiv:1601.04751 [hep-ph].
  
 \bibitem{binned}
J.~H.~Davis, M.~Fairbairn, J.~Heal and P.~Tunney,
  arXiv:1601.03153 [hep-ph].
  B.~J.~Kavanagh,
  arXiv:1601.07330 [hep-ph].



  
\end{thebibliography}
\end{document}